\documentclass[fleqn,usenatbib]{mnras}
\usepackage{newtxtext,mathptmx}
\usepackage[T1]{fontenc}
\DeclareRobustCommand{\VAN}[3]{#2}
\let\VANthebibliography\thebibliography
\def\thebibliography{\DeclareRobustCommand{\VAN}[3]{##3}\VANthebibliography}
\usepackage{amsmath}	% Advanced maths commands

\usepackage{graphicx} % Required for inserting images
\usepackage{txfonts}

\usepackage{mathtools}
\usepackage{empheq}
\usepackage[separate-uncertainty=true,detect-weight,retain-explicit-plus]{siunitx}
\usepackage[]{hyperref}
\usepackage{cleveref}
\usepackage{caption}
\usepackage{subcaption}
\usepackage{cuted}
\usepackage{microtype}
\usepackage{booktabs}
\usepackage{threeparttable}
\usepackage{enumerate}
\usepackage{multicol}
\usepackage{multirow}
\usepackage{array}
\usepackage{mdwtab}
\usepackage{comment}
\usepackage{longtable}
\usepackage{threeparttablex}
\usepackage{xcolor}
\usepackage{url}
\usepackage[normalem]{ulem} % To strike out text
\usepackage{tikz}

\newcommand{\LGalaxies}{\texttt{L-Galaxies}\,}
\newcommand{\Ha}{${\rm H{\alpha}}\,$}
\newcommand{\Hb}{${\rm H{\beta}}\,$}
\newcommand{\oii}{{\rm [\ion{O}{II}]}}
\newcommand{\oi}{{\rm [\ion{O}{I}]}}
\newcommand{\oiii}{{\rm [\ion{O}{III}]}}

\newcommand{\HeI}{{\rm [\ion{He}{I}]}}
\newcommand{\FeII}{{\rm [\ion{Fe}{II}]}}
\newcommand{\CIII}{{\rm [\ion{C}{III}]}}
\newcommand{\MgII}{{\rm [\ion{Mg}{II}]}}
\newcommand{\CIV}{{\rm [\ion{C}{IV}]}}
\newcommand{\SiIV}{{\rm [\ion{Si}{IV}]}}
\newcommand{\NV}{{\rm [\ion{N}{V}]}}
\newcommand{\lrdAGN}{{\tt LRD\,AGNs}}
\newcommand{\lrdGal}{{\tt LRD\,Gals}}
\newcommand{\nolrdAGN}{{\tt non-LRD\,AGNs}}
\newcommand{\lrd}{{\tt LRD}}
\newcommand{\lrds}{{\tt LRDs}}
\newcommand{\nolrd}{{\tt non-LRD}}
\newcommand{\nolrds}{{\tt non-LRDs}}
\newcommand{\msun}{\,\mathrm{M_\odot}}
\newcommand{\lgbh}{{\texttt{L-Galaxies}\textit{BH}}}

\usepackage{color}
\usepackage[dvipsnames]{xcolor}
\definecolor{myorange}{rgb}{0.8, 0.3, 0.0}
\newcommand{\davcoment}[1]{{\color{myorange}{\bf David:}[#1]}}
\newcommand{\diegcomment}[1]{{\color{magenta} #1}}
\definecolor{mygreen}{rgb}{0.1, 0.55, 0.25}

\graphicspath{{./figs}}

\hypersetup{draft=false,
  colorlinks=true,
  allcolors=blue,
  pagebackref=true,
  bookmarks=true,
}

\title[Simulating LRDs with \lgbh{}]{Back to basics: Little Red Dots as galaxies and dust-obscured AGNs in a synthetic NIRCam sky simulated with \lgbh{}}
\author[D. Herrero-Carrión et al.] {Diego Herrero-Carrión,$^{1,2}$\thanks{diego.herrero@dipc.org} 
Daniele Spinoso,$^{3}$
David Izquierdo-Villalba,$^{4,5}$
\newauthor Tong Su,$^{6,7,8,9}$
Silvia Bonoli,$^{1,10}$
and Pablo Renard$^{11,12}$\\
$^{1}$ Donostia International Physics Centre (DIPC), Paseo Manuel de Lardizabal 4, 20018 Donostia-San Sebastian, Spain\\
$^{2}$ University of the Basque Country UPV/EHU, Department of Physics, Barrio Sarriena s/n. 48940 Leioa, Bizkaia, Spain\\
$^{3}$ Como Lake Center for AstroPhysics,  University of Insubria, 22100, Como, Italy\\
$^{4}$ Dipartimento di Fisica ``G. Occhialini'', Universit\`{a} degli Studi di Milano-Bicocca, Piazza della Scienza 3, I-20126 Milano, Italy\\
$^{5}$ INFN, Sezione di Milano-Bicocca, Piazza della Scienza 3, 20126 Milano, Italy\\
$^{6}$ Key Laboratory for Computational Astrophysics, National Astronomical Observatories, Chinese Academy of Sciences, Beijing 100012, China\\
$^{7}$ School of Astronomy and Space Sciences, University of Chinese Academy of Sciences, 19A Yuquan Road, Beijing 100049, China\\
$^{8}$ Institute for Frontiers in Astronomy and Astrophysics, Beijing Normal University, Beijing 102206, People’s Republic of China\\
$^{9}$ School of Physics and Astronomy, Beijing Normal University, Beijing 100875, People’s Republic of China\\
$^{10}$ IKERBASQUE, Basque Foundation for Science, E-48013, Bilbao, Spain\\ 
$^{11}$ Institute of Space Sciences (ICE, CSIC), Campus UAB, Carrer de Can Magrans, s/n, 08193 Barcelona, Spain \\
$^{12}$ Institut d’Estudis Espacials de Catalunya (IEEC), E-08860 Castelldefels (Barcelona), Spain
}

\date{Accepted XXX. Received YYY; in original form ZZZ}
\pubyear{XXX}

\begin{document}
\label{firstpage}
\pagerange{\pageref{firstpage}--\pageref{lastpage}}
\maketitle

\begin{abstract}
\noindent
The enigmatic Little Red Dots (LRDs) discovered by the James Webb Space Telescope (JWST) exhibit properties challenging their interpretation as common galaxies or Active Galactic Nuclei (AGN). Understanding their nature is key to placing them within our picture of early galaxy and massive black hole (MBH) evolution. To this aim, we build a realistic comparison between LRD observations with photometric properties of galaxies and AGN simulated by the L-GalaxiesBH model in a NIRCam mock sky. We model stellar continua and emission lines, the MBH emission from accretion disk, infrared radiation from dusty torus, and lines from narrow and broad line regions, accounting for dust attenuation and obscuration. Using realistic photometric cuts, we select a population of LRDs including both AGN and galaxies. The LRD fraction peaks at 40\% ($\sim10^{-4}\rm Mpc^{-3}$) at $z\sim4$. Our LRDs are central galaxies spanning $M_\star=10^8-10^{10.5}\rm M_\odot$. A population of galaxies with $M_\star<10^9\rm M_\odot$ appear as LRDs due to older stellar populations. At higher masses, LRDs dominate the halo and stellar mass functions ($M_{\rm vir} > 10^{11.5}\rm M_\odot$, $M_\star > 10^{9.5}\rm M_\odot$), and the interplay between AGN and galaxy emission drives the LRD selection. AGN dominate rest-frame UV–optical emission, while dust obscuration is secondary. LRDs host lighter MBHs ($\sim 10^{6.5}\rm M_\odot$) than non-LRDs ($\sim 10^{7.5}\rm M_\odot$), with fainter emission unable to balance their hosts Balmer breaks. We find no evidence for dominant heavy-seed origin of MBHs. LRD Galaxies (97\% hosting MBHs) and LRD AGNs are disk-dominated, with LRD AGNs showing larger bulges formed mainly via disk instabilities.

\end{abstract}


%\maketitle

%\tableofcontents

%---------------------------------------------------------------
%---------------------------------------------------------------
\section{Introduction}
\label{sec:intro}

%Since its earliest data releases, the James Webb Space Telescope (JWST, \citealt{JWSTMission2023}) has spectacularly transformed our view of the infrared (IR) Universe, largely thanks to its Near Infrared Camera and Spectrograph \citep[NIRCam and NIRSpec respectively, see e.g.][]{rieke2023_nircam_performance,jakobsen2022_nirspec_performance} and its Mid-Infrared Instrument \citep[MIRI, see][]{wright2023_miri_performance}

The James Webb Space Telescope (JWST, \citealt{JWSTMission2023}) has spectacularly transformed our view of the infrared (IR) Universe. % since its earliest data releases. 
This has been possible largely thanks to its Near Infrared Camera and Spectrograph \citep[NIRCam and NIRSpec respectively,][]{rieke2023_nircam_performance,jakobsen2022_nirspec_performance} and its Mid-Infrared Instrument \citep[MIRI, see][]{wright2023_miri_performance}. From detailed studies of local galaxies \citep[e.g.][]{haidar2024,Bortolini2024,lopez-rodriguez2025}, to investigations of Active Galactic Nuclei (AGN) and galaxies at cosmic noon ($z\,{\sim}\,2$, \citealt{vanDokkum2024,Setton2024,Davies2024,Liboni2025}), and extending to the most distant galaxies observed to date \citep[][see also \citealt{stark2025} for a recent review]{morishita2024,Chemerynska2025,Naidu2025}, JWST is transforming our understanding of galaxy and AGN evolution. %From the observations of local galaxies \citep[e.g.][]{haidar2024,Bortolini2024,lopez-rodriguez2025}, Active Galactic Nuclei (AGN) and galaxies at cosmic noon ($z\,{\sim}\,2$, \citealt{vanDokkum2024,Setton2024,Davies2024,Liboni2025}), up to the farthest galaxies known to date \citep[][and \citealt{stark2025} for a recent review]{morishita2024,Chemerynska2025,Naidu2025}, the JWST is revolutionizing our understanding of galaxy and AGN evolution.

%--- general properties
Among its most significant discoveries, JWST has unveiled the so-called Little Red Dots \citep[LRDs,][]{matthee2024_lrds}, a remarkably common class of high-$z$ sources with point-like apparent morphology and red observed photometric colors \citep[][]{matthee2024_lrds,kokorev24_lrds,Akins2025}. Owing to a combination of peculiar spectro-photometric properties, LRDs have challenged our understanding of early black hole and galaxy co-evolution since their early discoveries \citep[][]{labbe2023_lrds,matthee2024_lrds,greene2024_lrds,Akins2025}. Generally speaking, LRDs appear as point-like, UV-faint sources \citep[$\rm -14\, <\, M_{UV}\, <\, -18$, see][]{labbe2023_lrds} with photometric redshifts spanning $3\,{<}\,z\,{<}\,9$ \citep{kokorev24_lrds,Akins2025}, and some detections and/or candidates extending to even higher redshifts \citep{Taylor2025, tanaka2025a}. Their most distinctive photometric signature is a ``V-shaped'' spectral energy distribution (SED), characterized by simultaneously blue rest-frame UV–optical colors ($\rm 0.15\,{-}\,0.4\mu m$) and red rest-frame optical–IR colors ($\rm0.4\,{-}\,1\mu m$) \citep[see e.g.][]{Kocevski2025}, a combination that was rarely identified before JWST. %Photometric redshifts of LRDs show a wide distribution \citep[$3\,{<}\,z\,{<}\,9$,][]{kokorev24_lrds,Akins2025}, possibly extending up to even higher redshifts \citep{Taylor2025}. %the highest redshifts currently observed \citep{Taylor2025}. %Nevertheless, their most noticeable photometric feature is likely a ``V-shaped'' spectral energy distribution (SED). Indeed, this shows at the same time blue rest-frame UV-optical colors ($\rm 0.15\,{-}\,0.4\mu m$) and red rest-frame optical-IR colors ($\rm0.4\,{-}\,1\mu m$) \citep[see e.g][]{Kocevski2025}, a feature only rarely observed before the discovery of LRDs.% \textbf{papers about LRD-type objects before JWST}.

%--- physical properties inferred from photometry of LRDs
SED fitting of LRDs photometric data provides stellar mass estimates of $10^9\,{\lesssim}\,M_\star/\msun\,{\lesssim}\,10^{11}$ (\citealt{PerezGonzalez2024,Leung2024,Akins2025,Chen2025}, although these values may be affected by biases in the subtraction of AGN light or stellar age inference, see \citealt{Berger2025,Harvey2025}) and strong Balmer breaks \citep[$\rm{\gtrsim}\,1.5\,dex$][]{baggen2024_lrdsarenotagn}. At the same time, the very compact sizes inferred by their photometric light profiles \citep[$\rm{\sim}\,10-100\,pc$, see][]{baggen2024_lrdsarenotagn,furtak2024_lrd} imply surprising compactness and densities for galaxies of these masses. In addition, if the photometric ``V-shape'' is due to stellar continua, then LRDs may harbor evolved (i.e. ``red'') stellar populations \citep{Setton20245_Vshape}. This appears in contrast with their blue rest-frame UV-optical colors, which suggest young (i.e. ``blue'') stellar populations. Reconciling these two aspects requires that at least a fraction of the light emitted by young stellar populations is absorbed and reprocessed (into the IR) by dust in the interstellar medium (ISM) \citep[see e.g][]{perezgonzalez2024_lrdsarenotagn}. Nevertheless, the large dust masses needed to support this picture are at stake with the reported lack of FIR detections for LRDs \citep{Setton2025,Casey2025,Xiao2025}. %\textbf{ALMA REFs}.

%--- physical properties inferred from spectroscopy
%Interestingly, spectroscopic observations add further complexity to this picture, with a large fraction of LRDs reportedly showing broad Balmer $\rm H\alpha$ and $\rm H\beta$ lines \citep[$\rm FWHM \,{\sim}\,10^3\,{-}\,10^4\,km/s$, see e.g.][]{Larson2023,Kokorev2023,Lin2024,Akins2025}.
Interestingly, spectroscopic observations add further complexity to this picture, as a significant fraction of LRDs exhibit broad Balmer $\rm H\alpha$ and $\rm H\beta$ emission lines, with typical full widths at half maximum (FWHM) of ${\sim}\,10^3\, km/s$ \citep[e.g.,][]{Larson2023, Kokorev2023, Lin2024, Akins2025}. Recent works have tentatively explained these line widths as an effect of galactic dynamics within ultra-compact, star-forming galaxies \citep[][]{baggen2024_lrdsarenotagn}. Alternatively, by interpreting broad lines as indicative of AGN activity, numerous works inferred the mass and luminosity of MBHs responsible for their emission. Currently reported values lie in the $10^7\,{\lesssim}\,M_{\rm BH}/\msun\,{\lesssim}\,10^{9}$ and $10^{44}\,{\lesssim}\,L^{\rm AGN}_{\rm bol}/{\rm erg}{\rm / s}\,{\lesssim}\,10^{46}$ ranges, typical of a mature AGN population \citep[][]{Kokorev2023,matthee2024_lrds,harikane2023_highzgals,Greene2025,Taylor2025,Tripodi2024}. However, LRDs do not show significant variability in the rest-frame UV-optical, as would be common for typical AGN of similar luminosity and mass. More in detail, the variability time-scales inferred for LRDs appear to be larger than ${\sim}\,1\, \rm yr$ \citep{Kokubo2024,Tee2025,Furtak2025}. %\textbf{REFs -> check; Kokubo \& Harikane 2024; Zhang et al. 2024; Tee et al. 2024 Zhang et al. 2024; Ji et al. 2025; Furtak et al.2025)}. %--- observational interpretation of LRDs down to this moment
Furthermore, AGN with masses and luminosity comparable to those reported for LRDs would typically appear as X-ray bright sources \citep{Weedman2012,Gupta2024,Laurenti2024}. Nevertheless, LRDs have so far eluded X-ray detections. This fact, together with their measured low variability, has fueled speculation that LRDs may be MBHs accreting at super-Eddington rates \citep{Pacucci2024,kido2025,inayoshi2025b,Madau2025}. On the other hand, the X-ray weakness may be attributed to significant dust obscuration of the central source, which could also help explain the red optical–IR colors observed in LRDs \citep[][]{InayoshiAndIchikawa2024,Yue2024,Pacucci2024_Xrays}. Nevertheless, this picture is again in contrast with the observed blue spectra at UV-optical wavelengths as well as with the lack of strong FIR detections \citep{Setton2025,Casey2025,Xiao2025}. In addition, spectroscopic observations of LRDs showing only narrow lines have also been reported, suggesting that the origin of red photometric colors may be due to the contribution of strong galactic emission lines \citep[see e.g.][]{zhang2025}.

The difficulty to fit together all these observed features in a single, coherent picture leaves open the possibility that LRDs may represent a complex, multi-face population of objects belonging to a previously unobserved stage of early galaxy formation and galaxy-AGN co-evolution. Therefore, addressing the challenge posed by these objects to current models is an effort which has attracted the keen interest of research communities focusing on galaxy and AGN evolution, both on the observational and theoretical sides. So far, a wide array of possible explanations have been proposed within the rapidly evolving context provided by JWST observations. Early works based on NIRCam and NIRSpec focused on interpreting LRDs properties as produced by either dusty star-forming galaxies \citep[e.g,][]{Gentile2024,Xiao2025}, heavily obscured AGN \citep[e.g,][]{Casey2024,InayoshiAndMaiolino2025} or a combination of both galaxies and AGN \citep[e.g,][]{killy2024,baggen2024_lrdsarenotagn,Leung2024,brooks2025}. In the latter case, the blue colors are typically ascribed to AGN components, either observed directly \citep[thanks to a favorable viewing angle, see e.g.,][]{greene2024_lrds} or reflected by dense gas surrounding the central engine \citep[whose direct view would be obscured by a dense dusty torus, see][]{greene2024_lrds}. As discussed above, each of these interpretations has its advantages and shortcomings, with the combined model possibly joining the best of the galaxy-only and AGN-only explanations. Nevertheless, even the combined galaxy–AGN model struggles to account for key observational features, such as the lack of X-ray and FIR detections or the low UV–optical variability. More recently, models which deviate from standard galaxy/AGN-based explanations have been put forward to address the uncommon observed properties of LRDs. These alternatives generally focus on the interaction between a growing MBH and dense gas surrounding it. Within these systems, the emission due to mass accretion onto the MBH would be heavily absorbed by the surrounding gas, naturally producing some of the observed properties typical of LRDs, such as a prominent Balmer break, X-ray weakness and a lack of significant dust emission  \citep[see e.g][]{kido2025,InayoshiAndMaiolino2025,Rusakov2025}. %\textbf{REFs blackTHUNDER}

%--- modeling efforts of LRDs so far
The impressive amount of observational data with such a difficult interpretation has given a strong impulse to the development of numerical simulations and theoretical models aimed at interpreting LRDs. One common approach is to compare the results of current numerical simulations with the overall constraints provided by JWST regarding the properties of LRDs, such as their global number densities, typical luminosity as well as BH or galaxy masses \citep[e.g.][]{Habouzit2022,Trinca2024,Habouzit2025,LaChance2025,Inayoshi2025a,Bonoli2025}. Alternatively, recent works have tackled the direct simulation of LRD photometric properties, offering a more direct comparison with JWST observations. \cite{volonteri2024_lrdsinobelisk} recently analyzed the simulated photometry of AGN and galaxies extracted from the OBELISK simulation. This work showed that attenuated AGN SEDs can explain the ``V-shaped'' photometric profiles of LRDs, especially in the case of luminous AGN and moderate dust attenuation. This work employs a detailed framework to compute realistic photometric properties of LRDs, based on theoretical models for the emission of AGN and galaxies \citep[see similar results in][by using the ASTRID simulation]{LaChance2025}. On the other hand, recent works have tried to expand these efforts by modeling the photometric counterparts of less standard objects. For instance, \cite{Cenci2025} followed the first phases of gas infall and star formation within halos that produce massive BH-seeds, showing that this process may go through initial phases characterized by a dense, central gas formation with photometric properties similar to those typical of LRDs. On similar lines, the recent work of \cite{inayoshi2025b} explored the possibility that LRDs correspond to the first MBHs forming and accreting within a dense gas obscurer, with the concurring formation of a nuclear stellar cluster, and \cite{zwick2025} picture LRDs as supermassive stars surrounded by massive self-gravitating accretion disks. 
%\textbf{mention Astrid  --> https://arxiv.org/pdf/2505.20439}\\
%\textbf{add more simulations here...}

%--- what is missing in the modeling efforts so far
Overall, the accurate replication of JWST observations is made difficult by several factors. On one side, the inherent uncertainty about LRDs nature requires to push current models to the limit in order to mimic their photometry and inferred physical properties. On the other hand, simulating at the same time the wide dynamic ranges and physical processes involved in the evolution of the first astrophysical objects over large cosmological volumes is out of reach for current hydrodynamical simulations. Therefore, the efforts to replicate JWST observations over a wide cosmological volume while taking into account the details of BH formation and galaxy evolution models have remained largely unexplored to date.

%--- our approach: LGBH, lightcone, addition of physically-motivated AGN SEDs...
Our work tries to tackle this aspect by building a detailed simulation of the photometric properties of large LRDs populations at different redshifts. To this aim, we use the \lgbh{} galaxy-formation model, which was specifically developed to track in detail the formation and evolution of MBHs within a refined galaxy evolution model, applied to N-body merger-trees over cosmological volumes \citep[][]{Bonoli2025}. Furthermore, \lgbh{} offers the possibility to build simulated mock-data (i.e. lightcones) of any photometric survey, such as those carried out by the JWST \citep[][]{david2019_jpas}. In particular, we implement a detailed model to simulate the photometric properties of AGN within the context of \lgbh{} and build a JWST-like lightcone populated with realistic galaxies and accreting MBHs. %--- our goals: contextualize LRDs in current galaxy/BH-formation models
With this approach we aim at contextualizing LRDs within state-of-the-art models for the formation and evolution of galaxies and BHs. In particular, under the working hypothesis that LRDs are the observational counterparts of galaxies and AGN, we aim to provide a realistic comparison between JWST observations and the result of a state-of-the-art model. The main aim of this comparison is to gain insights about which physical processes drive the emergence of LRDs in the high-$z$ universe and what is the context of LRDs within populations of objects not appearing as LRDs at similar redshifts. %--- disclaimer: we don't aim at being complete, other models can be valid
We underline that this effort is the first step into building a realistic comparison of the properties obtained by current theoretical models and JWST observations, over significantly large cosmological volumes. For this reason, we focus on the assumption that the photometry of LRDs (and, in general, high-$z$ objects) can be explained by the combination of galaxy and AGN contributions. For simplicity, we refrain to directly explore more complex models which may explain the photometric properties of LRDs \citep[e.g.,][]{inayoshi2025b}, leaving this task to future works.

This paper is organized as follows: In Section~\ref{sec:model} we describe our methods, which include the build-up of JWST mock data catalog with the \lgbh{} model. In Section~\ref{sec:model_diego} we describe how we associate a SED to galaxies and active MBHs. In Section~\ref{sec:photometric_selection_of_LRDs} we detail the procedure to identify LRDs within our sample of galaxies and AGN. We present our main results in Section \ref{sec:results}. Finally, we discuss the shortcomings and caveats of our model in Section~\ref{sec:caveats} and present our conclusions Section~\ref{sec:conclusions}. Throughout this paper, we use the cosmological parameters determined by the \cite{planckcolab_cosmoparams}:  $\Omega_{\rm m} \,{=}\,0.315$, $\Omega_{\rm \Lambda}\,{=}\,0.685$, $\Omega_{\rm b}\,{=}\,0.045$, $\sigma_{8}\,{=}\,0.9$ and $h \, {=} \, \rm H_0/100\,{=}\,67.3/100\, \rm km\,s^{-1}\,Mpc^{-1}$.

%---------------------------------------------------------------%---------------------------------------------------------------
\section{The \lgbh{} formation model} \label{sec:model}
We base our analysis on the galaxy and MBH populations generated using the \lgbh{} semi-analytical model (SAM, recently presented in \citealt{Bonoli2025}). Briefly, the \lgbh{} framework builds on the version of \LGalaxies{} introduced by \citet{henriques2015_lgalaxies101}. The latter was specifically designed to self-consistently follow the cosmological assembly of galaxies. It  incorporates a wide range of astrophysical processes within dark matter structures defined by merger trees extracted from N-body simulations. Crucially for this work, the recent developments presented in \citet{david2020_bhgrowth, david2022_bhgrowth} and \citet{spinoso23_bhseeds} enable the detailed modeling of the formation and evolution of both single and binary MBHs within a cosmological context, leading to the \lgbh{} model.
%with the modifications of \cite{david2022_bhgrowth}, \cite{spinoso23_bhseeds} and \cite{david2024_bhgrowth} \diegcomment{add REF. to Silvia's paper}. In short, the SAM self-consistently tracks a wide range of astrophysical processes within structures belonging to dark matter (DM) merger trees derived from N-body simulations. Specifically, \lgbh{} can be run on top of the merger trees from the \texttt{Millennium} simulation suite, which includes \texttt{Millennium-I} \citep[MS,][]{Springel2005}, \texttt{Millennium-II} \citep[MSII,][]{Boylan-Kolchin2006}, and \texttt{Millennium-XXL} \citep[MXXL,][]{Angulo2012}. The different box sizes and DM mass resolutions across these simulations allow us to explore a wide range of baryonic processes across various scales and environments.
%---------------------------------------------------------------
\subsection{The dark matter merger trees} \label{subsec:DM_merger_trees}
\lgbh{} was developed to run on both the Millennium \citep[MS,][]{Springel2005} and Millennium-II \citep[MSII,][]{millenium2_nbodysimul} merger trees \citep[see][]{Bonoli2025,IzquierdoVillalba2025}. In this work we use the latter, as our primary interest lies in relatively low-mass, high-redshift galaxies. MSII follows the cosmological evolution of $2160^3$ DM particles with mass $6.885 \times 10^6 \, \mathrm{M_\odot}/h$, within a periodic box of $100\, \mathrm{Mpc}/h$ on a side \citep{millenium2_nbodysimul}. The MSII outputs were recorded at 68 discrete \textit{snapshots}, at which DM halos were identified.
%using the \texttt{SUBFIND} algorithm. 
These structures are defined as gravitationally bound groups of particles with a minimum mass of ${\sim}\,10^7\, \mathrm{M_\odot}/h$ (20 particles). The simulation was originally run under the WMAP1 + 2dFGRS ``concordance'' $\Lambda$CDM cosmology \citep{Spergel2003}. However, we apply the technique of \citet{AnguloandWhite2010} to rescale the results to the cosmological parameters inferred from the Planck first-year data release \citep{PlanckCollaboration2014}, adopted in this work. This allows us to trace the cosmological assembly of galaxies hosted in halos with masses between ${\sim}10^8 {\rm M_{\odot}}/h$ and ${\sim}10^{14} \, \mathrm{M_\odot}/h$, within an effective box size of $96\, \mathrm{Mpc}/h$.

%---------------------------------------------------------------
\subsection{The baryonic treatment} \label{subsec:baryonic_physics}
\lgbh{} follows the framework presented in \cite{WhiteFrenk1991}, assuming that galaxy formation begins at the centers of DM halos. Specifically, once a dark matter halo collapses, a fraction of the surrounding baryonic matter falls into its gravitational potential well. During this infall, the baryons are shock-heated, leading to the formation of a hot gaseous atmosphere. The latter gradually cools down according to a metallicity-dependent cooling function, leading to the formation of a cold gas disk structure. The disk grows in mass until it becomes massive enough to trigger star formation (SF), building up a stellar disk. SF episodes self-regulate the galaxy growth by heating and ejecting cold gas through supernova explosions of massive and short-lived stars. Moreover, the central MBH also contributes to the regulation of galaxy assembly by releasing energy during the quiescent accretion of the galaxy hot gas. In addition to discs, galaxies can develop stellar bulges thanks to either galaxy mergers\footnote{The SAM uses the baryonic mass ratio of the interacting galaxies to differentiate between major and minor mergers (ratio $>0.2$ and $<0.2$, respectively). The former destroy the discs of the two galaxies, leading to a pure spheroidal remnant. In minor mergers, the disk of the larger galaxy survives and its bulge incorporates the entire stellar mass of the less massive galaxy.} or disk instabilities (DIs)\footnote{The DI are accounted by the analytic stability criterion of \cite{Efstathiou1982}. If satisfied, the disk stellar mass is transferred to the bulge to restore the disk stability.}. Besides these processes, \lgbh{} also accounts for environmental effects such as gas stripping and galaxy tidal disruptions.

We stress that the time resolution offered by the MSII snapshots\footnote{This time resolution corresponds to time intervals of about ${\sim}\,50\,{-}\,100\,\rm Myr$ at $z\,{>}\,6$ and ${\sim} \, 100\,{-}\,400\,\rm Myr$ at $z\,{<}\,6$.} is too coarse to accurately track the baryonic physics involved in galaxy evolution. To overcome this, \lgbh{} performs an internal time discretization between two consecutive snapshots by defining {\it substeps} of approximately $\rm {\sim}\,5\,{-}\,20 \, Myr$ of time resolution, depending on redshift.

\subsection{Massive black holes: Formation and growth} \label{subsec:mbh_formation_and_growth}
\lgbh{} includes a detailed description about the genesis (usually referred to as ``BH-seeding'') and growth of MBHs. Regarding the seeding, \lgbh{} uses the model presented in \cite{spinoso23_bhseeds} and Spinoso et al. (in prep), which describes the formation of BH seeds either within the first collapsing halos at $z\,{>}\,6\,{-}\,8$ or within the remnant of major galaxy mergers. The former case encompasses three plausible scenarios for MBH formation, which are highly dependent on the local physical conditions of the intergalactic medium (IGM), traced self-consistently \citep[see][]{spinoso23_bhseeds}. Seeding scenarios in \lgbh{} include: i) Light-seeds, namely black holes formed from the first generation of PopIII stars \citep[see e.g.][for a review]{klessen_glover2023}, ii) Intermediate-mass seeds, i.e. MBHs with initial masses of $\rm\sim10^{3-4}\msun$ which are predicted to form within the first, compact, stellar clusters \citep[e.g.][]{sakurai2017,reinoso2018,askar_davies_church2021a}, and finally iii) Heavy seeds, i.e. compact objects with masses of $\rm{\sim}\,10^{4-5}\,\msun$ which are thought to form via the direct collapse of pristine gas clouds which did not undergo early gas cooling and fragmentation \citep[see e.g.][for recent reviews]{latif_ferrara2016,smith_bromm_loeb2017,inayoshi_visbal_haiman2020}. On the other hand, massive BH seeds of $\rm8\times10^{4}\,\msun$ forming in the aftermaths of galaxy mergers are modeled as in \cite{bonoli2014}, although recent works showed that these objects can reach masses of up to $\rm\,{\sim}\,10^8\,\msun$ \citep[see e.g.][]{zwick2023,mayer2024}. These extreme objects may form thanks to the strong, multi-scale inflows produced by gravitational torques during gas-rich mergers \citep[see e.g.][]{capelo_dotti2017,mayer_bonoli2019}. %We refer the reader to Figure~6 of \cite{Bonoli2025} for the relative number density of all these co-existing BH-seeds in \lgbh{}.%\dani{there were some inconsistencies and some missing citations in the above paragraph, so I directly reviewed it. It should be fine now}\\

Once the MBH is seeded in a (proto-)galaxy, the model assigns it a random spin $\chi$. The evolution of the latter is then tracked after MBHs coalescence and gas accretion as in \cite{david2020_bhgrowth}. Among these two processes, cold gas accretion largely drives the growth of MBHs. This channel is triggered by either galaxy mergers or DIs, each of them driving a specific fraction of the galaxy cold gas towards the galactic center. This enables the formation of a gas reservoir around the MBH ($\rm M_{res}$, \citealt{david2020_bhgrowth}) which gets progressively consumed in time according to the two-phases model presented in \cite{david2024_bhgrowth}. In brief, during the first accretion phase, cold gas can be accreted at Eddington-limited rates \citep[][]{eddington1917} or at higher rates (super-Eddington), depending on the environment in which the MBH is embedded. The environmental dependence of gas accretion is evaluated in terms of: i) the ratio between the MBH mass and the amount of gas in the reservoir around it (i.e. $\mathcal{R} =  M_{\rm res}/M_{\rm BH}$) and ii) the strength of gas inflows induced by galaxy mergers or DIs, namely: $ \dot{M}_{\rm inflow} = \Delta M_{\rm BH}^{\rm gas} / t_{\rm dyn}$, where $t_{\rm dyn}$ is the dynamical time and $\Delta M_{\rm BH}^{\rm gas}$ is the fraction of gas added to the MBH gas reservoir. When $\mathcal{R}$ and $\dot{M}_{\rm inflow}$ are larger than $2\,{\times}10^4$ and $\rm 10\, \rm M_{\odot}/yr$, respectively a super-Eddington event is triggered. This is characterized by an $f_{\rm Edd} \,{=}\,L^{\rm AGN}_{\rm bol}/L_{\rm Edd}$:
\begin{equation}
    f_{\rm Edd} = B(\chi) \left[\frac{0.985}{\dot{M}_{Edd}/\dot{M}+C(\chi)}+\frac{0.015}{\dot{M}_{Edd}/\dot{M}+D(\chi)}\right],
\end{equation}
\noindent where $L^{\rm AGN}_{\rm bol}$ ($L_{\rm Edd}$) are the MBH bolometric (Eddington) luminosity, $\dot{M}$ ($\dot{M}_{Edd}$) are the (Eddington) accretion rates and the $B(\chi)$, $C(\chi)$, $D(\chi)$ functions are taken from \cite{madau2014_supereddington}. In case the super-Eddington conditions are not fulfilled, an Eddington-limited accretion episode (i.e. $f_{\rm Edd} \,{=}\, 1$) is triggered instead. The second accretion phase starts when the MBH consumes a fraction, $\mathcal{F}=0.7$, of the initial $M_{\rm res}$ (i.e. the one determined at the beginning of the first accretion phase). During this stage, the evolution of $f_{\rm Edd}$ is characterized by:
\begin{equation}
    f_{\rm Edd} = \left[1 + \left((t - t_0)/t_Q\right)^{1/2}\right]^{-2/\mathcal{B}},
\end{equation}
where $ t_Q \,{=}\, t_d\,\xi^{\mathcal{B}}/(\mathcal{B} \ln 10)$, with $t_d \,{=}\, 1.26{\times}10^8 \, \rm yr $, $\mathcal{B} \,{=}\, 0.4$ and $\xi \,{=}\, 0.3$ \citep[see][for further details about the parameter choice]{david2020_bhgrowth}.% The specific value of these variables are based on \cite{hopkins2009_eddingtonlimited} which showed that models of \textit{self-regulated} MBH growth require $0.3\,{<}\,\mathcal{B}\,{<}\,0.8$ and $0.2\,{<}\,\xi\,{<}\,0.4$. \dani{This level of detail is definitely nice, but in the interest of shortening the paper I would simply refer to David's papers where these details are introduced and justified}

\subsection{Building the lightcone} 
\label{subsec:lightcone}

In this work, we employ the method described in \cite{david2019_jpas} to convert the cosmological comoving box generated by \lgbh{} into a lightcone, i.e.: a simulated universe in which galaxies are positioned self-consistently in right ascension, declination, and redshift \citep[we refer to][for details not included in this section]{david2019_jpas}.

In short, we generate a lightcone by exploiting the periodic boundary conditions of the MSII. This involves replicating the box simulated by \lgbh{} at different redshifts in such a way that a custom set of replicas is built. The specific number of replicas required along each Cartesian direction to cover the whole lightcone volume depends on the area, redshift extend and specific orientation of the lightcone viewing angle. The latter is determined in order to minimize the box repetitions within the lightcone volume, following the method of \cite{KitzbichlerWhite2007}. By defining the lightcone orientation as $\hat{u}\,{=}\,(n,m,nm)/|(n,m,nm)|$, the two co-prime integers $n$ and $m$ can be chosen to minimize the intersection of the lightcone volume with the box replicas. We fix $n \,{=}\, 5$ and $m \,{=}\,9$, which provide $\rm (RA,DEC) \,{=}\, (60.95,77.1) \, \rm deg$ as the lightcone line of sight. 
%This sight-line definition minimises the inclusion of box replicas inside the lightcone volume.
We choose a squared shape for our footprint, with extension $\rm (\delta RA,\delta DEC) \,{=}\, (5,5) \, \rm deg$, corresponding to $5.62\, \rm deg^2$ \citep[i.e.: a factor of $\sim60$ larger than what analyzed in][]{kokorev24_lrds}. This is a compromise between the simulated sky area and the execution time of \lgbh{} when producing our lightcone.

\begin{figure}
    \centering
    \includegraphics[width=1.6\columnwidth, angle=90]{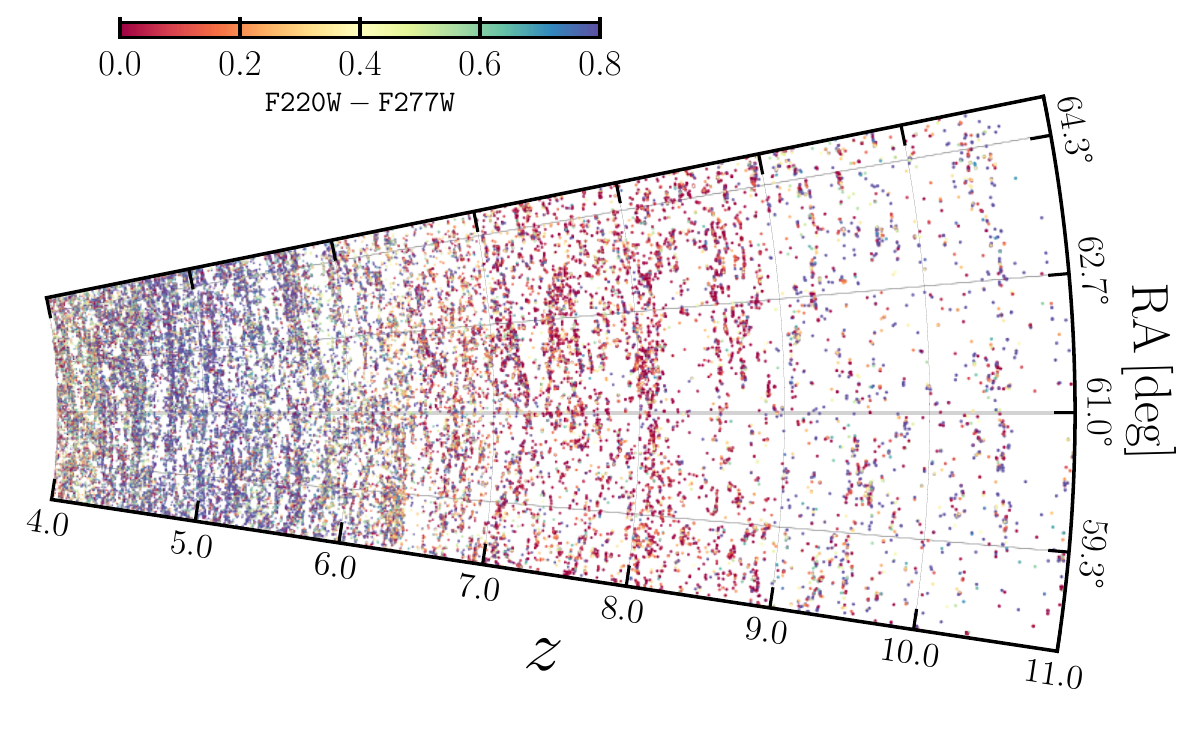}
    \caption{Distribution of galaxies in redshift ($z$) and right ascension (RA) for a thin slice in declination (DEC) of the JWST lightcone used in this work. 
    % For clarity, only galaxies within a narrow declination slice are shown. 
    Each point is color-coded by its {\tt F220W - F277W} color to provide a qualitative idea of the color shift induced at $z\,{\sim}\,6$ by different spectral features affecting the {\tt F220W} and {\tt F277W} filters.}
    \label{fig:lightcone_plot}
\end{figure}

Finally, having fixed the lightcone geometry, during the \lgbh{} execution, galaxies are selected to be part of the mock data catalog by determining the \textit{substep} (see sec. \ref{subsec:baryonic_physics} for the definition of {\it substep}) at which they cross the observer past lightcone. With this method, we build a mock catalog which includes objects simulated by \lgbh{} at $z \,{>}\,4$. In Fig. \ref{fig:lightcone_plot} we showcase the distribution of galaxies within our lightcone.

\subsection{The galaxy and MBH populations: model validation}
\begin{figure}
    \centering
    \includegraphics[width=\columnwidth]{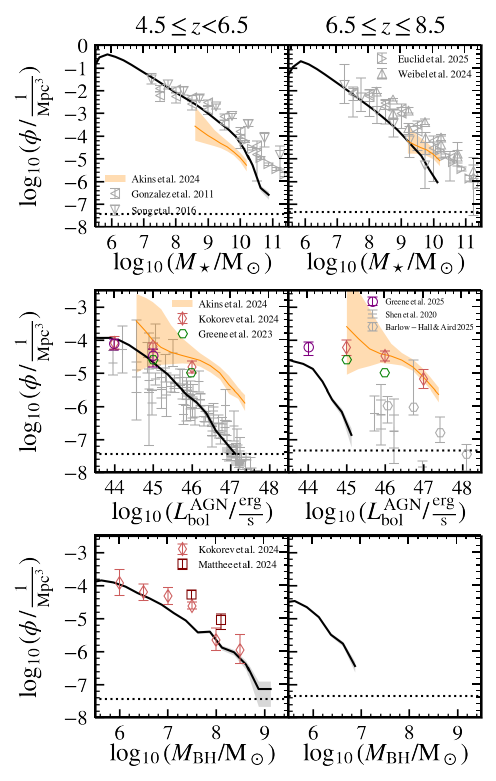}
    \caption{Population of galaxies and MBHs in the simulated lightcone at $4\,{\leq}\, z\,{<}\,6.5$ (left column) and $6.5\,{\leq}\, z \,{\leq}\, 8.5$ (right column). {\bf Upper row}: Comparison between the predicted stellar mass function (black line) and the observational constraints of \protect\cite{Gonzalez2011,Song2016} and \protect\cite{Akins2025}. {\bf Middle row}: Predicted AGN bolometric luminosity function (black line) and observations from \protect\cite{shen20_agnsathighredshift,greene2024_lrds,Akins2025,kokorev24_lrds,BarlowHall2025} and \protect\cite{Greene2025}. {\bf Lower row}: Predicted black hole mass function. The results are compared with \protect\cite{matthee2024_lrds,kokorev24_lrds}. Colored literature data relate to LRDs.}
    \label{fig:lgalaxies_validation}
\end{figure}

In this section, we validate the populations of galaxies and MBHs which make part of our lightcone. Fig.~\ref{fig:lgalaxies_validation} shows three global statistics computed within two broad redshift bins: $4.5\,{\leq}\,z\,{<}\,6.5$ and $6.5\,{\leq}\,z\,{\leq}\,8.5$ (left and right column). The top panels display the predicted stellar mass function (SMF), which shows good agreement with observational data. The discrepancies between our predictions and observations at the high-mass end are due to the limited volume of the MSII box, which hinders the sampling of rare, high-density peaks of the DM field where the most massive galaxies are expected to reside.

The middle panels display the AGN bolometric luminosity ($L_{\rm bol}^{\rm AGN}$) function (AGN LF). As with the SMF, the model provides a good match to the observed data, in particular to the luminosity function of \cite{shen20_agnsathighredshift}. 
Recent results from JWST observations (including broad-line  LRDs) and from deep X-ray data \citep[][]{BarlowHall2025}, point to higher normalization than previously estimated, in particular at very high-$z$. Indeed, the largest deviations with respect to our models occur at $z\,{>}\,6.5$, where the model under-predicts the luminosity reported by \citet{greene2024_lrds} and \citet{kokorev24_lrds}. Note that the \citet{greene2024_lrds} data were re-analyzed in \cite{Greene2025} using a different bolometric correction, providing measurements fainter by 1 dex, which are in better agreement with our predictions.
% This highlights the need for more accurate bolometric luminosities estimates. 
%This shows that the agreement we report between our results and observational constraints may change with more updated constraints.
Finally, at the highest redshifts and luminosity, we are again limited by the MSII box volume: \cite{Bonoli2025} used the combined trees from the MSII and Millennium simulation and obtained a broader dynamical range also in $L_{\rm bol}^{\rm AGN}$. 

Finally, the bottom panels show the predicted MBH mass function (BHMF), which also shows overall good agreement with respect to the values reported by \citet{kokorev24_lrds} and \citet{matthee2024_lrds}, derived from virial MBH-mass estimates from LRDs. We observe a slight deviation from these data at $M_{\rm BH}\,{\sim}\,10^7M_\odot$, although recent evidence suggests that measured values in the literature may be overestimated \citep{Rusakov2025, Greene2025,lupi2024}. 
%We stress that the picture of high-$z$ galaxy evolution is a rapidly evolving field, and that newer and updated data sets are necessary to further constrain and improve our model calibration. Nevertheless, the model generally shows good agreement with the available observational constraints.

We underline that current constraints regarding the properties of high-$z$ galaxies and AGN are rapidly evolving fields. Therefore, updated observations may provide different constraints, which may eventually require further model calibration. Additionally, advanced techniques in handling DM simulation outputs, like the {\it grafting} technique explored in \cite{Bonoli2025} can help alleviating the tension found in the highest redshift bins,  increasing the number of massive and bright MBHs, as mentioned above.
Up to this point, we have compared the outputs of \lgbh{} to literature data directly, that is: without replcating the exact selection effects under which the latter are obtained. In the following sections, we will detail the MBH SED modeling introduced and photometric selections applied to the \lgbh{} galaxies and MBHs. This will allow us to perform a realistic comparison with observations as well as to disentangle the contribution of active MBHs within the LRD population.

\section{Synthetic spectra and photometry of AGN and galaxies} \label{sec:model_diego}
%In this section, we detail the procedure used to construct the spectral energy distribution of galaxies and accreting MBHs. Specifically, the combination of six different features builds up the SED of any of our simulated objects: i) galaxy continuum, ii) galaxy emission lines, iii) AGN continuum, iv) AGN emission lines, v) AGN torus and vi) AGN light obscuration and attenuation. First, we will start with the description of the galaxy spectra, and then we move to the model for accreting MBHs.
Accurate modeling of the emission from both galaxies and AGN is essential for studying photometrically selected LRDs. In this section, we describe our method to construct the SEDs of galaxies and accreting MBHs within our simulated lightcone. The SED of each object is composed of five main components: galaxy continuum, galaxy emission lines, AGN continuum, AGN torus emission, and AGN emission lines (both broad and narrow). All components include the effects of attenuation due to absorption by dust and atomic gas within the ISM along the line of sight.

\subsection{The spectral energy distribution of galaxies} \label{subsec:galaxies_and_mbhs_seds}
Here we outline the procedure followed to simulate galaxy SEDs. This includes models for i) the galaxy stellar continuum, ii) emission lines from star-forming \ion{H}{II} regions and iii) the effect of dust and gas absorption on both these components. Fig.~\ref{fig:Building_SEDs} shows the resulting SEDs (including emission lines) of two $z \,{\sim}\, 6$ galaxies in our lightcone, each with $M_\star{\sim}\,10^8\,\msun$. To illustrate the impact of the ISM, two different lines of sight (LOS) are considered.% For reference, the modeled emission lines are also shown.

\subsubsection{The galaxy continuum: emission and attenuation} \label{subsubsec:galaxy_cont}
\lgbh{} predicts a wide range of intrinsic properties for each simulated galaxy, such as its stellar mass, SF history, age, and metallicity. To convert these intrinsic quantities into an observed SED, the SAM uses evolutionary population synthesis prescriptions. % and dust prescriptions. 
These are implemented via Simple Stellar Populations (SSPs) models, which predict the time-evolution of the SED emitted by a single star-formation burst of a given mass and metallicity, after assuming an initial mass function (IMF). Consequently, the simulated SEDs of galaxies in a given redshift are obtained by a superposition of SSPs along the galaxy star formation history. Following \cite{henriques2015_lgalaxies101}, \lgbh{} uses the \cite{Maraston2005} synthesis model with the \cite{Chabrier2003} IMF.

%Once the stellar continuum emission is determined, \lgbh{} attenuates it according to a specified dust attenuation model. Following the approach of \textcolor{red}{De Lucia et al. 2007}, the model assumes that extinction arises from two main components: the diffuse interstellar medium (ISM) and molecular clouds. Specifically, the light emitted by stars is absorbed and scattered by dust grains distributed throughout the ISM of a galaxy. The optical depth of diffuse dust in galactic discs follows (see \textcolor{red}{Devriendt et al. 1999} for details) Once the galactic continuum emission is computed, accounting for the galaxy evolutionary pathway, the attenuation of this stellar light by dust is accounted for by following the approach of \cite{DeLucia2007}

Once the galactic continuum emission is computed based on the galaxy evolutionary pathway, the attenuation of this stellar light by dust is accounted following the approach of \citet{DeLucia2007}. As in \cite{henriques2015_lgalaxies101}, \lgbh{} assumes that extinction arises from both the diffuse interstellar medium (ISM) and molecular clouds. The ISM optical depth $\tau_{\lambda}^{\rm ISM}$ is computed following the prescription of \cite{Devriendt1999}, which relates the dust column density to key galaxy properties:
\begin{equation}
    \tau_\lambda^{\rm ISM} = (1+z)^{-1} \left(\frac{A_\lambda}{A_{\rm V}}\right)_{Z_\odot} \left(\frac{Z_{\rm gas}}{Z_\odot}\right)^s \left(\frac{\langle N_H \rangle}{2.1 \times 10^{21}}\right),
    \label{eq:tau_ISM}
\end{equation}
where $Z_{\rm gas}$ is the gas metallicity\footnote{Note that the metallicity dependence is motivated by \cite{Guiderdoni1987} and sets $s \,{=}\,1.35$ for $\lambda \,{<}\, 2000\ \AA$ and $s \,{=}\, 1.6$ for $\lambda \,{>}\, 2000\ \AA$.} and $ \langle N_H \rangle$ is the mean column density of hydrogen defined as:
\begin{equation}
    \langle N_H \rangle = \frac{M_{\rm cold}^{\rm gas}}{1.4\,m_p\,\pi \left(aR_{\rm d,gas}\right)^2}\ ,
    \label{eq:nh_ism}
\end{equation}
with $R_{\rm d,gas}$ being the scale-length of the cold gas disk. The 1.4 factor accounts for the presence of helium, and $a$ is set to $1.68$ in order for $\langle N_H \rangle$ to represent the mass-weighted average column density of an exponential disk. We note that the redshift dependence in Eq.~\eqref{eq:tau_ISM} is chosen to reproduce luminosity functions and extinction estimates for Lyman-break galaxies at $z \,{>}\, 5$ \citep[see][]{henriques2015_lgalaxies101,Clay2015}. Regarding the dust attenuation law, we tested both the one provided by \cite{gaskell07_attcurve} and by \cite{Gordon2003}, with an $R_V=2.505$ \cite[see similar approach in][]{Jones2025,Ji2025_blackthunder}. Both of these are common choices in the recent literature focusing on the modeling of high-$z$ galaxies and LRDs. Based on our BHMF, AGN LF and SMF we decided to use the one of \cite{gaskell07_attcurve}, as it fits better the observational data, although we find very little difference.

Taking into account Eq.~\eqref{eq:tau_ISM} and Eq.~\eqref{eq:nh_ism}, and assuming a slab geometry for the disk, the mean absorption coefficient can be written as:
\begin{equation} 
     A_\lambda^{\rm ISM} = -2.5\log_{10} \left(\frac{1-e^{-\tau_\lambda^{\rm ISM}\csc{\alpha_{\rm LOS}}}}{\tau_\lambda^{\rm ISM}\csc{\alpha_{\rm LOS}}}\right),
    \label{eq:att_ism_lgalaxies}
\end{equation}
where $\alpha_{\rm LOS}$ is the angle between the observer (placed at the corner of the first replica of the MSII box) and the angular momentum vector of the galaxy disk. The variable \texttt{csc} refers to the co-secant of the $\alpha_{\rm LOS}$ angle. %. To guide the reader, $\rm LOS\,{=}\,0 \deg$ and \texttt{csc} refers the co-secant of the LOS angle.
For the molecular cloud component, the model assumes that dust grains embedded within these dense regions attenuate the light emitted by newly formed stars. To account for this effect, \lgbh{} assumes that stars with lifetimes shorter than the typical lifetime of their birth clouds (${\sim}\, 10^7$ yr) are subject to additional attenuation, following the prescription of \cite{Charlot2000}. Specifically, the mean absorption coefficient is given by:
\begin{equation}
    A_\lambda^{\rm BC} = -2.5\log_{10}\left(e^{-\tau_\lambda^{\rm BC}}\right).
\end{equation}
where
\begin{equation}
    \tau_\lambda^{\rm BC} = \tau_{\lambda}^{ISM}\left( \frac{1}{\mu} \,{-}\, 1 \right) \left( \frac{\lambda}{5500 \AA}\right)^{-0.7},
\end{equation}
and $\mu$ is given by a random Gaussian deviate with mean 0.3 and standard deviation 0.2, truncated at 0.1 and 1 \citep[see][for further details]{henriques2015_lgalaxies101}.

We note that \lgbh{} separately tracks the stellar emission produced by stars in the galactic bulge and in the disk. As the SAM assumes that only the latter component contains cold gas, we only apply ISM attenuation to the disk stellar emission. In contrast, molecular cloud attenuation affects young stars in both the disk and the bulge\footnote{Following \cite{henriques2015_lgalaxies101}, birth cloud attenuation for the stars in the bulge is applied as a fixed value of $A_\lambda^{\rm BC} \,{=}\,-2.5\log_{10}(0.5)$.}. We note that this distinction in ISM attenuation has a negligible impact on our results, as the majority of high-$z$ galaxies in our sample are either bulgeless or have only a minor bulge component.

\subsubsection{The galaxy emission lines} \label{subsubsec:galaxy_lines}
Following \cite{david2019_jpas}, we also include the contribution of  9 different nebular emission lines in the galaxy SED: $\rm Ly{\alpha}$ (1216\AA), \Hb (4861\AA), \Ha (6563\AA), {\oii} (3727\AA, 3729\AA), {\oiii} (4959\AA, 5007\AA),  $\rm [\ion{Ne}{III}]$ (3870\AA), {\oi} (6300\AA), $\rm [\ion{N}{II}]$ (6548\AA, 6583\AA), and $\rm [\ion{S}{II}]$ (6717\AA, 6731\AA). To this end, we use the model presented in \cite{Orsi2014} which uses as a foundation the \cite{LevesqueKewleyLarson2010}\footnote{\cite{Orsi2014} obtain theoretical SEDs of $\rm \ion{H}{II}$ regions by pairing \texttt{Starburst99} \citep{Leitherer1995} together with the \texttt{MAPPINGS-III} photoionization code \citep{Dopita1995,Dopita1996,Groves2004}.} model grid of $\rm \ion{H}{II}$ regions. Based on that grid, the luminosity of each emission line is fully characterised by four different parameters: i) age of the stellar cluster that provides the ionising radiation ($t_\star$), ii) density of the ionised gas ($\mathrm{n}_{e}$), iii) galaxy gas-phase metallicity ($\rm Z_{cold}$), and iv) ionization parameter ($q$). For $t_\star$ and $\rm n_e$ we assume constant values ($t_{*}\,{=}\,0$ and $\mathrm {n}_{e} \mathrm{{=}\, 10 \, cm^{-3}}$, see \citealt{Orsi2014} for further details). %\dani{\\ \textit{All the details between ``To this end'' until this point were already discussed in Orsi+2014 and Izquierdo-Villalba+2019. All this can be strongly summarized. Suggestion:}\\ The flux of these emission lines is modeled as in Orsi+2014 and included in \lgbh{} using the same method presented in Izquierdo-Villalba+2019. In brief, Orsi+2014 obtained the emission lines luminosity from the galaxies star-formation rate, based on assumptions on the stellar population age, ionized gas density, gas metallicity and ionization parameter. For $t_\star$ and $\rm n_e$ we assume...as in Orsi+2014...\\}
Conversely, the values of $q$ and $\rm Z_{\rm cold}$ are directly set by the cold gas metallicity predicted by \lgbh{} adopting:
\begin{equation}
    q\left(\mathrm{Z}\right) = q_0 \left(\frac{\mathrm{Z}_{\rm cold}}{\mathrm{Z}_0}\right)^{-\gamma} \,\, \rm [cm/s],
    \label{equation:RelationZq}
\end{equation}
\noindent where $q_0$, $\rm Z_0$, and $\gamma$ are free parameters set to $\rm 2.8\,{\times}\,10^7\, cm/s$, 0.012 and 1.3, respectively, to match observational measurements of {\Ha}, {\oii}, and {\oiii} luminosity functions \citep{Orsi2014,david2019_jpas}. Finally, following \citet{david2019_jpas}, we assume that the profiles of galactic emission lines can be approximated as Dirac delta functions.
 %By using the predicted line fluxes, the luminosity of a given line, $\rm L(\lambda_j)$, is given by:\begin{equation}\label{equation:luminosities}
%\rm  L(\lambda) \,{=}\,  1.37\times 10^{-12} Q_{H^o} \frac{F(\lambda_j|\,q,Z_{cold})}{F(H_{\alpha}|\,q,Z_{cold})} \;\;\; [erg/s],
%\end{equation}
%\noindent where $\rm F(H_{\alpha}|\,q,Z_{cold})$ and $\rm F(\lambda_j|\,q,Z_{cold})$ are, respectively, the flux of \Ha and $\lambda_j$ line in a galaxy with ionization parameter q and metallicity $\rm Z_{cold}$. $\rm Q_{H^o}$ is the ionization photon rate in units of $\rm s^{-1}$ calculated from the \lgbh{} predictions.\\% Here, we assume that all the emitted photons contribute to the production of emission lines.\\ 

As shown by \citet{david2019_jpas}, it is necessary to consider dust attenuation onto the line-emission luminosity in order to correctly recover the observed line-luminosity functions at $z\,{<}\,1.5$. This model closely follows Eq.~\eqref{eq:att_ism_lgalaxies} but with a different assumption for the optical depth, $\tau_{\lambda}^{\rm line}$:
\begin{equation}
    A_\lambda^{\rm line} = -2.5\log_{10} \left(\frac{1-e^{-\tau_\lambda^{\rm line}\csc{\alpha_{\rm LOS}}}}{\tau_\lambda^{\rm line}\csc{\alpha_{\rm LOS}}}\right)\,,
\end{equation}
where $\tau_{\lambda}^{\rm line}$ depends on the cold gas metallicity as:
\begin{equation}
    \tau_{\lambda}^{\rm line} {=} C(z) \frac{A_V}{A_B} \frac{A_{\lambda}}{A_V},
\end{equation}
\noindent where the values of $A_v/A_B$ and $A_{\lambda}/A_V$ are computed based on the same attenuation law chosen for the ISM (see \ref{subsubsec:galaxy_cont}). Finally, $C(z)$ is a free parameter set to $C(z)\,{=}\,161.46\,{e}^{-(0.46\ z)}$ calibrated to reproduce the redshift evolution of \oii, \oiii, \Hb and \Ha  observed luminosity functions \citep[see the appendix of][for further details]{david2019_jpas}.

\begin{figure*}
    \centering
    \includegraphics[scale=0.28]{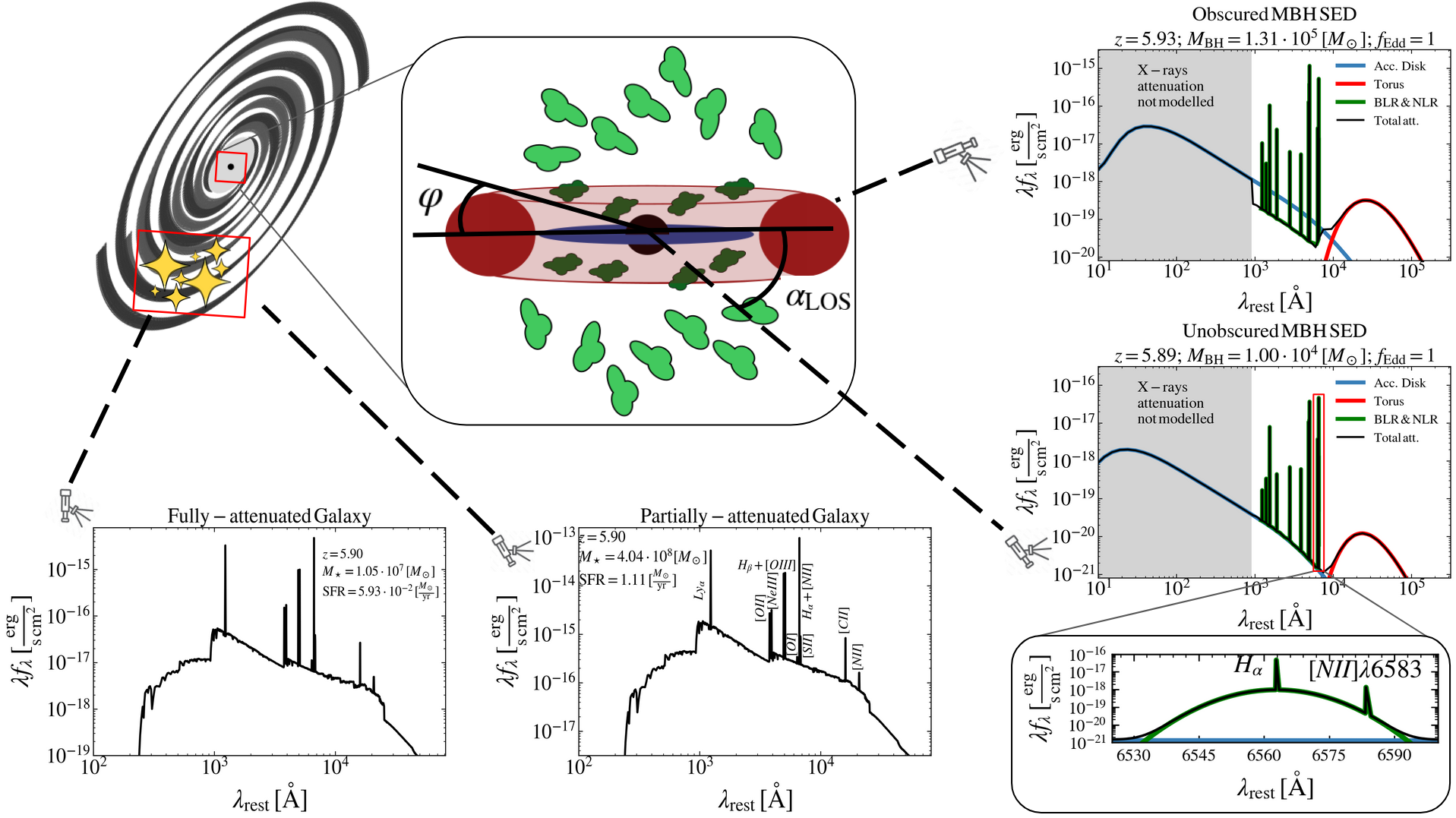}
    \caption{Schematic illustration of the SED components in our galaxy and AGN modelling. On the  bottom left we show the SEDs of two galaxies corresponding to different lines of sight ($\alpha_{\rm LOS}$): one directly towards the ISM and another with a more direct view of the disk plane. On the right, we show a zoom into the nuclear MBH configuration adopted. Different colours in the spectra correspond to different regions of the AGN structure: blue for the accretion disk, red for the torus, and green for the BLR \& NLR. In black we show the total observed SED (i.e the MBH emission obscured and/or attenuated). Two $\alpha_{\rm LOS}$ are provided, corresponding to the regime of obscuration ($\varphi\leq\alpha_{\rm LOS}$) and another to non-obscuration ($\varphi>\alpha_{\rm LOS}$). On the bottom right, we show a zoom-in to the line profiles of \Ha{} (6563 \AA) and  [\ion{N}{II}] (6583\AA) lines.}
    \label{fig:Building_SEDs}
\end{figure*}

\subsection{The spectral energy distribution of active MBHs} \label{subsec:agnsed}

%\textcolor{red}{Should this be a section or a subsection??} {\color{red} David: To me, I think a sub-section would be better. We include the photometry of galaxies and MBHs in the same section. But feel free to change it. We can shape the structure later, but let's write all the info. I already wrote a bit from the LGalaxies model. It is missing the MBH part. }

The standard version of \lgbh{} \citep[][]{Bonoli2025} does not include any SED modeling for the emission of accreting MBHs. In this section, we introduce an AGN emission model to overcome this limitation. Particularly, our AGN SED model includes four main components: i) the continuum emission from the accretion disk surrounding the accreting MBHs, ii) the IR emission due to the reprocessing of rest-frame UV-optical light by dust. iii) emission lines from both narrow and broad line regions (NLR and BLR, respectively), iv) the effect of dust and gas absorption on all the above components. One key component of the model is the development of a physically informed density profile for the accreted gas, which ultimately determines torus and line emission. To guide the reader, in Fig.~\ref{fig:Building_SEDs}, we present a schematic overview of the SED assigned to two active MBHs of $10^4\,{-}\,10^5\, \msun$, both accreting at the Eddington limit. These objects are seen from two different viewing angles: one with a direct view of the central engine (unobscured), and the other with the line of sight intersecting the torus (obscured). A zoom-in between $6530 \,{<}\, \lambda \,{<}\, 6600\, \AA$ highlights the resulting line profiles generated by our model (see following sections for details).
%As of now, \lgbh{} lacks a full implementation of the AGN SED. For this work, we have developed a first approach to this intricate problem to tackle some of the questions surrounding high-z AGNs. 

%The model is described in Fig. \ref{fig:model_scheme} and Table \ref{tab:model_scheme}. The complete SED is computed in different steps. First, the emission from the accretion disk is modeled using {\tt CLOUDY}, in a method presented in \textcolor{red}{Su et al. 2025}, and discussed in section \ref{ssec:tong}. This model is then complemented with the modeling of an IR bump, which would correspond to the reprocessing of UV light from the AGN from a dusty-like structure surrounding it, that is, a torus. In an effort to be as agnostic as possible, we do not make any assumptions about the geometry of the torus, and test different models, from no dust to the point where the torus fully covers the central source. More details can be found in section \ref{ssec:torus}. Finally, we provide a detailed treatment of UV light, which accounts for galaxy-scale, orientation-dependent obscuration of UV light that is not extinct by the torus. This is further discussed in section \ref{subsubsec:att_uv}.

\subsubsection{AGN continuum} \label{subsubsec:mbh_acc_disk}

The continuum of an accreting MBH is modeled using the method presented in \cite{Su2025}\footnote{ \url{https://github.com/SuTong1999/agnSED}}, which computes AGN SEDs as a function of the accretion rate $\dot{m} \,{=}\, \dot{M}/\dot{M}_{\rm Edd}$. In detail, different emission regimes are modeled depending on how $\dot{m}$ compares to the critical threshold %\diegcomment{$\dot{m}_{\rm crit} \,{=}\, 0.38\, \alpha^{2.34}\,\beta^{-0.41}$ \citep[as in][where $\alpha \,{=}\, 0.05$ is the disk viscosity and $\beta \, {=} \, 0.95$ the magnetic parameter]{Taam2012}}
$\log_{10}\, \dot{m}_{\rm crit} \,{=}\, -2.9$\footnote{This is chosen to ensure continous radiative efficiency between different accretion regimes, see \citet{Su2025} for details}:
\begin{itemize}
    \item At $\dot{m} \,{<}\, \dot{m}_{crit}$ a disk evaporation model is adopted \citep{Meyer1994}, where the disk inner region is described by an Advection-Dominated Accretion Flow (ADAF), surrounded by an outer thin disk truncated at the radius $r_t$. The truncation radius is given by \citep{Taam2012}:
    \begin{equation}
        r_t \, {=} \, 17.3\ \dot{m}^{-0.886}\,\alpha^{0.07} \beta^{\,4.61} \, [R_s],
    \end{equation}
    where $R_s$ is the MBH Schwarzschild radius. The accretion disk is optically thin and geometrically thick, and its emission, $L_{\nu,{\rm disk}}$, is characterized by multi-color blackbody emission \citep{shakura1973_accretion}:
    \begin{equation}
        L_{\nu,{\rm disk}} = 2 \pi h R_S \frac{\nu^{3}}{c^{2}}
        \ \int_{r_{\rm in}}^{r_{\rm out}}{4 \pi r \left[{\rm exp}\left(\frac{h \nu}{k_{\rm B} \, T(r)}\right)-1 \right]^{-1} dr},
        \label{eq:EmissionDisc}
    \end{equation}
    with an effective temperature, $T(r)$: 
    \begin{equation}
        T(r) \,{=}\, \left[ \frac{3\dot{M}_{\rm BH}c^6}{64\pi \sigma_{\rm B} G^2 M^{2}_{\rm BH}} \right]^{1/4} \left[ \frac{1}{r^{3}} \left(1 - \sqrt{\frac{3}{r}}\right)\right]^{1/4} \, [K],
        \label{eq:TemperatureSS}
    \end{equation}
    where $r \, {=}\, R/R_{S}$, $R$ is the radial distance to the MBH, $k_{\rm B}$ the Boltzmann constant, $h$ the Planck constant, $c$ the speed of light and $\nu$ the rest-frame frequency. Conversely, the ADAF geometry is geometrically thick and optically thin, whose emergent spectra is computed using the self-similar solution of \cite{Narayan1994,Narayan1995} with the procedure presented in \cite{Manmoto1997} and \cite{Qiao2010,Qiao2013}.\\

    \item %\diegcomment{\bf At $\dot{m} \,{\geq}\, \dot{m}_{crit}$, a magnetic-reconnection heated disk-corona model proposed by \cite{Liu2002,Liu2003} is adopted. In brief, this model assumes that a thin accretion disk extends inward to the innermost stable orbit and is compressed by a plane-parallel corona. At $\dot{m}_{crit} \,{<}\,\dot{m}\,{<}\,0.3$ (\textit{hard state} or gas dominated) the emission of the corona prevails over the disk one and it is characterized by $L_{\nu} \,{\propto}\, \nu^{-1.1}$. Conversely, at $\dot{m}\,{>}\,0.3$ (\textit{soft state} or radiation dominated), the emission is governed by the disk. While at $\dot{m}\,{<}\,0.3$ Eq.~\eqref{eq:EmissionDisc} is used to model the disk emission, at larger accretion rates the slim disk solution presented in \cite{Watarai2006} is employed with an effective temperature}: 
    At $\dot{m} \,{\geq}\, \dot{m}_{crit}$, a magnetic-reconnection heated disk-corona model is adopted. In brief, this model assumes that a thin accretion disk extends inward to the innermost stable orbit and is compressed by a planeparallel corona. In the \cite{Su2025} model, the energy equilibrium equations are solved for each annulus individually, rather than globally as in the original model (\cite{Liu2002,Liu2003}).
    The modified model further considers super-Eddington accretion. As the BH accretion rate increases, an optically thick, radiative pressure-dominated, slim disk-like region would emerge in the inner part of the disk, replacing the disk-corona configuration, whose effective temperature adopts a modified version of the self-similar solution presented in \cite{Watarai2006}, namely:
    \begin{equation}
       T(r) \,{=}\,4.956 \,{\times}\,10^7 \mathcal{C}^{1/8} f^{1/8} \left(\frac{\rm M_{BH}}{\rm M_{\odot}}\right)^{-1/4} r^{-1/2} \, [\rm K],
       \label{eq:Teff_SuperEddington}
    \end{equation}
    where $f\,{\approx}\,1$ represents the fraction of the viscously dissipated energy cooled by advection, $\mathcal{C} \,{=}\, 1 \,{-}\, \sqrt{3/r}$, $\rm M_{BH}$ the MBH mass, $r \,{=}\, R/Rs$ and $R$ the distance from the central MBH.
\end{itemize}

\subsubsection{The infrared emission: The torus structure} \label{subsubsec:torus}

The SED described in the previous sections lacks the torus-like component, i.e a dusty structure around the MBH, which is thought to be responsible for the infrared (IR) bump seen in the spectra of an accreting MBH  \citep{antonucci1993_torus,urry1995_torus}. To account for that emission, a hot dusty torus is modeled following \cite{barvainis87_torusmodel}, which describes a smooth structure composed of graphite grains. These have a fixed radius of $a\,{=}\,0.05\ {\rm \mu m}$, with a near-IR absorption efficiency curve characterized by $Q_\nu \,{=}\, q_{\rm IR}\,\nu^{\gamma}$, where $q_{\rm IR} \,{=}\, 1.4 \,{\times}\, 10^{-24}$ and $\gamma \,{=}\, 1.6$. Each grain emits as a blackbody with an IR luminosity of:
\begin{equation}
   L^{\rm gr}_{\nu, \rm IR} = 4\pi^2 a^2 Q_\nu B_\nu(T_{\rm gr})\ [{\rm erg/(s\, Hz)}],
   \label{eq:dustlgrain}
\end{equation}
where $B_\nu$ is the standard Planck law for the black-body spectrum. Under the ultraviolet (UV) radiation field generated by the MBH accretion disk, the dust grains reach an equilibrium, and the temperature profile of the overall dust-grain distribution can be described as:
\begin{equation}  \label{eq:dusttemperature}
    %$T_{\rm gr} = 1587.1 \left(\frac{{\rm L}_{\rm UV,46}}{r^2_{\rm pc}}\right)^{1/5.6} e^{-\tau_{\rm uv}/5.6}\ [{\rm K}]\, ,  
    T_{\rm gr} \,{=}\, 1587.1 \left(\frac{{L}_{\rm ion}}{\rm 10^{46} erg/s}\right)^{1/5.6}  \left(\frac{r}{\rm pc}\right)^{11.2} e^{-\tau_{\rm torus}/5.6}\ [{\rm K}]\, , 
\end{equation}
where ${L}_{\rm ion}$ represents the total incident ionizing luminosity, expressed in $\rm erg\,s^{-1}$. This radiation corresponds to the portion of the spectrum most effective at heating dust grains which we defined in our model as the integrated luminosity over the wavelength range from $912$ \AA{} to $6564$ \AA\footnote{At lower wavelengths, photodisociation effects play a role, which is beyond the scope of this work. On the contrary, longer wavelengths are not efficient for dust heating. In fact, one could argue that 6564 $\AA$ is already a conservative value, but we did not find any strong arguments to stop the ionization earlier.}. The variable $r$ represents the distance from the central MBH in parsecs, and $\tau_{\rm torus}$ is the optical depth of the torus, defined as:
\begin{equation}
    \tau_{\rm torus} = \pi a^2 \int_{r_1}^r n_{\rm gr}(r') dr',
    \label{eq:dusttau}
\end{equation}
The variable $n_{\rm gr}$ corresponds to the radial density profile of the torus smoothly distributed as:
\begin{equation}
    n_{\rm gr} (r) = n_{\rm gr,1} \left(\frac{r}{r_1}\right)^{-\beta},
    \label{eq:dustn}
\end{equation}
 with $n_{\rm gr,1}$ being the grain density at the innermost torus radius $r_1$, and $\beta$ the slope of the distribution. The value of $r_1$ coincides with the sublimation radius and is defined as:
\begin{equation}  \label{eq:dustrsubl}
    r_1 = r_{\rm subl} = 1.17\ \left(\frac{{L}_{\rm ion}}{10^{46}\, \rm erg/s}\right)^{1/2}\ \left(\frac{T_{\rm subl}}{1500\, {\rm K}}\right)^{-2.8}\ [{\rm pc}]\,,
\end{equation}
where $T_{\rm subl}$ corresponds to the innermost temperature of the torus fixed at $\rm 1500\, K$ \citep[see][for further details]{barvainis87_torusmodel}. Note that the factors in Eq.~\eqref{eq:dusttemperature} and Eq.~\eqref{eq:dustrsubl} differ from those of \cite{barvainis87_torusmodel}, as we have recomputed them to account for rounding inconsistencies in the original publication. % \davcoment{What does it mean inconclusive results?}.

Taking into account Eq.~\eqref{eq:dustlgrain}, Eq.~\eqref{eq:dusttemperature}, Eq.~\eqref{eq:dusttau}, and Eq.~\eqref{eq:dustn}, the total IR emission emitted by the tours can be written as:
\begin{equation} \label{eq:Trous_InfrarredEmission}
    {L}_{\nu,\rm IR} = \Omega \int_{r1}^{r2} n_{\rm gr} {L}^{\rm gr}_{\nu,\rm IR} r^2 dr\ [{\rm erg/(s\, Hz)}],
\end{equation}
\noindent where $r_2$ corresponds to the outer extension of the torus and $\Omega \,{=}\,2\pi\left[1\,{-}\,\cos\left(2\varphi\right)\right]$ is the solid angle subtended by the torus as seen from the central source. The value of $\varphi$ varies between $0\,{<}\,\varphi\,{<}\,\pi/2$, implying that at $\varphi\,{=}\,0$ ($\Omega\,{=}\,0$) there is no to torus obscuring the system. In contrast, when $\varphi\,{=}\,\pi/2$ ($\Omega\,{=}\,4\pi$), the torus has a full coverage of the central MBH.

Consistent with \citet{gravity_torussize}, we assume that the ratio $r_2/r_1$ is a random variable uniformly distributed between 2 and 20, implying a maximum outer radius of $r_2 \,{=}\,20\,r_{\rm subl}$. Additionally, \citet{gravity_torussize} reported that the slope of the torus density profile is typically around $0.5$ \citep[see also][]{guise2016_betaparam}. Motivated by these results, we sample the parameter $\beta$ from a normal distribution with a mean of $0.5$ and a standard deviation of $0.1$, restricting values to be strictly positive to avoid nonphysical torus configurations. In the following section, we introduce the model for the gas density profile required to determine the value of $n_{gr,1}$.

\subsubsection{The gas density profile} \label{subsubsec:gasdensityprofile}

%\davcoment{If I do not remember wrong, this was another way to compute $n_g$, right? I would move to the appendix and show there the results for this model. What do you think?} -> \textcolor{magenta}{{\bf Diego:} The thing is that this model is also used to derive the densities for the NLR} \davcoment{ok!} \\

Motivated by the recent works of \cite{DongPaez2023} and \cite{volonteri2024_lrdsinobelisk} we model the gas density profile around the MBH to determine the value of $n_{\rm gr,1}$. %\beta$. %in an independent way and inform the torus model with a more physically motivated value (see left column of table \ref{tab:model_scheme}).
To this end, we assume that the gas distribution $\rho_{\rm gas}$ follows the form:
\begin{equation}\label{eq:MBH_Gas_Distribution}
    \rho_{\rm BH,gas} \,{=}\, \rho_{0,\rm gas} \left(\frac{r}{r_{sg}}\right)^{-\beta_g},
\end{equation}
where $\rho_{0,\rm gas}$ is the central gas density, $r_{\rm sg}$ is the edge of the accretion disk, assumed to correspond with the self-gravity radius \citep{laor1989_selfgravrad}:
\begin{equation}
    r_{sg} \,{=}\, 3586  \alpha^{2/9} \left( \rm \frac{M_{BH}}{10^8 \, M_{\odot}}\right)^{-2/9} \left( \frac{R_s}{2}\right) 
 f_{\rm Edd}^{4/9} \, \, \, \rm [pc],
\end{equation}
where $\alpha \,{=}\, 0.1$. Finally, we assume that the steepness of the gas distribution is the same as for the torus profile, i.e.: $\beta_g \,{=}\,\beta$. This ensures that the resulting densities are consistent with those expected for the narrow-line region (NLR), and aligns with the interpretation that dust and gas are accreted (or at least influenced) by the gravitational potential of the central source, suggesting a non-nuclear origin for the material funneled towards the MBH from larger galactic scales. However, we note that if dust were produced in situ by stars evolving within the nuclear region, this assumption would no longer hold. Exploring this scenario lies beyond the scope of this work. To find the value of $\rho_{0,\rm gas}$, we consider that 
\begin{equation}
   \dot{M}_{\rm BH} \cdot t_{\rm ff}(r_{\rm Bondi}) = \int_{r_{\rm sg}}^{r_{\rm Bondi}} 4\pi r^2 \rho_{\rm BH, gas}\,,
\end{equation}
where $\dot{M}_{\rm BH}$ is the black hole accretion rate computed by \lgbh{}, $t_{\rm ff}$ the free fall time and $r_{\rm Bondi}$ the Bondi radius, defined as $r_{\rm Bondi}\,{=}\ 2 \, \mathrm{G\, M_{BH}}/c_s^2$ where $G$ is the gravity constant and $c_s\,{\propto}\, \sqrt{T_{\rm gas}}$ is the sound speed. 
%This latter quantity depends on the gas temperature, $T_{\rm gas}$. 
Since \lgbh{} does not provide explicit predictions for $T_{\rm gas}$, we adopt a simplified approach by assuming a fixed temperature of $T_{\rm gas} \,{=}\, 10^6\,\mathrm{K}$. This prevents nonphysically large $r_{\rm Bondi}$ relative to the galaxy sizes at $z \,{>}\, 4$, and the resulting gas densities remain consistent with those found in NLRs of AGN (see Section~\ref{subsubsec:AGN_emission_lines}).

After setting all the gas properties around the MBH, we can determine the $n_{gr,1}$ value as:
\begin{equation}
    n_{gr,1} \,{=}\, {\rm DtG} \cdot \frac{\rho_{\rm BH, gas}(r_1)} {m_{\rm gr}},
\end{equation}
where $m_{\rm gr}$ is the mass of the grain (with a density value of 2.26 g/cm$^3$) and $\rm DtG$ is the dust-to-gas ratio. Unlike other branches of \LGalaxies{} \citep[see e.g][]{Yates2021,Parente2024}, the \lgbh{} branch does not trace dust. Thus, we determine the $\rm DtG$ randomly between 10$^{-2}$ and 10$^{-3}$ for each object (see \citealt{volonteri2024_lrdsinobelisk}).

\subsubsection{The AGN emission lines} \label{subsubsec:AGN_emission_lines}

%\dani{repetita iuvat: I wouldn't add blank spaces between paragraphs of related topics, as this just adds space to the text. If you really want some space, I suggest to use the ``vspace'' command and set the space to 2 millimeters at most}

Broad and narrow emission lines are commonly observed in the SEDs of accreting MBHs. In this work, we include 11 prominent UV and optical emission lines, namely:
\Ha{} (6563 \AA), \Hb{} (4861\AA), {\HeI} (3888\AA), {\FeII} (1215\AA), [\ion{N}{II}] (6583\AA), {\CIII} (1908\AA), {\oiii} (4959\AA, 5007\AA), {\MgII} (2795\AA, 2802\AA), {\CIV} (1548\AA, 1550\AA), {\SiIV} (1393\AA, 1402\AA), and {\NV} (1238\AA, 1242\AA). These lines are modeled leveraging the method developed in Su et al. (in prep.), which we shortly summarize below.

Su et al. (in prep.) employs the AGN continuum model described in Section~\ref{subsubsec:mbh_acc_disk} as input to the \texttt{CLOUDY} photoionization code \citep{Ferland2017_CLOUDY} to model the emission from the 11 spectral lines listed above. To simulate the narrow-line region (NLR), Su et al. (in prep) assumes a uniform, constant density throughout the NLR \citep{Feltre2016}. The adopted density values in Su et al. (in prep) are computed by finding an empirical relation linking the gas density to $M_\star$ and SFR, enabling a physically motivated parameterization of the NLR conditions. The relation is constrained with the MPA-JHU catalog (\citealt{tremonti04_mpa-jhu}). However, we highlight that with the gas density profile introduced in Section~\ref{subsubsec:gasdensityprofile} we get hydrogen densities in good agreement with those derived from the MPA-JHU relation (of the order of $10^{-1} - 10^5\, {\rm1 1/cm^3}$). Therefore, those are the ones used to compute our emission for the NLR. Beyond the density, the model requires to define an inner radius of the NLR. The definition adopted is,
\begin{equation} \label{eq:Narrow_Line_Region}
r_{\rm NLR} = 100 \sqrt{\frac{L_{\rm bol}^{\rm AGN}}{10^{43} \, \rm erg/s}} \ [{\rm pc}]\,,
\end{equation}
where $L^{\rm AGN}_{\rm bol}$ is the bolometric luminosity of the MBH. To avoid non-physical values at both low and high luminosity, $r_{\rm narrow}$ is restricted to lie within the range $10 - 400\, \rm pc$. The outer radius of the NLR is instead defined by \texttt{CLOUDY} as the point where the gas kinetic temperature drops below $4000\,\rm K$. For the broad-line region (BLR), Su et al. (in prep.) assumes the same uniform gas density structure as for the NLR, but with densities scaled up by a factor of $10^{9.5}$ to reflect the higher typical densities in this internal region. The inner radius of the BLR is set following the empirical relation from \citet{Panda2018}:
\begin{equation}
r_{\rm BLR} = 10^{(1.55 + 0.542 \log L_{5100,44})} \ [\rm pc], \label{eq:rbroad}
\end{equation}
where $L_{5100,44}$ is the monochromatic luminosity at 5100 \AA{} in units of $10^{44}\,\rm erg/s$. We note that this definition implies $r_{\rm broad} \,{<}\, r_{\rm subl}$, meaning that the BLR lies within the dust sublimation radius\footnote{We have tested that the values derived for $r_{\rm subl}$ in  Eq.~\ref{eq:dustrsubl} and $r_{\rm broad}$ with Eq.~\ref{eq:rbroad} are consistent with $r_{\rm broad} \,{<}\, r_{\rm subl}$.}. As a result, the lines generated in this region will be subject to attenuation by the torus surrounding the MBH (see next section for details). The outer radius of the BLR is computed by \texttt{CLOUDY} and is defined as the point where the total hydrogen column density reaches $10^{24}\, \rm cm^{-2}$. Finally, the metallicity of the BLR is adjusted to be twice that of the NLR.

Once the key physical parameters are defined, Su et al. (in prep.) runs a suite of \texttt{CLOUDY} simulations to predict emission line luminosity across a broad range of AGN conditions. The resulting outputs are organized into a multidimensional grid that characterizes the emission of the 11 selected lines as a function of four parameters: i) black hole mass, ii) accretion rate, iii) gas metallicity, and iv) ionization parameter. Thus, for each accreting MBH in our simulated lightcone, the luminosity of the NLR and BLR emission lines is computed by interpolating this grid using the corresponding values of these four parameters. These are all tracked self-consistently by \lgbh{} except the ionization parameter, which we compute as:
\begin{equation}
    U \,{=}\, \frac{Q_s}{4\pi \, c \, n_{H} \, r_{i}^2},
\end{equation}
where $n_{\rm H}$ is the hydrogen number density, computed as $n_{\rm H,NLR}=\rho_{\rm BH, gas}(r_{\rm NLR})/m_{\rm H}\, \rm 1/cm^3$ for the NLR (see Eq.~\eqref{eq:MBH_Gas_Distribution}), and as $n_{\rm H,BLR}=n_{\rm H,NLR}\cdot10^{9.5}\, \rm 1/cm^3$, following the modelling of Su et al. (in prep). Finally, $Q_s$ is the ionization photon emission rate derived from the AGN continuum, $L_{\nu}$, as:
\begin{equation}
   Q_s \,{=}\, \int_{13.6\, \rm  eV}^{0.1 \, \rm GeV} \frac{L_{\nu}}{h \nu} \, dh\nu.
\end{equation}

Regarding the line profiles, we assume that narrow lines originating in the NLR can be approximated as Dirac delta functions. In contrast, broad lines from the BLR are modeled with Gaussian profiles whose full width at half maximum (FWHM) is determined by the virial velocity at the BLR radius:
\begin{equation}
{\rm FWHM}_i \, {=} \,  \left(\frac{\lambda_i}{c}\right) \sqrt{\frac{G M_{\rm BH}}{r_{\rm BLR}}} \,\,\,\, [{\rm \AA}],
\end{equation}
where $\lambda_i$ is the central rest-frame wavelength of the $i$-th line. We tested our AGN emission line model by comparing our $H_{\alpha}$ AGN luminosity function with the one measured by recent observational works, finding overall good agreement (see Appendix~\ref{appendix:EmissionLineTest}). 

\subsubsection{Attenuation of AGN spectra: Tours and ISM} \label{subsubsec:att_uv}

As we did for galaxies, it is essential to model the obscuration of the SED of accreting MBHs. In particular, the spectrum emitted by the central accreting MBH may pass through the torus before reaching the ISM of the host galaxy. Consequently, the final observed AGN SEDs may be obscured and attenuated by both the torus and the ISM. Here we outline the method we use to address both types of dust attenuation.\vspace{1mm}

\noindent -- \textbf{Torus obscuration}: this depends on the opening angle of the torus ($\varphi$) and the orientation $\alpha_{\rm LOS}$ of the observer LOS (defined in Section~\ref{subsubsec:galaxy_cont}). In particular, if 
% $\alpha_{\rm LOS}$ is smaller than the torus aperture ($\alpha_{\rm LOS} \,{\leq}\, \varphi$),
$\alpha_{\rm LOS} \,{\leq}\, \varphi$,
then the intrinsic emission from the accreting MBH is subject to torus obscuration. Thus, the incident radiation is attenuated according to the optical depth at the outer edge of the torus: $\tau_{\rm torus}(r_2)$. Conversely, if 
% $\alpha_{\rm LOS}$ falls beyond the torus aperture ($\alpha_{\rm LOS} \,{>}\, \varphi$), 
$\alpha_{\rm LOS} \,{>}\, \varphi$,
the incident radiation is not affected by the dusty torus. Consequently, the emerging spectrum $f_{\rm \lambda}^{\rm\, out}$ is given by:
\begin{equation}
    f_{\rm \lambda}^{\rm\, out} \,{=}\, 
    \begin{cases}
        f_{\rm \lambda}^{\rm inc} \quad\quad\quad\quad\quad\, {\rm if} \,\,\,\,\,\,\, {\alpha_{\rm LOS}} \,{\geq}\, \varphi,\\
        f_{\rm \lambda}^{\rm inc} \, {\rm e}^{-\tau_{\rm torus}(r_2)}\quad\ \ \, {\rm if} \,\,\,\,\,\,\, {\alpha_{\rm LOS} \,{<}\, \varphi,}
    \end{cases}     
\end{equation}
\noindent where $f_{\rm \lambda}^{\rm inc} \,{=}\, {\rm L_{\lambda}} / (4 \pi d_{L}^{2})$ is the incident radiation. We assume that the torus attenuation only affects $f_{\rm \lambda}^{\rm inc}$ at 912\,\AA{}$\,{<}\,\lambda\,{<}\,$6564 \AA{}, as these wavelengths have been used to model the torus emission (see Eq.~\ref{eq:dusttemperature}). Shorter wavelengths can photodissociate dust, therefore modeling attenuation at $\lambda<912$\AA{} requires a detailed treatment which lies beyond the scope of this work. Finally, we underline that NLR emission is not subject to torus attenuation as it originates beyond the torus boundaries, in contrast with the BLR light originating within the torus boundaries.\vspace{1mm}\\
\noindent --  \textbf{ISM attenuation}: Regardless of whether $f_{\rm \lambda}^{\rm \,out}$ has been reprocessed by the torus ($\alpha_{\rm LOS} \,{<}\, \varphi$) or not ($\alpha_{\rm LOS} \,{\geq}\, \varphi$), it may still undergo additional attenuation as it propagates through the galaxy ISM before reaching the observer. To determine whether ISM attenuation occurs, we define the angular extent of the galaxy disk as:
\begin{equation}
    \alpha_{\rm crit} \,{=}\, \arctan \left(\frac{h_d}{R_d}\right),
\end{equation}
where $R_d$ and $h_d$ represent the radial and vertical scale lengths of the galaxy disk, respectively. While \lgbh{} self-consistently tracks the evolution of $R_d$ \citep[see][]{Guo2011}, it does not explicitly model $h_d$. To estimate the vertical scale height, we follow the approach of \cite{Alonso-Tetilla2024} and assume $h_d\,{=}\,R_d/8$. Based on this, the final 
% radiation emitted by the accreting MBH 
observed AGN spectrum
% which reaches the observer 
is given by:
\begin{equation}
    f_{\rm \lambda}^{\rm\, out, Final} \,{=}\, 
    \begin{cases}
        f_{\rm \lambda}^{\rm\, out} & {\rm if}\ {\alpha_{\rm LOS}} \geq\alpha_{\rm crit},\\
        f_{\rm \lambda}^{\rm\, out} \, \left(\frac{1-e^{-\tau_\lambda^{\rm ISM}\csc{\alpha_{\rm LOS}}}}{\tau_\lambda^{\rm ISM}\csc{\alpha_{\rm LOS}}}\right)\ & {\rm if}\ {\alpha_{\rm LOS}} < \alpha_{\rm crit},
    \end{cases}
\end{equation}
where 
% we assume that 
the ISM optical depth $\tau_{\lambda}^{\rm ISM}$ is the same as the one in Eq.~\eqref{eq:att_ism_lgalaxies}.% for the galaxy component.

\subsection{Apparent magnitude: The JWST filter system}

As extensively described in previous sections, the resulting SED for a simulated object is obtained by combining its AGN and stellar SEDs. From these composite SEDs we then obtain our simulated JWST photometry. In detail, we apply the standard AB magnitudes definition \citep[see][]{OkeGunn1983} to the convolution between the composite SEDs (redshifted according to each source redshift) and the transmission curves of JWST filters.

% Since this study aims to characterise photometrically selected LRDs, we will conduct this procedure using the JWST filter system. Specifically, among the 18 filters equipped in the JWST, we will focus on the following: ${\tt F115W}$, ${\tt F150W}$, ${\tt F200W}$, ${\tt F277W}$, ${\tt F356W}$, and ${\tt F444W}$ (NIRCam filters), which correspond to the most typically used for photometric selections of LRDs (see \citealt{Labbe2024,kokorev24_lrds}).
Since this study aims to characterise photometrically-selected LRDs while maximizing the comparability with recent observations, we conduct this procedure using the ${\tt F115W}$, ${\tt F150W}$, ${\tt F200W}$, ${\tt F277W}$, ${\tt F356W}$, and ${\tt F444W}$ NIRCam filters, which correspond to those typically used for photometric selections of LRDs (see \citealt{Labbe2024,kokorev24_lrds}).

%Given the rest-frame luminosity of an object, $L_{\nu}$, the corresponding observed flux density, $f_{\lambda}$, can be calculated as:
%\begin{equation} \label{eq:Flux_From_Luminosity}
%    f_{\lambda} \,{=}\, (1+z) \frac{L_{\nu}}{4\pi d_L^2}\cdot\frac{c}{\lambda_{\rm obs}^2},
%\end{equation}
%where $d_L$ is the luminosity distance of the source, $c$ the light speed and $\lambda_{\rm obs}$ the observed wavelength. By integrating the flux density over a specific wavelength range, it is possible to calculate the magnitudes ($m$) in the AB magnitude system using:
%\begin{equation}
%    m \,{=}\, -2.5 \log_{10} \left[\frac{\int S(\lambda)\,\lambda\,f_\lambda\,d\lambda}{c \int S(\lambda)/\lambda \,d\lambda}\right] - 48.6\,.
%    \label{eq:abmag}
%\end{equation}
%where $0\leq S(\lambda)\leq1$ is the transmission curve of a given filter.}

\subsection{The light profile of galaxies and AGN}
\label{sec:light_profile_model}
Besides photometry, we also construct a toy model to determine the light profile of objects within a given filter$j$, i.e.: $f_{\lambda, j}({<}\, r)$. To do this,  we use a standard Sérsic law \citep{Sersic1963}:
\begin{equation} \label{eq:SersicIntensity}
    I(r) \,{=}\, I_e \, e^{-b_n\left[ \left(\frac{r}{r_e}\right)^{1/n} \,{-}\, 1\right]},
\end{equation}
where $b_n\,{=}\,1.9992n\,{-}\,0.3271$, whil $I_e$ is the intensity at the effective radius $r_e$ and $n$ the Sérsic index. For simplicity, we set $n\,{=}\,4$ for the galaxy bulge and $n\,{=}\,1$ for the galaxy disk. Therefore the total luminosity within a given radius $r$ in a filter $j$ can be determined as:

\begin{equation} \label{eq:Luminosity_within_Radius}
\begin{aligned}
    L_j(\,< r) &= 2\pi \int_{0}^{r} I(r') \, r' \, dr' \\
    &= 2\pi \int_{0}^{r} \left( I_{disk}(r') + I_{bulge}(r') \right) r' \, dr' \\
    &= 2\pi  \int_0^{r} r' \, I_e^{\rm disk} e^{-b_1 \left[ \left(\frac{r'}{r_e}\right) - 1 \right]} \, dr' \\
    &+ 2\pi  \int_0^{r} r' \, I_e^{\rm bulge} e^{-b_4 \left[ \left(\frac{r'}{r_e}\right)^{1/4} - 1 \right]} \, dr'
\end{aligned}
\end{equation}
%where $I_e$ is obtained independently for each component, disk or bulge, by doing the integral for $r=\infty$, where $L_j=L_{j,\rm bol}$, and we take advantage that \lgbh{} computes the photometry of each filter for the disk and bulge components of a galaxy \davcoment{"where $I_e$ is obtained [...] is not clear how you compute $I_e$ from the Luminosity."} \davcoment{Suggested rephrasing:
where $I_e$ is obtained independently for the disk and bulge components. \lgbh{} determines the luminosity of bulge and disk separately for each given filter $j$: $L_{j}^{\rm disk}$ and $ L_{j}^{\rm bulge}$. Consequently, we can define $I_e^{\rm bulge/disk}\,{=}\, \rm \mathit{L}_{\it j}^{bulge/dsic} / [ \mathit{r_e b_n^{-2/n}}\Gamma(2/n) ]$, where $r_e$ is the effective disk/bulge length computed by \lgbh{}. Finally, if a source has an AGN component, we add it to the galactic light profile as a point-like source. The final light profile $f_{\lambda,j}(< r)$ is then:
%\begin{equation}
%\begin{aligned}
%    f_{\lambda,j}(< r) & \,{=}\,f^{\rm stars}_{\lambda,j}(< r)+f^{\rm MBH}_{\lambda,j}(< r) \\
%    & = f^{\rm disk}_{\lambda,j}({<}\,r)+f^{\rm bulge}_{\lambda,j}(< r)+f^{\rm MBH}_{\lambda,j}(< r)
%\end{aligned}
%\end{equation}
\begin{equation}
\begin{aligned}
    f_{\lambda,j}({<}\,r) & \,{=}\,f^{\rm disk}_{\lambda,j}({<}\,r) \, + \, f^{\rm bulge}_{\lambda,j}(< r) \, + \, f^{\rm MBH}_{\lambda,j}
\end{aligned}
\end{equation}

\subsection{Classification of galaxies and AGN}
\label{sec:Classification_Galaxies_AGNs}

One key aspect of our work is the possibility to classify our simulated objects as galaxies or AGN. This is not a trivial choice, as the stellar emission and that of accreting MBHs can concur to produce the final SED of each object. To address this, different methods can be employed, based on photometric considerations. Specifically, an object may be classified as an AGN if the nuclear emission from the accreting MBH dominates the total observed flux over a set of filters. However, given the limited number of filters we use, this method is incomplete and does not account for all information on the final SED. To overcome this limitation, we employ an alternative classification based on the total bolometric luminosity of the MBH \cite[see][for a similar approach]{volonteri2024_lrdsinobelisk}. In detail, we classify an object as AGN when its accreting central MBH has an associated $L^{\rm AGN}_{\rm bol}\,{\geq}\,10^{44}\, \rm erg/s$. Otherwise, the object is classified as a galaxy.

\begin{figure}
    \centering
    \includegraphics[width=0.85\columnwidth]{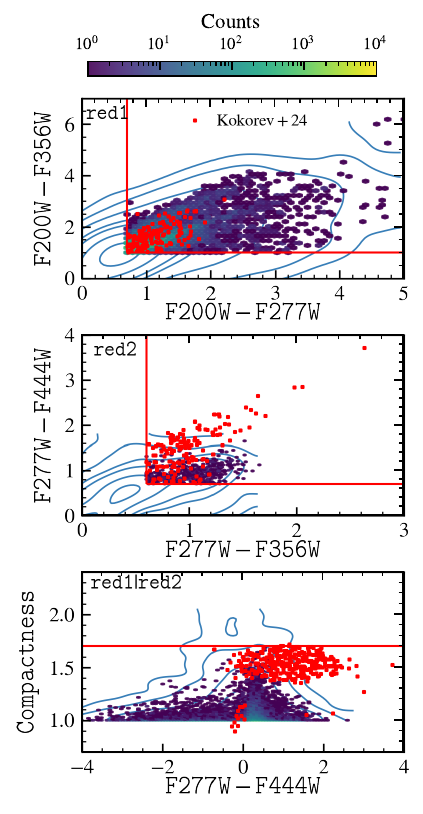}
    \caption{Color-color (upper and middle panels) and compactness versus color (lower panel) diagrams used to define our photometrically selected \lrds{}. The blue contours correspond to all detected objects. The color map show the distribution of simulated objects from our lightcone (the color code represents the number of selected objects in each bin). The red dots represent the sample from \protect\citep{kokorev24_lrds}, with the vertical and horizontal red lines indicating the photometric cuts we apply to replicate their selection.}
    \label{fig:Color_Color_Selection}
\end{figure}

%-----------------------------------------------------------------------------
%-----------------------------------------------------------------------------
% \section{The photometrically selected Little Red Dot population}
\section{Photometric selection of simulated Little Red Dots}
\label{sec:photometric_selection_of_LRDs}
Here we present the photometric cuts we use to define our population of \lrds{}, which are a close replica of those introduced by \cite{labbe2023_lrds} and used in \citealt{kokorev24_lrds}:
\[
{\rm \tt red 1} \,{=}\, \begin{cases}
{\tt F115W} \,{-}\, {\tt F150W} \,{<}\, 0.8 \\
{\tt F200W} \,{-}\, {\tt F277W} \,{>}\, 0.7\, , \\
{\tt F200W} \,{-}\, {\tt F356W} \,{>}\, 1.0
\end{cases}
\]
\[
{\rm \tt red 2} \,{=}\, \begin{cases}
{\tt F150W} \,{-}\, {\tt F200W} \,{<}\, 0.8 \\
{\tt F277W} \,{-}\, {\tt F444W} \,{>}\, 0.7\, , \\
{\tt F277W} \,{-}\, {\tt F356W} \,{>}\, 0.6
\end{cases}
\]
\[
{\rm \tt compactness} \,{=}\, \frac{f_{\lambda\,,\,{\tt F444W}}({<}\,0.4'')}{f_{\lambda\,,\,{\tt F444W}}({<}\,0.2'')} < 1.7,
\]
The ${\rm \tt red1}$ and ${\rm \tt red2}$ cuts are expected to select LRDs at $z\,{<}\,6$ and $z\,{>}\,6$ respectively \citep[see][]{kokorev24_lrds}. Interestingly, we find that our simulated colors agree well with these expected redshifts intervals (see Appendix~\ref{appendix:RedshiftSelection_red1_red2} for the redshift distribution of each color cut). We underline that, following the results from \cite{kokorev24_lrds}, we exclude all objects selected by the above criteria but which show $z\,{<}\,4$. Moreover, we also exclude objects selected by the {\tt red2} cut with $z<6$, given their negligible contribution to the overall properties of the {\tt red2}-selected population of objects. On top of the previous cuts, we add the extra cut of:
\begin{equation}
    {\tt bd\_removal} \,{=}\, {\tt F115W} \,{-}\, {\tt F200W} \,{>}\, -0.5\,
\end{equation}
which is expected to reduce the contamination of brown dwarfs from the selected sample. We underline that, even if we do not model brown dwarfs, we apply this cut to maximize the comparability of our results with recent observational works.

For analogous reasons, we also apply realistic detection limits to all our samples. First of all, we impose a detection cut based on the deepest magnitudes detectable by NIRCam in the fields typically used for LRDs searches (namely, the GOOD-S field depth values reported in Table 1 of \citealt{kokorev24_lrds}). In detail, we only consider objects that satisfy:  ${\tt F150W} \,{<}\, 29.6$, ${\tt F200W} \,{<}\, 29.5$, ${\tt F277W} \,{<}\, 29.8$, ${\tt F356W} \,{<}\, 29.6$,\, ${\tt F444W} \,{<}\, 29.3\, \rm mag$. Then, we explicitly replicate the approach of \cite{kokorev24_lrds} by only selecting objects with $\rm {\tt F444W} < 27.7\, mag$.
%We also impose a detection cut that requires that the object must be brighter than the deepest magnitudes NIRCam can detect for the fields used to build the observational sample of LRDs, which corresponds to the GOODS-S field. To be explicit, this conditions means that we will only include objects in the analysis that meet the following conditions: ${\tt F150W} < 29.6,\, {\tt F150W} < 29.6,\, {\tt F200W} < 29.5,\, {\tt F277W} < 29.8,\, {F356W} < 29.6,\, {F444W}< 29.3$.

The color distributions of our selected samples are presented in Fig.~\ref{fig:Color_Color_Selection}. For comparison, we have included a collection of data from \cite{kokorev24_lrds} as red points. 
% As shown, our simulated colors are in good agreement with observations, despite some discrepancies. 
Overall, in the {\tt red1} color space (upper panel), we find 30.05\% of our simulated objects within the photometric colors selection, although with a larger spread than observational LRDs. More in detail, we find that 89.71\% are compatible with the colors of observed LRDs in \cite{kokorev24_lrds} ($0.7\,{<}\,$ {\tt F200W}-{\tt F277W} $\,{<}\,1.7$ and $1\, {<}\,$ {\tt F200W}-{\tt F356W} $\,{<}\, 2.4$). Analogously, for the {\tt red2} cut (middle panel) the 15.93\% of our simulated population pass our color selection, with 67.45\% in agreement with the colors observed by \cite{kokorev24_lrds} (we note how the maximum {\tt F277}-{\tt F444W} color we get is 1.66 in comparison to the $\sim\!2\,\rm dex$ values of \cite{kokorev24_lrds}). Although this being a qualitative assessment, it shows that our model is able to produce reliable simulated colors up to $z>6$, compatible with those of the observed LRDs presented by \cite{kokorev24_lrds}. Overall, these discrepancies are likely due to the different area surveyed by \cite{kokorev24_lrds}, smaller by a factor of 0.017 with respect to the large $5\,\rm deg {\times}\,5\,\rm deg$ area we model with our lightcone (which translates to $0.094\, \rm deg^2$ in \cite{kokorev24_lrds} against the $5.62\, \rm deg^2$ we get). Indeed, this may impact the effective extend of the color-space region which can be sampled observationally. On the other hand, it is possible that biases in the simulation of our photometric colors, may produce larger variations than those observed in real data. Nevertheless, since the bulk of our simulated objects show color compatible with the observed ones, we consider these discrepancies as negligible.

\begin{table}
    \caption{Number (fractions) of objects selected by the color and compactness cuts (upper and middle row) as well as by the combination of the two (i.e. our \lrds{} definition, bottom row). Each column shows a different redshift bin. The numbers and fractions are applied only to the sample of \textit{detected} objects, after removing those identified by the {\tt bd\_removal} cut. Red text shows the numbers and fractions for AGN.}
    \begin{tabular}{c r|c|c}
        \multicolumn{2}{l|}{Photometric cut}  & $4\leq z <6$    & $z\geq6$\\
        \hline
        \hlx{v}
                                      & All & 13859 (30.26\%) & 621 (15.93\%) \\
        \hlx{v}
        \multirow{-2}{*}{Color-color} & AGN & {\color[HTML]{9A0000} 5816 (75.32\%)} & {\color[HTML]{9A0000} 151 (82.07\%)}  \\
        \hline
        \hlx{v}
                                      & All & 45744 (99.88\%) & 3896 (100.00\%) \\
        \hlx{v}
        \multirow{-2}{*}{Compactness} & AGN & {\color[HTML]{9A0000} 7719 (99.97\%)} & {\color[HTML]{9A0000} 184 (100.00\%)} \\
        \hline
        \hlx{v}
                                      & All & 13763 (30.05\%) & 621 (15.93\%) \\
        \hlx{v}
        \multirow{-2}{*}{\lrds{}}     & AGN & {\color[HTML]{9A0000} 5761 (74.61\%)} & {\color[HTML]{9A0000} 151 (82.07\%)} 
    \end{tabular}
    \label{tab:lrd_numbers}
\end{table}

Finally, the compactness selection is shown in the bottom panel of Fig.~\ref{fig:Color_Color_Selection}. Our selected objects tend to be relatively more compact than those in the \cite{kokorev24_lrds} sample. This is likely due to our simplified light-profile modeling (see Sect.\ref{sec:light_profile_model}). In particular, our normalization of the Sérsic light profile does not account for the variation of its spatial extent across different filters ($r_e$ in Eq.~\ref{eq:SersicIntensity}). The lack of modeling of the JWST point spread function (PSF) may also be affecting this result although only to a minor extent, being the {\tt F444W} filter PSF $\sim0.14''$. Overall, our compactness cut is the less restrictive among those we impose, including $>99\%$ of objects at $z>4$ (see Table~\ref{tab:lrd_numbers}). Consequently, the effect of possible compactness biases on our work are negligible. We also note that \cite{kokorev24_lrds} does not provide a number of objects selected with this cut amongst his detected sample, so assessing the effectiveness of our cut is not as straight forward. Including a more detailed model of the Sérsic profile and JWST PSF may improve the agreement between the compactness of our sources and those in \cite{kokorev24_lrds}. Nevertheless, addressing this point is beyond the scope of this work.

Hereafter, we define detected objects as those that meet the depth limit described above. Among these, objects that also satisfy the color and compactness criteria will be referred to as \lrds{}. Conversely, objects that meet the detection limit but do not fulfill the color–color and compactness criteria will be referred to as \nolrd{}.
%This classification allows us to form three samples: 1) a galaxy {\tt gal} sample, of objects whose BH has not been assigned an SED (i.e.: they do not satisfy the above conditions); 2) an AGN sample, comprised of MBHs with SED assigned that also satisfy that their bolometric luminosity $L_{\rm bol.}\geq10^{44}\, {\rm [erg/s]}$; and 3) a composite sample, of those SED-ed MBHs with $L_{\rm bol.}<10^{44}\, {\rm [erg/s]}$ to tackle the properties of the possibly yet to discover faint AGN population. For now, we include this ``hybrid'' group in the galaxy sample unless otherwise specified, since observationally they will be likely considered as galaxies.
%\dani{\textbf{IMPORTANT} here we need to define the AGN selection, because we immediately need it (already in Fig.\ref{fig:RedshiftDistribution}, for instance, but also in the Results section)}. \davcoment{We did it in Section~\ref{sec:Classification_Galaxies_AGNs}: Classification between galaxies and AGNs.}

%-----------------------------------------------------------------------------
%-----------------------------------------------------------------------------
\section{Results} \label{sec:results}
In this section, we present our simulated \lrds{} and \nolrds{} populations, starting from the analysis of their properties across different redshifts (Section~\ref{sec:physical_properties_of_phot_selected_LRDs}). In Section~\ref{sec:the_role_of_gals_and_MBHs_in_LRD_photometry} we analyze how our selection of objects as \lrds{}, based on realistic photometric cuts (see Section~\ref{sec:photometric_selection_of_LRDs}), depends on the evolution of their galaxy and MBH properties across redshift. Finally, in Section~\ref{sec:Gals_and_AGN_in_the_LRD_population}, we present a deeper look into the \lrds{} population, separating them into galaxies or AGN classes (respectively \lrdGal{} and \lrdAGN{}). Our findings will help to contextualize our simulated \lrds{} within the current paradigm of early galaxy and MBH evolution, providing a tentative interpretation of recent JWST observations.

%-----------------------------------------------------------------------------
\subsection{The properties of photometrically selected LRDs}
\label{sec:physical_properties_of_phot_selected_LRDs}
Here we analyze the overall properties of our \lrd{} sample, showing their redshift distribution (Section~\ref{sec:redshift_evolution_of_LRDs}) and comparing their properties against the sample of \nolrds{} (Section~\ref{sec:global_properties_of_LRDs}).

\subsubsection{Redshift distribution of \lrds{}}
\label{sec:redshift_evolution_of_LRDs}
\begin{figure}
    \centering
    \includegraphics[width=\columnwidth]{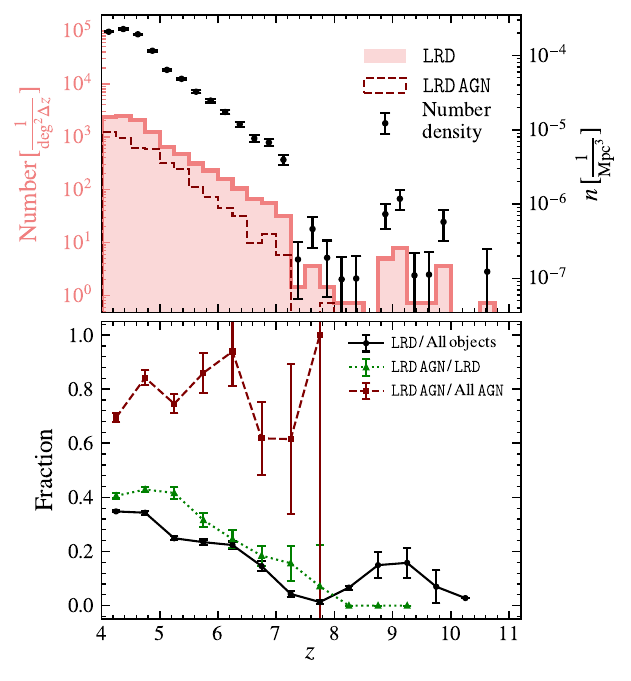}
    \caption{\textbf{Upper panel}: Redshift distribution (red histograms) and number density (black points) of the detected \lrds{}. \textbf{Lower panel}: Fraction of \lrds{} over the total number of detected systems (black points), fraction of \lrds{} classified as AGN over the total number of detected objects classified as AGN (red points) and fraction of \lrds{} classified as AGN over the number of \lrds{} (green points). In all the plots, the error bars correspond to the Poisson error.}
    \label{fig:RedshiftDistribution}
\end{figure}
The redshift distribution of our \lrds{} is shown in Fig.~\ref{fig:RedshiftDistribution}. The photometric selection we apply produces a sample that spans over $4\,{<}\,z\,{<}\,10$, with a peak at $z\,{\sim}\,4$. In the same figure, we also show the evolution of the \lrds{} number density,%\footnote{We associate a simple $1\!/\!\sqrt{N_{}}$ error to the number counts of our \lrds{} in each redshift bin.}, 
which exhibits an increasing trend with redshift. More in detail, at $z\,{\sim}\,7$ the \lrds{} number density is ${\sim}\,8\,{\times}\,10^{-6}\, \rm Mpc^{-3}$, while at $z\,{\sim}\,4$ it reaches ${\sim}\,2\,{\times}\,10^{-4}\, \rm Mpc^{-3}$ where the number of our \lrds{} flattens. These values are smaller than the value of ${\sim}\,2\,{\times}\,10^{-4}\, \rm Mpc^{-3}$ at $z\, {\sim}\, 7$ reported by \cite{PerezGonzalez2024}, which they claim to be constant between $4\,{<}\,z\,{<}\,9$. In contrast, we find a $\sim\!1\,\rm dex$ drop between $z=5$ and $z=7$, although then we find that the number density sets at at constant value of ${\sim}\,8\,{\times}\,10^{-6}\, \rm Mpc^{-3}$ (within error bars). These differences likely arise as a consequence of their different color selection and survey depth. The incising trend of \lrds{} towards low-$z$ can be expected from the combination of completeness effects of flux-limited samples and the fixed mass-resolution of our underlying simulation. Specifically, by applying magnitude cuts (see Section~\ref{sec:photometric_selection_of_LRDs}), we remove from our samples an increasing number of faint objects at higher redshifts and with the progress of structure growth towards lower redshifts, an increasing number of objects can be resolved and tracked along their redshift evolution within the MSII merger trees. 
%variations in color selection and depth limits used in the two studies. 

Although \lrds{} have a relatively high number density in our samples, they do not represent the full population of detected objects in our lightcone. This is analyzed in the lower panel of Fig.~\ref{fig:RedshiftDistribution} where we present the redshift evolution of the fraction of \lrds{} over the total number of detected objects (black line). We find that at $z\,{>}\,6$ the population of LRDs only accounts for the $10\%\,{-}\,20\%$ of the whole detections. This increases at $4\,{<}\,z\,{<}\,6$ where \lrds{} account for $30\%-40\%$ of objects in our lightcone. Notably, \cite{PerezGonzalez2024} find that at $z\sim\!7$, the incidence rate of LRDs in their observed sample is $14\pm 3\%$, in agreement to what we report. 
\begin{figure*}
    \centering
    \includegraphics[width=1.9\columnwidth]{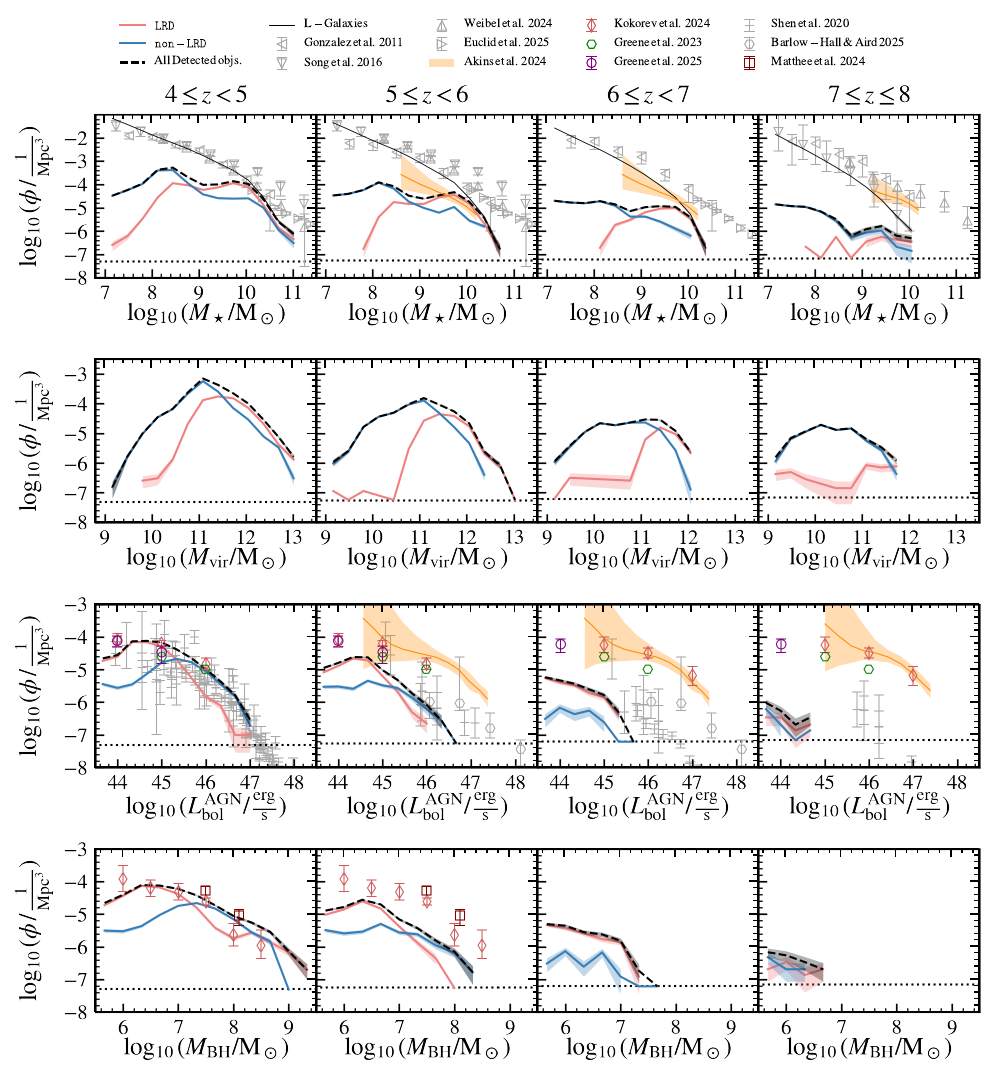}
    \caption{Properties of \lrds{} and \nolrds{} samples (respectively, pink and blue lines) across different redshifts. Different columns correspond to different redshift bins, as shown by the column titles. From top to bottom, on each row we show: the stellar mass function, halo mass function, the AGN bolometric luminosity function and the BH mass function. A collection of recent observational data is shown as points (either gray or colored) and shaded areas (in orange), as specified by the legend (on top of the figure). Colored literature points refer to LRD-derived quantities, while the rest correspond to the same as Fig.~\ref{fig:lgalaxies_validation} but with the inclusion of \protect\cite{Weibel2024} and \protect\cite{EuclidStellarMassFunction2025} on the stellar mass function. Finally, in each panel the dashed lines show the distributions of all detected objects in our lightcone (i.e. \lrds{} and \nolrds{} together), while the horizontal dotted line shows the number density we would measure if we had only 1 object in the whole lightcone at that redshift. The thin black line in the top row correspond to the \lgbh{} line without applying any cuts, as in Fig.~\ref{fig:lgalaxies_validation}.}
    \label{fig:Properties_Mstellar_Mvir_Lbol_MBH_LRDs_no_LRDs}
\end{figure*}

To further contextualize our \lrds{}, we also show which fraction of this sample is classified as AGN (AGN: $\L_{\rm bol}^{\rm AGN}\geq 10^{44}\, \rm erg/s$, green line). Our results show a clear redshift evolution: less than 20\% of \lrds{} are AGN at $z\,{>}\,7$, rising up to $30\%\,{-}\,40\%$ at $z\,{<}\,7$. Although this indicates that our \lrds{} are primarily classified as galaxies (hereafter \lrdGal{}), those classified as AGN (henceforth \lrdAGN{}) make up the vast majority of all detected AGN sample in our lightcone. This is also shown in the lower panel of Fig.~\ref{fig:RedshiftDistribution} where \lrdAGN{} make up for ${>}\,70\%$ of the entire population of detected AGN in our lightcone, independently of redshift.

We stress that estimating these fractions is a key result of our work, which allows us to infer the role played by LRDs within  the broader population of high-$z$ objects, as well as their prevalence in photometrically selected (AGN) samples. We also acknowledge that the fractions reported are dependent on the definitions of each sample. Most notably, relaxing the cut on the {\tt AGN} definition increase the incidence of \lrdAGN{} in the \lrd{} by 10 to 30\% (see Appendix~\ref{appendix:alternative_agn_definition}). This information is generally challenging to obtain from high-$z$ surveys, due to limitations in either their sampled area, their achieved depth, or their ability to reliably classify high-$z$ sources. In this regard, our study helps fill this observational gap, providing a view of photometrically selected LRDs and LRD AGN within their full cosmological context over a broad redshift interval.

% \subsubsection{Properties of \lrds{} and \nolrds{} samples}
\subsubsection{Physical properties of \lrds{} and \nolrds{} samples}
\label{sec:global_properties_of_LRDs}
Understanding \lrds{} requires examining their properties and comparing them with those of \nolrds{}. In Fig.~\ref{fig:Properties_Mstellar_Mvir_Lbol_MBH_LRDs_no_LRDs} we show a comprehensive picture of \lrds{} and \nolrds{}\footnote{As a reminder for the reader, we select \lrds{} and \nolrds{} amongst our sample of objects that meet the detection limit based on if they meet the photometric and compactness cut (\lrds{}) or not (\nolrds{})} properties, comparing these two classes of objects in different redshift bins. From top to bottom, each row shows: the SMF, the halo mass function (HMF), the AGN LF and finally the BHMF.

Regarding the SMF, \lrds{} represent at any redshift the dominant galaxy population at $M_\star\,{>}\,10^{10}\, M_{\odot}$, with their distribution exhibiting a prominent peak at $M_{*} \,{\sim}\, 10^{10}\, \msun$ \citep[see similar results in the simulations of][]{LaChance2025}. In contrast, the SMF of \nolrds{} peaks at significantly lower masses, around $M_{*} \,{\sim}\, 10^{8}\, M_{\odot}$. The SMF comparison also shows that \nolrds{} are a factor of ${\sim}\,5$ less abundant than \lrds{} in galaxies with $\,10^9\,{<}\,M_{*}/\msun\,{<}\,10^{10}$ and become as abundant as \lrds{} only for $M_{*} \,{>}\, 10^{10.5} \, \msun$ at $4\!<\!z\!<\!5$. These trends align with recent observational studies, which suggest that \lrds{} typically are galaxies with stellar masses of $M_{*} \,{\sim}\, 10^{9-10}\, \msun$ \citep[see][]{Baggen2024, PerezGonzalez2024, Leung2024, Akins2025}. Note that the population of \lrds{} with $M_\star<8\, {\rm M_{\odot}}$ is minimal, especially at $z>5$ (indeed it is missing between $6\leq z < 7$). For the SMF we also provide the SMF derived from \lgbh{} without applying any cut as in Fig.~\ref{fig:lgalaxies_validation}. When comparing the SMF of detected objects with the total SMF, we see that most galaxies above $M_\star \sim 10^{10}\, \rm M_\odot$ are within the detection limit, while the number decreases BY $1.5-2.5\, \rm dex$ at lower masses. As for the HMF, \lrds{} are found within halos spanning a wide mass range of $5 \,{\times}\, 10^{10} \,{<}\, M_{\rm vir} \,{<}\, 10^{13}\, M_{\odot}$, with a characteristic peak at $M_{\rm vir} \,{\sim}\, 5 \,{\times}\, 10^{11}\, \msun$ \citep[see similar findings in][]{Schindler2025,CarranzaEscudero2025}. In contrast, \nolrds{} are consistently associated with lower-mass halos, ranging from $10^{9} \,{<} \, M_{\rm vir} \,{<}\, 10^{12} \, \msun$, with a typical halo mass around $M_{\rm vir} \,{\sim}\, 10^{11} \, \msun$.

Regarding the population of MBHs, it is clearly dominated by \lrds{}at $z\,{>}\, 6$, as shown by our BHMF. \lrds{} span a broad mass range, from $10^6 \,{<}\, M_{\rm BH} \,{<}\, 5 \,{\times}\,10^8 \, \msun$. However, this trend shifts at $4 \,{<}\, z \,{<}\, 6$, where the number of \lrds{} with $M_{\rm BH} \,{>}\, 10^7 \, \msun$ declines sharply. In this higher-mass regime, \nolrds{} become more prominent, with their number density increasing by roughly one order of magnitude compared to \lrds{}. We underline that this shift in BH properties has a significant impact on the photometric selection of our simulated objects as \lrds{}, as we will analyze below in greater detail (Section~\ref{sec:imprint_of_galaxy_properties_on_LRD_photometry} and \ref{sec:imprint_of_BH_properties_on_LRD_photometry}). Despite this shift, \lrds{} continue to dominate the mass function at the low-mass end ($M_{\rm BH} \,{<}\, 10^7 \, \msun$), maintaining their significant presence within the overall MBH population. Interestingly, the trends we observe in the BHMF are reflected in the AGN LFs. At $z\,{>}\, 6$, \lrds{} dominate the LFs, spanning over two orders of magnitude in luminosity ($10^{44} \,{<}\, L^{\rm AGN}_{\rm bol} \,{<}\, 10^{46} \, \rm erg/s$). However, at lower redshifts, the \lrds{} population experiences a significant decline in the number of luminous AGN ($L^{\rm AGN}_{\rm bol} \,{>}\, 10^{46} \, \rm erg/s$), occupying primarily the intermediate-luminosity range ($10^{44} \,{<}\, L^{\rm AGN}_{\rm bol} \,{<}\, 10^{45} \, \rm erg/s$). In contrast, \nolrds{} increasingly dominate the LF bright end at z<6, becoming the primary contributors to the high-luminosity AGN population (see \citealt{ma2025_LFcutoff} for similar findings).

The results presented in this section reveal two key characteristics of our photometrically selected \lrds{}. First, they are consistently found at the most massive end of the stellar and halo mass function ($M_\star\,{>}\, 10^{9.5} \, M_{\odot}$, and $M_{\rm vir} \,{>}\, 10^{11.5} \, M_{\odot}$), regardless of redshift. From the SMF, in addition, we observe the emergence of a low-mass population of \lrds{} with $M_\star\, {\sim}\,10^{8.5}\, M_{\odot}$ at $z<6$. At the same time, MBHs in the \lrd{} sample tend to be moderately luminous ($L^{\rm AGN}_{\rm bol}\,{\sim}\,10^{45}\, \rm erg/s$) and massive ($M_{\rm BH}\, {\sim}\,10^{7}\, M_{\odot}$). These results highlight a notable distinction in the evolution of simulated \lrds{} versus \nolrds{} in the high stellar mass regime: the former are typically massive galaxies, where stellar growth tends to precede that of their central MBHs. In contrast, at $z\,{<}\,6$, the MBHs hosted in \nolrds{} tend to be more massive than that of \lrds{}, thereby populating the bright end of the AGN LF. As shown in the next sections, this directly affects the photometry of our simulated objects and their selection as \lrd{}. We stress that our findings do not portray \lrds{} and \nolrds{} as \textit{monolithic} populations, but rather as complex and heterogeneous classes, each comprising a variety of objects.% In the following sections, we will explore the photometry of both \lrds{} and \nolrds{}, focusing on how the evolving stellar and MBH properties shape their photometric signatures.

\subsection{The LRD photometry: interplay between galaxy and MBH}
\label{sec:the_role_of_gals_and_MBHs_in_LRD_photometry}
\begin{figure*}
    \centering
    \includegraphics[width=2\columnwidth]{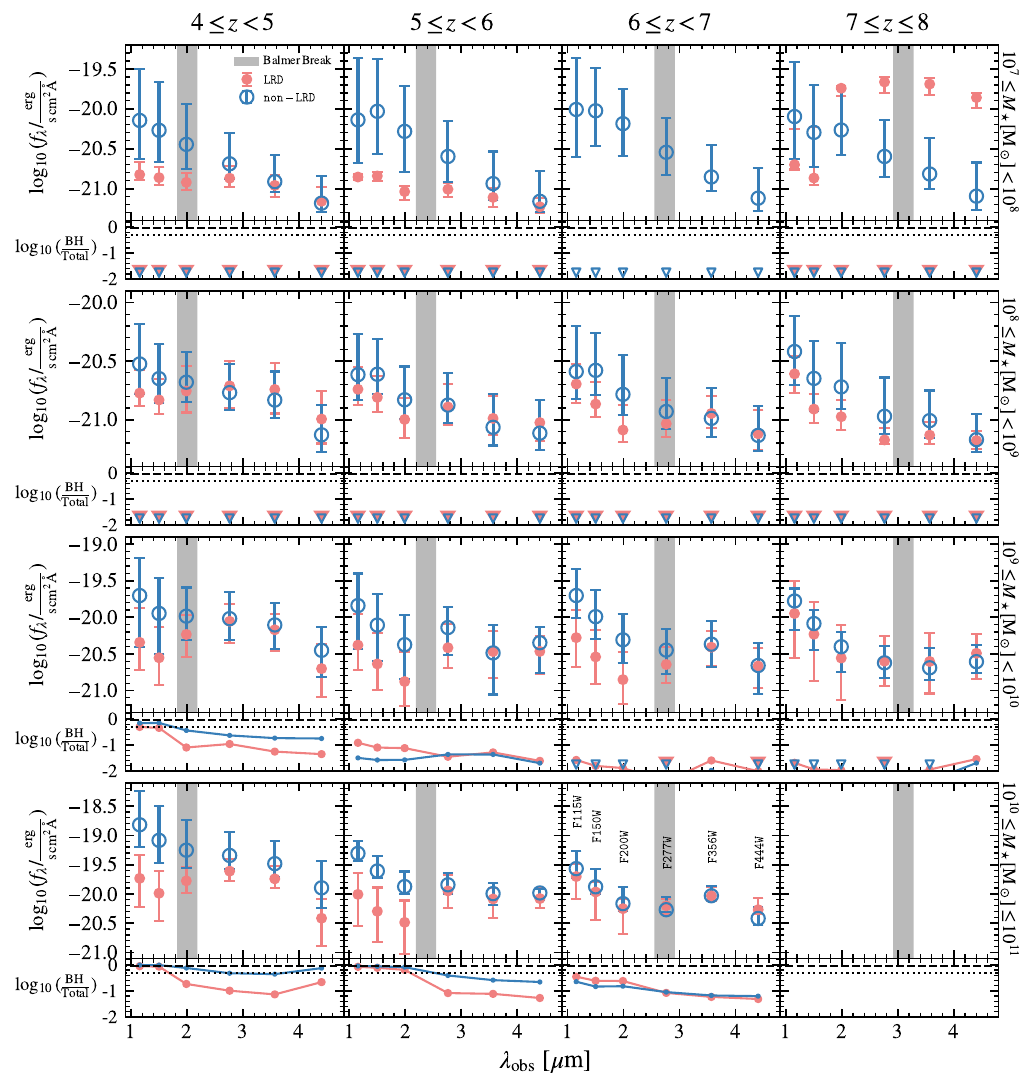}
    \caption{Median photometry of the \lrd{} (red) and \nolrd{} (blue) samples at different redshifts. The error bars correspond to the $\rm 16^{th}\,{-}\,84^{th}$ percentiles while the vertical gray shaded areas mark the Balmer break position. The small panels show the fractional contribution of the flux provided by AGN (i.e. accreting MBHs) within each JWST filter. The triangles represent the upper limits of this contribution. To guide the reader, the horizontal dashed lines highlight the 90\% and 50\% value (half of the flux is provided by the AGN). Different panels show different $M_\star$ bins, from top to bottom: $10^7 \,{\leq}\, M_\star \,{<}\, 10^8 \, M_\odot$, $10^8 \,{\leq}\, M_\star \,{<}\, 10^{9} \, M_\odot$ and $10^9 \,{\leq}\, M_\star \,{\leq}\, 10^{11} \, M_\odot$.}
    \label{fig:lrd_photometry}
\end{figure*}

To better understand the selection of objects classified as LRDs, in Fig.~\ref{fig:lrd_photometry} we present the comparisons between the median photometry of \lrd{} and \nolrd{} samples.
%As shown, regardless of the host galaxy $M_\star$ and redshift, the \lrd{} population exhibits a characteristic ``V-shaped'' photometric profile, with the ${\tt F115W}$ and  ${\tt F150W}$ filters showing lower flux densities compared to the ${\tt F277W}$ and ${\tt F356W}$ filters, though we see a decline in flux in the ${\tt F444W}$ filter.
Generally speaking, at $z\,{<}\,7$ and independently of the host galaxy's stellar mass, the \lrd{} population exhibits a characteristic ``V-shaped'' photometric profile. This is visible as a decrease in flux immediately after the Balmer break. The exceptions to this feature are the lowest $M_\star$ bin at $6\,{<}\,z\,{<}\,7$ and the whole $7\,{<}\,z\,{<}\,8$ redshift bin. In the first case, we cannot identify any \lrd{} object in this mass and redshift bin. In the case of $7\,{<}\,z\,{<}\,8$, we find an overall very small number of detected objects (see also Figs.~\ref{fig:RedshiftDistribution} and \ref{fig:Properties_Mstellar_Mvir_Lbol_MBH_LRDs_no_LRDs}), hence any features in our median photometry are erased by the low sample statistics. Regarding the \nolrd{} sample, their median photometry does not display any ``V-shaped'' feature. Indeed, at any $z$ and $M_\star$, it generally shows a decreasing trend with wavelength, indicative of blue photometric colors for \nolrd{}.

To investigate the origin of the ``V-shaped'' feature in \lrds{}, the small panels of Fig.~\ref{fig:lrd_photometry} show the percentage of flux within a given filter that comes from an AGN (accreting MBH). Focusing on the lowest and intermediate stellar mass bins ($10^7 \,{\leq}\, M_\star \,{<}\, 10^8 \, \msun$ and $10^8 \,{\leq}\, M_\star \,{<} \, 10^9 \, \msun$, i.e. top and second rows in Fig.~\ref{fig:lrd_photometry}), we find that the observed flux is dominated entirely by the stellar component, regardless of redshift. This implies that, for \lrds{} at this low-mass, the ``V-shaped'' spectral feature originates solely from the Balmer break (${\gtrsim}\,3000\,{-}\,4000\, \AA$, rest-frame) produced by the stellar continuum. Notably, this is also valid at all redshifts for \nolrds{} in the lowest mass bin. The strong prevalence of the stellar continuum in the photometry, coupled with the different spectral shapes at fixed stellar mass, suggests that low-mass \lrds{} and \nolrds{} occupy distinct evolutionary stages. As we will show in Section~\ref{sec:imprint_of_galaxy_properties_on_LRD_photometry}, low- and intermediate-mass \lrds{} exhibit systematically larger metallicity and generally lower specific star formation rates (sSFR) when compared to their \nolrd{} counterparts. These properties are indeed consistent with \lrds{} showing a stronger Balmer break than \nolrds{}.
%the reduced flux in the bluer filters.
%For systems with , we observe similar trends, where the "V-shaped" photometry is entirely driven by the galaxy, with no contribution from the AGN.

Moving to the photometry of objects in the most massive galaxies ($10^{9} \,{\leq}\, M_\star \,{<}\, 10^{10} \, \msun$ and $10^{10} \,{\leq}\, M_\star \,{\leq} \, 10^{11} \, \msun$, third and bottom row in Fig.~\ref{fig:lrd_photometry}), we also observe different trends depending on redshift and stellar mass, with the AGN contribution to the photometry increasing with lower redshift and higher $M_\star$. At $6 \,{\leq}\, z \,{<}\, 7$ and $7 \,{\leq}\, z \,{\leq}\, 8$, the behavior is similar to that of low-mass galaxies, that is: the AGN component rarely exceeds $\sim\!1\%$ (except for the highest mass bin, where the MBH contribution is almost $\sim50\%$) while the galaxy component dominates the flux in all filters. 
However, at $4 \,{\leq}\,z \,{<}\, 5$ and $5 \,{\leq}\, z \,{<}\, 6$, the AGN contribution is appreciable across all filters (even dominating in some cases). At $5 \,{\leq}\, z \,{<}\, 6$, the photometry of \lrds{} with $10^{9} \,{\leq}\, M_\star \,{<}\, 10^{10} \, \msun$ shows a $\sim\!4\%-8\%$ contribution from AGN in all filters, while this raises to $\sim\!10\%-15\%$ in ${\tt F115W}$, ${\tt F150W}$ and ${\tt F200W}$. In that same $z$ and $M_\star$ bin, the MBH in \nolrds{} contributes $\lesssim 8\%$.
In contrast, at $4 \,{\leq}\, z \,{<}\, 5$, 50\% of the flux in ${\tt F115W}$ and ${\tt F150W}$ is due to the AGN emission for \lrds{}, while this contribution sits at $\sim\!10\%$ for ${\tt F200W}$, ${\tt F277W}$, ${\tt F356W}$, and ${\tt F444W}$. A similar pattern holds for the \nolrd{} sample although with an even stronger contribution from AGN. Indeed, this reaches 90\% in ${\tt F115W}$ and ${\tt F150W}$ at $4 \,{\leq}\, z \,{<}\, 5$ and it accounts for $>\!40\%$ in ${\tt F200W}$, ${\tt F277W}$, ${\tt F356W}$, and ${\tt F444W}$. In the highest mass bin, $10^{10} \,{\leq}\, M_\star \,{\leq} \, 10^{11} \, \msun$, these patters are also seen in a more extreme version. Indeed, both \lrds{} and \nolrds{} are dominated by the MBH emission by $\gtrsim\! 60-90\%$ in the {\tt F115W}, {\tt F150W} (in the $4\,{\leq}\, z\, {<} \, 5$ and $5\, {\leq} \, z \, {<} \,6$ bins) and {\tt F200W} ($5\, {\leq} \, z\, {<}\, 6$ bin) filters, while this decreases to $\sim\!10\%$ for the rest of the filters in \lrds{} but stays at $\gtrsim\! 40\%$ for \nolrds{}.

All the results presented above show that the classification of an object as an \lrd{} is closely linked to either the properties of the galaxy stellar population (in low-mass systems) or the interplay between AGN activity and stellar emission (in high-mass systems). In the following sections, we examine these two regimes in detail, investigating both the galaxy and MBH properties that differentiate the \lrd{} and \nolrd{} samples.

\subsubsection{Imprint of galaxy properties on \lrds{} photometry} \label{sec:imprint_of_galaxy_properties_on_LRD_photometry}

\begin{figure}
    \centering
    \includegraphics[width=1\columnwidth]{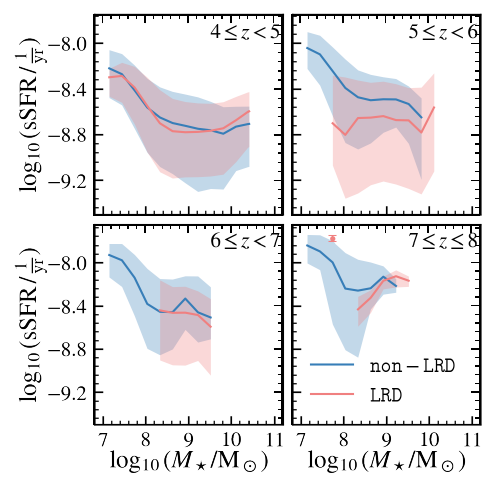}
    \caption{Specific star formation rate (sSFR) and stellar mass plane for the \lrd{} and \nolrd{} samples (red and blue, respectively). The lines and the shaded areas correspond to the median and $\rm 16^{th}\,{-}\,84^{th}$ percentiles of the sSFR at a fixed stellar mass. Each panel represents a redshift bin. Dots are shown for cases where only one object is found.}
    \label{fig:galaxy_main_sequence}
\end{figure}

\begin{figure}
    \centering
    \includegraphics[width=1\columnwidth]{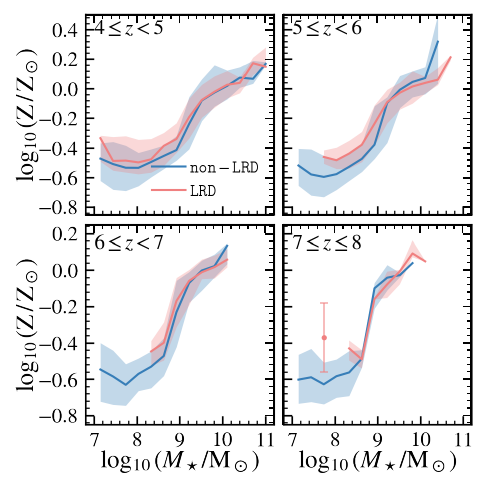}
    \caption{Galaxy metallicity (Z) and stellar mass plane for the \lrd{} and \nolrd{} samples (red and blue, respectively). The lines and the shaded areas correspond to the median and $\rm 16^{th}\,{-}\,84^{th}$ percentiles of metallicity at a fixed stellar mass. Each panel represents a given redshift bin. Dots are shown for cases where only one object is found.}
    \label{fig:mstar_vs_metallicity}
\end{figure}

As shown in the previous section, the differences between the median photometry of \lrds{} and \nolrds{} are primarily concentrated in the {\tt F115W}, {\tt F150W} and {\tt F200W} filters at all redshifts (except the highest redshift and lowest $M_{*}$ bin). Moreover, we showed that the nuclear emission plays a negligible role in the photometry of objects with $M_{*} \,{<}\, 10^8 \,M_{\odot}$. Therefore, the photometric divergences we observe at low $M_{*}$ should arise exclusively from differences in the underlying stellar populations of our selected objects. To understand the origin of these differences in terms of galaxy physical properties, Fig.~\ref{fig:galaxy_main_sequence} and Fig.~\ref{fig:mstar_vs_metallicity} compare the specific star formation rate (sSFR) and metallicity of \lrds{} and \nolrds{} as a function of their stellar mass. At $4 \,{\leq}\, z \,{\leq}\, 5$, \lrd{} systems with $10^7 \,{<}\, M_{*} \,{<}\, 10^8 \,M_{\odot}$ exhibit generally lower sSFR (${\lesssim}\, 0.25$ dex) and higher metallicities (${\sim}\, 0.1$ dex) than their \nolrd{} counterparts. In the higher mass range of $10^8 \,{<}\, M_{*} \,{<}\, 10^9 \,M_{\odot}$, the difference in sSFR is less evident although \lrds{} still tend to have slightly higher metallicities than \nolrds{}. At $5 \,{\leq}\, z \,{\leq}\, 6$, the low-mass \lrd{} population is only present in the $10^8 \,{<}\, M_{*} \,{<}\, 10^9,M_{\odot}$ range. As in the previous redshift bin, these \lrd{} galaxies display lower sSFRs (${\lesssim}\, 0.4$ dex) and higher metallicities (${\sim}\,0.2$ dex). At $6 \,{\leq}\, z \,{\leq}\, 7$, the differences in sSFR between the \lrd{} and \nolrd{} populations with $10^8 \,{<}\, M_{*} \,{<}\, 10^9 \,M_{\odot}$ largely disappear, leaving only a small metallicity offset (${\sim}\,0.1$ dex). Finally, at $7 \,{\leq}\, z \,{\leq}\, 8$, this trend is reversed: while \lrds{} and \nolrds{} show comparable metallicities, the former exhibit larger sSFR (${\sim}\, 0.25$ dex) than \nolrd{} systems.

The observed trends presented above indicate that low-mass \lrds{} tend to host more evolved stellar populations than their \nolrd{} counterparts. This naturally results in reduced flux contributions in the bluest JWST filters, accounting for the differences observed in Fig.~\ref{fig:lrd_photometry}. Moreover, even as differences in star formation activity diminish, the persistently higher metallicities of \lrds{} imply an earlier assembly of their stellar populations, which have experienced more prolonged chemical enrichment compared to the \nolrd{} sample. Consequently, this extended evolutionary history of \lrds{} offers a coherent explanation for the photometric and spectral distinctions between the two low-mass populations. In the following section, we explore the differences between the \nolrd{} and \lrd{} samples at higher stellar masses, with a particular focus on the properties of the MBH population.

%We note how, even if the \lrds{} are typically the most massive objects in the sample, at a given stellar mass, they present the same properties of {\tt non-LRDs}. In other words, {\tt non-LRDs}, at fixed stellar mass of \lrds{}, match the latter's properties. Indeed this is true for {\it every} property we have inspected (age of the stellar population, bulge size, stellar disk radius, SFR, cold gas mass, number of mergers...). This suggest that the \lrd{} photometric properties may arise from any combination of events that make it so that at the given time of observations, the spectra (in \lgbh{}: the stellar SED templates and MBH emission) meet the specific conditions for being classified as \lrd{}.

% \subsubsection{Imprint of MBH properties on JWST photometry}
\subsubsection{Imprint of MBH properties on \lrds{} photometry} \label{sec:MBHs_in_LRDsAGNs_noLRDsAGNs}
\label{sec:imprint_of_BH_properties_on_LRD_photometry}
\begin{figure}
    \centering
    \includegraphics[width=1\columnwidth]{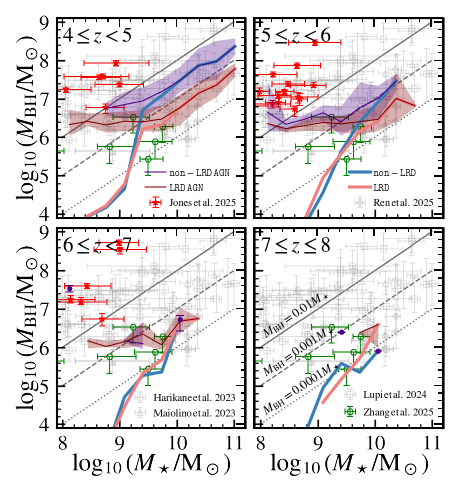}
     \caption{Massive black hole and stellar mass plane for the \lrd{}, \nolrd{}, \lrdAGN{} and \nolrdAGN{} samples (red, blue, brown and purple lines, respectively).  The lines and the shaded areas correspond to the median and $\rm 16^{th}\,{-}\,84^{th}$ percentiles of the sSFR at a fixed stellar mass. Each panel represents a redshift bin. Purple-filled dots are shown for cases where only one object is found. The trends are compared with the observational data of \protect\cite{harikane2023_highzgals} (grey circles) and \protect\cite{maiolino2023_highzagns} (grey triangles), \protect\cite{lupi2024} (grey squares), \protect\cite{zhang2025} (green circles) \protect\cite{Ren2025} (grey stars) \protect\cite{Jones2025} (red stars). Colors reffer to literature LRD samples.}
    \label{fig:bhmass_vs_starmass}
\end{figure}

As discussed in previous sections, AGN contributions to the photometry of our detected objects become relevant only at $z\,{<}\,6$ and $M_\star\,{>}\,10^9\,\msun$ for both \lrds{} and \nolrds{} (see Fig.~\ref{fig:lrd_photometry} and Appendix~\ref{appendix:Photometry_LRDAGN_noLRDAGN} for further details).
%Indeed, at these (relatively) low redshifts and high $M_\star$ the photometry of our simulated objects is the result of the interplay between AGN and galaxy components. 
% To clearly display the interplay between MBHs and their host galaxies, and understand the origin of photometric differences between \lrds{} and \nolrds{} seen in Fig.~\ref{fig:lrd_photometry}, it is necessary to analyze the role played by accreting MBHs. One direct way to analyze this is by inspecting the $M_{\rm BH}\,{-}\,M_\star$ relation of active black holes.
To better understand the interplay between MBHs and their host galaxies, we analyze the role played by accreting MBHs in the $M_{\rm BH}\,{-}\,M_\star$ relation. This will also help to understand the origin of photometric differences between \lrds{} and \nolrds{} seen in Fig.~\ref{fig:lrd_photometry}. 
In Fig.~\ref{fig:bhmass_vs_starmass}, we show the median scaling for both \lrds{} and \nolrds{} as well as for the sub-samples of these classes defined as AGN\footnote{We stress that our definition of AGN is based on the $L^{\rm AGN}_{\rm bol}$ associated to accreting MBHs, a property which is self-consistently tracked by our SAM along MBHs evolution \citep[see][]{david2020_bhgrowth,david2023_bhgrowth,david2024_bhgrowth,Bonoli2025}} %(i.e. objects with $L^{\rm AGN}_{\rm bol}\,{>}\,10^{44}\,{\rm erg/s}$), 
(hereafter, \lrdAGN{} and \nolrdAGN{}, red and purple lines). For reference, we show fixed $M_{\rm BH}$-to-$M_\star$ ratios as diagonal gray lines.

We start discussing Fig.~\ref{fig:bhmass_vs_starmass} from $4\,{\leq}\,z\,{<}\,5$. The results show that for $M_\star\,{\gtrsim}\,10^{9}\,\msun$ the median $M_{\rm BH}\,{-}\,M_\star$ relations for \nolrds{} and \nolrdAGN{} coincide, as do those of \lrds{} and \lrdAGN{}. This indicates that AGN accounts for the vast majority of objects at the highest $M_{\rm BH}$ and $M_\star$ bins in both \lrd{} and \nolrd{} samples and explains the strong contribution of AGN in the photometry of these two populations at $M_\star\,{\gtrsim}\,10^{9} \, \msun$.
%This indicates that AGN accounts for the vast majority of objects at the highest $M_{\rm BH}$ and $M_\star$ bins in both \lrd{} and \nolrd{} samples. This finding is consistent with the high AGN fractions measured in the \lrd{} population at $z\,{\leq}\,5$ ($\sim\!40\%$, see Fig.\ref{fig:RedshiftDistribution}) and explains the prevalence of AGN in the photometry of \lrds{} and \nolrds{} populations at $4\,{\leq}\,z\,{<}\,5$ for $M_\star\!\gtrsim\!10^{9}\msun$. 
Despite these overall similarities, the relations for \lrds{} and \nolrds{} show a remarkable difference. The former is systematically offset by $\sim 0.5$ dex, with MBHs in \nolrds{} being typically more massive than those in \lrds{}. Thus, combining the significant AGN contribution to the {\tt F115W} and {\tt F150W} filters seen in Fig.~\ref{fig:lrd_photometry} with the lower black hole masses of \lrds{} at fixed stellar mass, we conclude that the more pronounced spectral jump observed in $M_\star \gtrsim 10^{9}\, \msun$ \lrds{} at $\lambda_{\rm obs} \,{<}\, 2 \,{\times}\, 10^4$\,\AA\ is driven by a \textit{fainter} AGN population found in \lrds{} compared to that in \nolrds{} (see further discussion in Section~\ref{sec:AGN_and_MBH_properties}). As a result, the fainter AGN of \lrds{} at $M_\star \gtrsim 10^{9}\, \msun$ cause their UV-to-optical emission not to be bright enough to erase the spectral jump produced by the stellar Balmer break. Conversely, the more massive MBHs powering brighter AGN in \nolrds{} produce sufficiently strong UV-to-optical emission to compensate for the stellar Balmer break, effectively diluting the characteristic ``V-shaped'' feature in the photometric profiles of \nolrds{}. % By joining this finding with the large contribution of AGN photometry in the {\tt F115W} and {\tt F150W} filters (see bottom left panel of Fig.~\ref{fig:lrd_photometry}), we can conclude that the larger spectral jump exhibited by \lrds{} at $\lambda_{\rm obs}\,{<}\,2\times10^4$ \AA is due to \textit{fainter} AGN being hosted in \lrds{} than in \nolrds{} (see further discussion in Section~\ref{sec:AGN_and_MBH_properties}). Thus, the main reason why objects with $M_\star\!\gtrsim\!10^{9}\msun$ are selected as \lrds{} in our model is that at fixed stellar mass, they host relatively faint AGN, whose UV-to-optical emission is not bright enough to ``erase'' the spectral jump produced by the stellar Balmer break. Conversely, the \nolrds{} host brighter AGN, which can compensate for the stellar Balmer break with their bright UV-to-optical emission, hence removing the ``V-shaped'' feature from the photometric profiles of \nolrds{}.
These results indicate that, despite \lrdAGN{} constituting the majority of the AGN population (see bottom panel of Fig.~\ref{fig:RedshiftDistribution}), they are typically powered by lower-mass MBHs compared to their \nolrdAGN{} counterparts. This is a key result of our work and provides a framework to interpret the abundant population of \lrdAGN{} recently uncovered by JWST. It also implies that LRDs do not represent the entire AGN population at $z\,{>}\,4$, but are complemented by a less numerous population of “ordinary” bright AGN that lack the distinctive spectral features of LRDs. %This shows that, despite \lrdAGN{} being numerically dominant in the overall AGN sample (bottom panel of Fig.\ref{fig:RedshiftDistribution}), they tend to be powered by typically lower-mass MBHs in comparison to those powering \nolrdAGN{}. This is a major result of our work, which allows us to interpret the abundant population of \lrdAGN{} recently unveiled by the JWST as follows. In our model, \lrdAGN{} represent the large majority of the AGN population at $z\,{\geq}\,4$ ($\sim\!80\%$, see Fig.\ref{fig:RedshiftDistribution}) and they are powered by comparatively low-mass black holes relative to those found in their \nolrdAGN{} counterparts. This also suggests that LRDs may not account for the whole AGN population at $z\,{>}\,4$, being complemented by a less abundant population of ``ordinary'' bright AGN, lacking the typical LRD spectral features.
On the other hand, at $M_\star\,{<}\,10^{9}\msun$ and $4\,{\leq}\,z\,{<}\,5$ we observe a significant drop of the $M_{\rm BH}\,{-}\,M_\star$ relation in both \lrd{} and \nolrd{} samples, which is not observed for AGN. This is driven by the large number of low-mass MBHs populating this stellar mass range. The discrepancy between AGN-selected samples and the broader \lrd{} and \nolrd{} populations at $M_\star\,{\lesssim}\,10^{9.5}\msun$ points to the existence of a large population of small galaxies hosting low-mass black holes ($M_{\rm BH} \,{<}\, 10^5 \, \msun$), which may be inactive or shining as very faint AGN (i.e. with $L_{\rm bol}^{\rm AGN}<10^{44}\, \rm erg/s$).%as evidenced by the BHMF of Fig.~\ref{fig:Properties_Mstellar_Mvir_Lbol_MBH_LRDs_no_LRDs}.
This result suggests that the flattening of the $M_{\rm BH}\,{-}\,M_\star$ relations for \lrdAGN{} and \nolrdAGN{} at $M_\star\,{<}\, 10^9 \, \msun$ is driven by a small number of active MBHs that are rather overmassive with respect to the bulk of the MBH population and likely undergoing rapid growth. Due to their scarcity, they do not significantly impact the photometric properties of the overall \lrds{} and \nolrds{} populations. 
%Considering the overall predominance of \nolrd{} in the low mass end of the SMFs (see Fig.~\ref{fig:Properties_Mstellar_Mvir_Lbol_MBH_LRDs_no_LRDs}), this finding suggests that these potentially faint AGNs are common within the $M_\star\,{<}\,10^9\, \rm M_\odot$ \nolrd{} population.

%This result implies that the flattening of the $M_{\rm BH}\,{-}\,M_\star$ relations for \lrdAGN{} and \nolrdAGN{} at $M_\star\!<\!10^{9}\msun$ is due to rare ``overmassive'' MBHs which are likely in the process of quickly building up their mass and do not contribute significantly to the photometry of the overall \lrds{} and \nolrds{} samples. This large difference between AGN-selected samples and the broader \lrds{} and \nolrds{} populations hints at the presence of a significant population of low-mass galaxies ($M_\star\,{<}\,10^{9.5}\,\msun$) hosting small MBHs ($M_{\rm BH}\,{<}\,10^{5} \, \msun$) which are possibly shining as faint AGN. Considering the overall predominance of \nolrd{} at $M_\star\,{<}\,10^9 \, \msun$ (see the upper-left panel of Fig.~\ref{fig:Properties_Mstellar_Mvir_Lbol_MBH_LRDs_no_LRDs}) and the generally low fractions of \lrd{} in our lightcone (see Fig.~\ref{fig:RedshiftDistribution}), this finding suggests that these objects may typically appear as \nolrd{}, although they remain out of reach of current high-$z$ surveys.

Proceeding towards higher redshifts, at $5\,{\leq}\,z\,{<}\,6$ we observe the build-up of the trends observed in the $z\,{<}\,5$ $M_{\rm BH}\,{-}\,M_\star$ relation. Specifically, MBHs in \nolrds{} (either AGN or non-AGN) start to populate the highest bins of $M_{\rm BH} \,{-}\,M_\star$ plane, reaching $M_{\rm BH}$-to-$M_\star$ ratios of about ${\sim}\,10^{-3}$. \lrds{} (either AGN or non-AGN) follow a similar trend, but their $M_{\rm BH}$-to-$M_\star$ ratios only reach ${\sim}\,10^{-3.5}$. Although this offset starts to be visible only at $M_\star\,{\gtrsim}\,10^{10}\, \msun$, it will be fully developed down to $M_\star\,{\sim}\,10^{9}\,\msun$ by $z\,{<}\,5$. As at lower redshifts, the median $M_{\rm BH}\,{-}\,M_\star$ relation for \lrdAGN{} and \nolrdAGN{} with $M_\star\,{>}\, 10^{10} \, \msun$ closely tracks that of \lrds{} and \nolrds{} respectively, further supporting the enhanced AGN contribution to \lrds{} photometry observed at $5 \,{\leq}\, z \,{<}\, 6$ in  Fig.~\ref{fig:lrd_photometry}.
%These trends imply that by $5\,{\leq}\,z\,{<}\,6$, \nolrds{} and \nolrdAGN{} have begun developing the trends observed in the $M_{\rm BH}\,{-}\,M_\star$ relation at lower redshift %developing the offset with respect to \lrd{} and \lrdAGN{} in terms of $M_{\rm BH}$-to-$M_\star$ ratios. 
%In addition to these trends, it is noteworthy that the median $M_{\rm BH}\,{-}\,M_\star$ relation for \lrdAGN{} aligns with that of \lrds{} at $M_\star\!>\!10^{10}\,\msun$. This is consistent with the results shown in Fig.~\ref{fig:RedshiftDistribution}, where the AGN fraction among \lrds{} reaches 30–40\% at these redshifts. This supports the increase of AGN contribution to \lrds{} photometry shown in Fig.\ref{fig:lrd_photometry} at $5\,{\leq}\,z\,{<}\,6$ for the highest $M_\star$ bin. Similar conclusions hold true for \nolrds{} and \nolrdAGN{}.Nevertheless, also in this case \lrdAGN{} are largely dominant among all AGN (i.e. $\sim\!80\%$ at $5\,{\leq}\,z\,{<}\,6$, see Fig.\ref{fig:RedshiftDistribution}). This explains the smaller contribution of AGN to the photometry of \nolrds{} at this redshift and $M_\star\!>\!10^9\msun$. %Besides those trends, it is worth noticing that the median $M_{\rm BH}\,{-}\,M_\star$ relation for \lrdAGN{} coincides with that of \lrds{} at $M_\star\!>\!10^{10}\msun$, in line with the results shown in Fig.\ref{fig:RedshiftDistribution}, where the AGN fraction within \lrds{} rises to $30\%\!-\!40\%$ at this redshift. 
These trends break at $z \,{>}\, 6$ as MBHs in both \lrds{} and \nolrds{} exhibit similar $M_{\rm BH}\,{-}\,M_\star$ scaling relations, largely independent of stellar mass. This change of behavior is most evident at $6 \,{\leq}\, z \,{<}\, 7$, as at $7 \,{\leq}\, z \,{<}\, 8$ the interpretation becomes more uncertain due to the smaller number of systems with $M_{\rm BH} \,{>}\, 10^5 \, \msun$ (see bottom right panel of Fig.~\ref{fig:Properties_Mstellar_Mvir_Lbol_MBH_LRDs_no_LRDs}). %, while the minor deviations between the two classes at $7\,{\leq}\,z\,{<}\,8$ can be attributed to the small sample sizes of objects with $M_{\rm BH}>10^5\msun$ (see Fig.~\ref{fig:Properties_Mstellar_Mvir_Lbol_MBH_LRDs_no_LRDs}, bottom right panel). 
Independently of that, at these high redshifts the ratio between $M_{\rm BH}$ and $M_\star$ barely exceeds $\sim\!10^{-4}$, confirming that MBHs play a minor role in determining the photometric properties of \lrds{} and \nolrds{}.

In summary, the populations of \lrds{} and \nolrds{} display steep median scaling relations, which hover around typical ratios of $M_{\rm BH}\,{\sim}\,10^{-3}\,M_\star$ (${\sim}\,10^{-4}\,M_\star$) for $M_\star\,{>}\,10^{9.5} \, \msun$ (${<}\,10^9\,\msun$). This is significantly smaller than typical $M_{\rm BH}\,{-}\,M_\star$ values reported by recent observational works, especially for $M_\star\,{<}\,10^{9}\, \msun$ galaxies where observations point out $M_{\rm BH}\,{\sim}\,10^{-1}\,{-}\,10^{-2}\,M_\star$ \citep[see e.g.][]{harikane2023_highzgals,maiolino2023_highzagns}. Nevertheless, these discrepancies may be alleviated by considering that the common assumptions underlying observational estimates of MBH masses may induce significant bias. Indeed, as suggested by \cite{lupi2024}, uncertainties on the observed values can produce over-estimated $M_{\rm BH}$ by up to one order of magnitude, especially in high-$z$ systems which may be growing close to (or slightly above) the Eddington limit \citep[see][also see discussion on selection biases in \citealt{Ren2025}]{lupi2024}. Also, recent works started posing doubts on the $M_{\rm BH}$ estimates at high-$z$ (\citealt{Rusakov2025}), and recently \cite{Greene2025} proposed a different bolometric luminosity correction that lowered the $L_{\rm bol}$ estimates by 1 dex, pointing to a lower $M_{\rm BH}$. On the other hand, a fair comparison between observational data and our results should take into account that the former are likely AGN-dominated samples \citep[e.g.][]{greene2024_lrds,hviding2025_preprint,ronayne2025_preprint}. Overall, our \lrdAGN{} and \nolrdAGN{} populations show much flatter scalings than \lrds{} and \nolrds{} samples at all redshifts for  $M_\star\,{<}\,10^{9.5}\, \msun$. This is broadly in line with current observations, especially when considering the possible over-estimation of BH masses we discussed above \citep[as suggested by][]{lupi2024}. As expected, the most massive bins of \lrdAGN{} and \nolrdAGN{} relations coincide with the relations of the broader \lrds{} and \nolrds{} classes, showing that AGN fractions are close to unity for the most massive LRDs in our samples \citep[as suggested by recent observations, see e.g.][]{hviding2025_preprint,ronayne2025_preprint}. 

\subsection{Evolution of the AGN population}
\label{sec:AGN_and_MBH_properties}
\begin{figure}
    \centering
    \includegraphics[width=1\columnwidth]{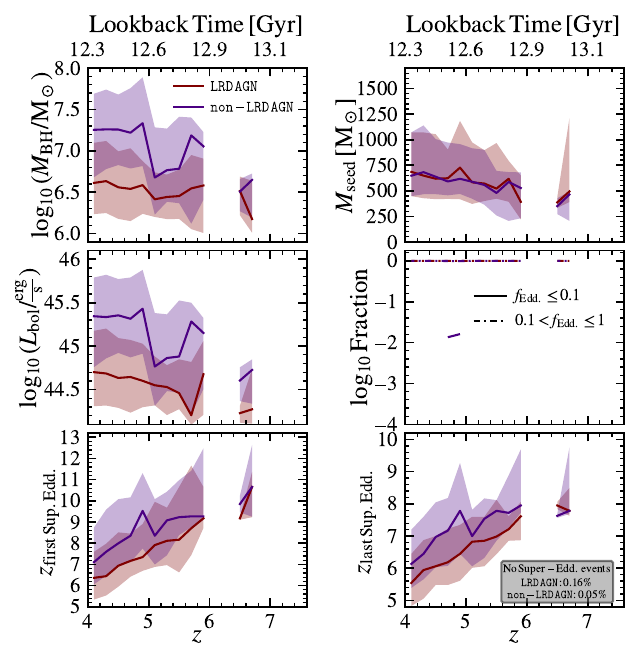}
    \caption{Black hole mass ($\rm M_{BH}$), initial seed mass ($\rm M_{seed}$), bolometric luminosity ($L_{\rm bol}^{\rm AGN}$), fraction of MBHs accreting a different $f_{\rm Edd}$, redshift of the last super-Eddington ($z_{\rm last,Sup.Edd.}$) and redshift of the first super-Eddington ($z_{\rm first,Sup.Edd.}$) for \lrdAGN{} and \nolrdAGN{} (red and purple, respectively). The lines and the shaded areas correspond to the median and $\rm 16^{th}\,{-}\,84^{th}$ percentiles. The values of $f_{\rm Edd}$ divides between ``quiescent'' AGN with $f_{\rm Edd}\,{\leq}\,0.1$ and Eddington-limited accretion with $f_{\rm Edd}\,{=}\,1$.}% and super-Eddington accretion with $f_{\rm Edd}\,{>}\,1$).}
    \label{fig:accreting_MBHs_props_in_lrdAGN_and_non-lrdAGN}
\end{figure}

As shown in Fig.\ref{fig:bhmass_vs_starmass}, the first \lrdAGN{} with $M_{\rm BH}\,{>}\,10^6\msun$ appear already at $z\,{>}\,7$. Given the high AGN fractions among LRDs reported in the recent literature \citep[see e.g][]{Kocevski2025,Durodola2025,CarranzaEscudero2025}, in this section we take a closer look at the redshift evolution of our  \lrdAGN{} and \nolrdAGN{} samples\footnote{We acknowledge that the results of this section are highly dependent on the definition of AGN. We explore in Appendix~\ref{appendix:alternative_agn_definition} a more relaxed AGN definition}. To control for potential differences arising from AGN being hosted in different galaxies, we build a control sample of \nolrdAGN{} that replicates the stellar mass distribution of the corresponding \lrdAGN{} population at each $z$. %AGN, either classified as \lrds{} or \nolrds{}, i.e. \lrdAGN{} and \nolrdAGN{}). To factor-out possible differences due to AGN being hosted in galaxies with different $M_\star$, for this analysis we match the \lrdAGN{} and \nolrdAGN{} populations in redshift. In detail, at each redshift we build a control-sample of \nolrdAGN{} with the same stellar mass distribution as \lrdAGN{}.
The comparison between $M_\star$–matched samples of \lrdAGN{} and \nolrdAGN{} is presented in Fig.~\ref{fig:accreting_MBHs_props_in_lrdAGN_and_non-lrdAGN}. On the left panels, from top to bottom, we present the evolution of $M_{\rm BH}$, $L_{\rm bol}^{\rm AGN}$ and of the redshift of the first super-Eddington accretion episode in the MBHs growth history (i.e. $z_{\rm first,Sup.Edd.}$). On the right panels, from top to bottom, we show the evolution of the total MBHs seed mass, of the fraction of AGN growing in different accretion regimes and of the median redshift of the last super-Eddington accretion episode (i.e. $z_{\rm last,Sup.Edd.}$). The absence of data at $6 \,{<}\, z \,{<}\, 7$ reflects the lack of \nolrdAGN{} objects available to construct a valid $M_\star$-matched control sample.  %On the left, from top to bottom we show the evolution of BH mass, $L^{\rm AGN}_{\rm bol}$ and $z_{\rm first\,Sup.Edd.}$, with the latter being the redshift at which the first super-Eddington accretion episode was registered in the past history of the MBH. On the right, from top to bottom we show the total BH-seed mass associated to AGN selected at a given redshift, the fraction of AGN accreting mass in specific regimes and finally the median $z_{\rm last\,Sup.Edd.}$ corresponding to the last super-Eddington accretion episode of the MBH. In all cases we show median quantities  and their associated $16^{\rm th}\,{-}\,84^{\rm th}$ percentiles (except for the $f_{\rm Edd}$ evolution).

As anticipated in the previous section, the comparison of black hole masses reveals that at $z\,{<}\,6$, \nolrdAGN{} typically host MBHs more massive by ${\sim}\,0.5\, \rm dex$ than \lrdAGN{}. These larger $M_{\rm BH}$ translate directly into higher bolometric luminosity, as seen in the middle left panel. This correlation arises because the vast majority of the AGN in both samples accrete at or near the Eddington rate ($f_{\rm Edd} \,{=}\, 1$), as illustrated in the middle right panel. Given these similar accretion rates, the observed differences in $M_{\rm BH}$ must be attributed to a combination of differences in the growth history of MBHs and variations in their initial seed masses. Disentangling this degeneracy is essential to understanding the origin of the mass discrepancy. %As anticipated in the previous section, the comparison between BH masses indicates that at $z\,{<}\,6$ \nolrdAGN{} tend to host more massive MBHs than \lrdAGN{}. Larger $M_{\rm BH}$ directly correspond to brighter $L^{\rm AGN}_{\rm bol}$, as shown in the middle left panel. This is because most of the AGN grow at the Eddington rate (i.e. $f_{\rm Edd}\,{=}\,1$), as shown in the middle right panel. Given the similar growth rates, to explain the larger $M_{\rm BH}$ we should disentangle the degeneracy between the growth history of MBHs and their initial BH-seed mass. 
To investigate this, we compare the redshifts of the first and last super-Eddington accretion episodes between \lrdAGN{} and \nolrdAGN{}. We find that \nolrdAGN{} selected at $z\,{<}\,6$ consistently exhibit higher values of both $z_{\rm first,Sup.Edd.}$ and $z_{\rm last,Sup.Edd.}$ compared to \lrdAGN{}. This indicates that the MBHs powering \nolrdAGN{} experienced phases of rapid growth earlier in cosmic time than those hosted by \lrdAGN{}. As a result, these MBHs had an earlier start in their mass assembly, giving them a significant advantage in reaching higher masses by $z\,{<}\,6$. %We explore this by comparing $z_{\rm first\,Sup.Edd.}$ and $z_{\rm last\,Sup.Edd.}$ between \lrdAGN{} and \nolrdAGN{}. This shows that \nolrdAGN{} selected at $z\,{<}\,6$ consistently tend to have larger $z_{\rm first\,Sup.Edd.}$ and larger $z_{\rm last\,Sup.Edd.}$ than \lrdAGN{}. Consequently, MBHs powering \nolrdAGN{} underwent phases of rapid assembly earlier than MBHs hosted in \lrdAGN{}. This provided them with a head-start in terms of $M_{\rm BH}$ with respect to \lrdAGN{}, which ultimately favored their typically larger BH masses at $z\,{<}\,6$.
\begin{figure}
    \centering
    \begin{subfigure}[t]{1\columnwidth}
        \centering
        \includegraphics[width=1\columnwidth]{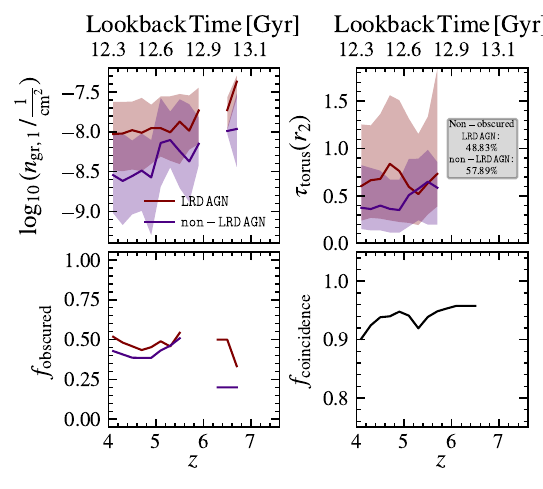}
        \caption{}
    \end{subfigure}
    \hfill
    \begin{subfigure}[t]{1\columnwidth}
        \centering
        \includegraphics[width=1\columnwidth]{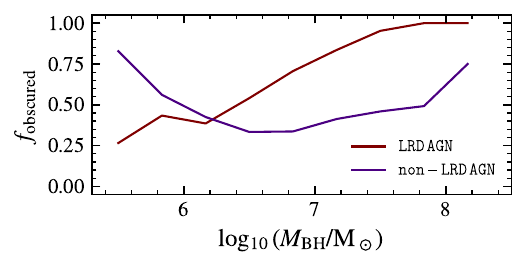}
        \caption{}
        \label{fig:obscuration_props2b}
    \end{subfigure}
    \caption{Obscuration properties of the \lrdAGN{} and \nolrdAGN{} samples (red and purple, respectively). The upper panel shows the grain number density at the innermost region of the torus ($n_{gr,1}$), the torus optical depth ($\tau_{\rm torus}(r_2)$), the fraction of obscured objects ($f_{\rm obscured}$) and the fraction of AGN that are still selected as \lrdAGN{} after removing the torus component ($f_{\rm conincidence}$). The lower panel represent the fraction of obscured objects as a function of black hole mass, independently of redshift. In the two upper panels, the solid lines represent the median values of the distributions, while the shaded areas correspond to the $16^{\rm th}\,{-}\,84^{\rm th}$ percentiles.}
    \label{fig:obscuration_props2}
\end{figure}

Regarding the seeding, our model indicates that the larger black hole masses of \nolrdAGN{} at $z\,{<}\,6$ cannot be attributed to differences in the initial seed masses. In fact, MBHs powering both \nolrdAGN{} and \lrdAGN{} originate from remarkably similar seeds, with typical total seed masses of $M_{\rm seed} \,{\sim}\, 500 \,M_\odot$\footnote{We emphasize that $M_{\rm seed}$ refers to the total black hole seed mass that contributes to the final $M_{\rm BH}$ at a given redshift. This explains its increasing trend with decreasing redshift, as mergers between MBHs combine the seed mass contributions of each progenitor.} %Regarding the seeding, our model shows that the larger black hole masses of \nolrdAGN{} at $z\,{<}\,6$ cannot be ascribed to the initial seed mass. Indeed, MBHs powering both \nolrdAGN{} and \lrdAGN{} originated from remarkably similar seeds with typical masses of $M_{\rm seed}\,{\sim}\,500\msun$. We stress that $M_{\rm seed}$ is the \textit{total} BH-seed mass which contributes to the $M_{\rm BH}$ measured at a given $z$. This explains its increasing trend with decreasing redshift, as MBHs merge to produce the final $M_{\rm BH}$ (each bringing its $M_{\rm seed}$ contribution).
By directly analyzing the formation channels of BH seeds (not shown here for brevity), we find that 99.5\% (98.3\%) of MBHs hosted in \lrdAGN{} (\nolrdAGN{}) originally formed as light seeds with masses ${\lesssim}\,100\, \msun$ \citep[i.e. remnants of PopIII stars, see][]{spinoso23_bhseeds,Bonoli2025} and never merged with any heavier seed type. This is in  contrast with recent studies that suggest a heavy-seed origin for LRDs  \citep[e.g.][]{InayoshiAndMaiolino2025,Cenci2025,Jeon2025}. However, the recent work by \citet{Bonoli2025}, within the framework of the \lgbh{} model, demonstrates that combining JWST observations with pulsar timing array constraints favors a scenario in which MBHs in \lrdAGN{} arise primarily from light seeds undergoing phases of super-Eddington accretion. Moreover, other studies do not exclude light seeds as plausible progenitors for LRDs, further supporting this interpretation \citep[see e.g,][]{Hu2022,Scoggins2024}. Finally, we note that in our model, heavy seeds tracked by \lgbh{} never grow efficiently enough to match the black hole masses inferred for \lrdAGN{} from JWST observations, further disfavoring their origin as descendants of heavy BH-seeds. We do not explore this point in further detail here, as it will be the focus of a forthcoming study (Spinoso et al., in prep). %Nevertheless, the recent work of \cite{Bonoli2025} showed that, within the context of \lgbh{}, joining recent JWST observations and Pulsar Timing Array experiments (\textbf{REFERENCES}) favors a scenario in which MBHs in \lrdAGN{} are explained primarily as light seeds undergoing phases of Super-Eddington accretion. In addition, \dani{refer to other seeding works which do not rule out light seeds as the origin of LRDs}. Finally, we anticipate that heavy seeds tracked by \lgbh{} never grow efficiently enough to reach the BH masses inferred by JWST observations for \lrdAGN{}, hence further disfavoring the explanation of \lrdAGN{} as descendants of heavy BH-seeds. We refrain to comment further on this point, as this is the subject of an upcoming work (Spinoso et al. in prep).

Finally, in Fig.~\ref{fig:obscuration_props2} we investigate the obscuration in the \lrdAGN{} and \nolrdAGN{} samples. The upper left panel presents the grain number density in the innermost region of the torus ($n_{\rm gr,1}$), i.e the area directly illuminated by radiation from the MBH. Although this quantity is challenging to constrain observationally, it offers valuable insight into the degree of AGN obscuration and the reprocessed radiation emitted at infrared wavelengths. The results show that \lrdAGN{} generally exhibit higher dust densities in their nuclear regions. This is not an unexpected result, as in the context of our modeled gas density profiles, higher black hole masses correspond to lower density normalisations. This trend is consistent with the broader picture discussed throughout this section: \lrdAGN{} are typically in earlier growth phases, retaining larger gas reservoirs that fuel ongoing accretion, while \nolrdAGN{} have already consumed much of their available gas during earlier evolutionary stages. The results in $n_{\rm gr,1}$ also contribute to the differences observed in the torus optical depth at the outer radius ($\tau_{\rm torus}(r_2)$) where the \lrdAGN{} sample exhibits values up to a factor of 1.4 higher than those found in the \nolrdAGN{} sample. To further assess the role of the torus in the detection of \lrdAGN{}, we quantify its contribution using the parameter $f_{\rm coincidence}$, shown in the middle right panel of Fig.~\ref{fig:obscuration_props2}. This quantity represents the fraction of \lrdAGN{} that would still be classified as such even if the torus were entirely removed\footnote{Given that the set of NIRCam filters used captures just the rise of the infrared emission (the {\tt F444W} filter peaks at $42500 \, \AA$, i.e. $8500\, \AA$ at $z\,{=}\,4$, and our torus emission usually peaks at $10^{4-4.5}\, \AA$, see Fig. \ref{fig:Building_SEDs}), the torus effect is mainly present as obscuration of the central source in the rest frame UV-optical.}. To guide the reader, a lower $f_{\rm coincidence}$ indicates a stronger dependence on torus-related effects for their selection. As shown, $f_{\rm coincidence}$ decreases slightly with redshift but remains relatively constant, with typical values around $f_{\rm coincidence} \,{\sim}\, 0.9$. This implies that obscuration plays a secondary role and is only relevant for ${\sim}\,10\%$ of the \lrdAGN{} sample, while the majority would still be identified as AGN based solely on their host galaxy and MBH accretion properties. Indeed, to assess obscuration, we also examine the quantity $f_{\rm obscured}$, which represents the fraction of \lrdAGN{} and \nolrdAGN{} sources that are obscured due to the torus (see Section~\ref{subsubsec:att_uv}). This is shown in the lower panels of Fig.~\ref{fig:obscuration_props2}, where we see how the fraction of obscured objects in both samples is constant regardless of redshift. Specifically, \lrdAGN{} typically exhibit $f_{\rm obscured} \,{\sim}\, 50\%$, while \nolrdAGN{} sit at slightly lower values of $f_{\rm obscured} \,{\sim}\, 40\,{-}\,50\%$. A more interesting behavior arises when the mass of the black hole is fixed (Fig.~\ref{fig:obscuration_props2b}): \lrdAGN{} consistently show a higher obscured fraction. Interestingly, 
%while $f_{\rm obscured}$ for \nolrdAGN{} shows little to no dependence on black hole mass, 
the \lrdAGN{} sample displays a clear trend of increasing obscuration fraction with increasing mass, going from 25\% to $\sim\!100\%$. This behavior is consistent with the previous finding that the MBHs hosted in \lrds{} (or \lrdAGN{}) need to not be bright enough for the v-shape to remain. Therefore, since the \lrd{} selection introduces a bias towards fainter MBHs, and the more massive MBHs are also brighter, it is required for them to be more obscured in order to be selected as \lrd{} (\lrdAGN{})

\subsection{Galaxies and AGN in the LRD population}
\label{sec:Gals_and_AGN_in_the_LRD_population}

As shown in Fig.~\ref{fig:RedshiftDistribution}, the photometrically selected \lrd{} population includes both objects classified as galaxies and AGN (\lrdAGN{} and \lrdGal{}, respectively, see Sect.\ref{sec:Classification_Galaxies_AGNs}). Our results show that the AGN sample accounts for 40\% of the population at $z \,{\sim}\, 4$, decreasing to 10\% at $z\,{\sim}\,7$. In this section, we characterize these two subpopulations and highlight their key differences.

\begin{figure}
    \centering
    \includegraphics[width=1\columnwidth]{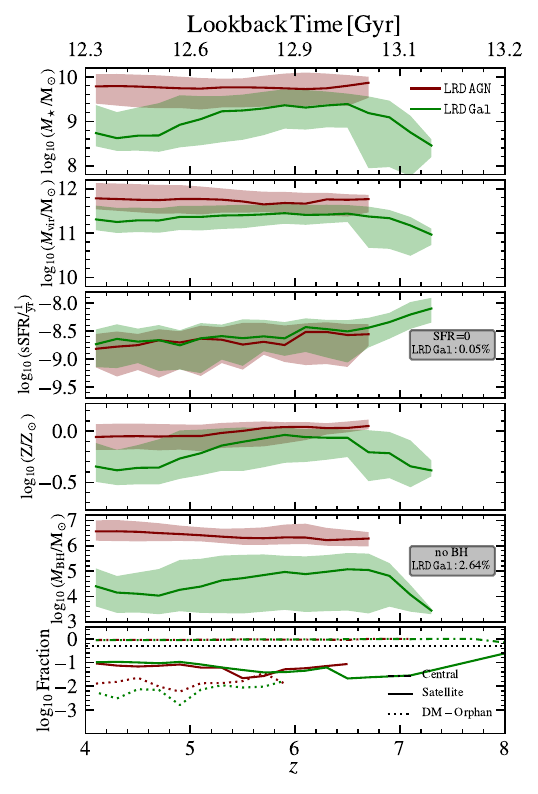}
    \caption{Properties for the \lrdAGN{} and \lrdGal{} samples (red and green, respectively). From top to bottom: $M_\star$, halo mass, specific star formation rate (sSFR), metallicity ($Z$), $M_{\rm BH}$ and fraction of galaxies which are centrals or satellites. The solid lines represent the median values of the distributions, while the shaded areas correspond to the $16^{\rm th}\,{-}\,84^{\rm th}$ percentiles. The text displayed inside gray boxes represents the fraction of galaxies with a null SFR and without a central MBH.}
    \label{fig:lrd_agn_vs_lrd_gal_props}
\end{figure}

In Fig.~\ref{fig:lrd_agn_vs_lrd_gal_props}, we present the evolution in redshift of the median (and $16^{\rm th}\!-\!84^{\rm th}$ percentiles) of the following properties, from top to bottom: stellar mass, halo mass, sSFR, metallicity, black hole mass, and the fraction of systems that are satellites. Regarding stellar mass, the \lrdAGN{} sample exhibits little to no redshift evolution, maintaining an approximately constant value of $M_\star \,{\sim}\, 10^{9.5} \, \msun$. In contrast, the \lrdGal{} sample shows lower stellar masses and a more pronounced redshift evolution. At $z \,{\sim}\, 9$, \lrdGal{} have $M_\star  \,{\sim} \, 10^7  \, \msun$, which increases to $M_\star  \,{\sim} \, 3  \,{\times} \, 10^8  \, \msun$ by $z  \,{\sim} \, 4$. These trends are consistent with the results presented in previous sections, where objects with a predominant galaxy (i.e., negligible AGN) contribution in their photometry are typically found in low-mass galaxies. Concerning the halo mass, the trends closely mirror those seen in stellar mass. The \lrdAGN{} sample exhibits only mild redshift evolution, with $M_{\rm vir} \,{\sim} \, 5  \,{\times} \, 10^{11}  \,{-} \, 10^{12}  \, \msun$, whereas the \lrdGal{} sample evolves from $M_{\rm vir} \,{\sim} \, 10^{10}  \, \msun$ at $z  \,{\sim} \, 9$ to $M_{\rm vir}  \,{\sim} \, 2  \,{\times} \, 10^{11}  \, \msun$ at $z  \,{\sim} \, 4$. The differences in both halo and stellar masses result in distinct metallicity profiles between the two populations, with \lrdAGN{} galaxies being more metal-rich ($Z  \,{\sim}\, 1  \,{-} \, 0.7  \, Z_{\odot}$) compared to \lrdGal{} galaxies ($Z  \,{\sim} \, 0.15  \,{-} \, 0.7  \, Z_{\odot}$). Interestingly, both samples show a drop in metallicity around $z \,{\sim} \, 7$, likely due to the infall of pristine gas into the galactic discs. In terms of sSFR, both samples exhibit a decreasing trend, but they remain consistent with star-forming galaxies ($\rm sSFR  \,{>} \, 10^{-9}  \, yr^{-1}$). Notably, the sSFR values of the \lrdGal{} and \lrdAGN{} samples are very similar, suggesting that any potential AGN feedback in the \lrdAGN{} sample is still too weak to quench star formation in their host galaxies. In this respect, the MBHs powering \lrdAGN{} objects are more massive than those in \lrdGal{} systems. While the former host MBHs with $M_{\rm BH} \,{\sim} \, 5  \,{\times} \, 10^6 \, \msun$, the latter host systems with MBHs of $M_{\rm BH}  \,{\sim} \, 10^3  \,{-} \, 10^5 \, \msun$. Finally, approximately 90\% of the objects in both the \lrdGal{} and \lrdAGN{} populations are central galaxies, implying that they are the most massive objects in their host dark matter halos, with only about 10\% corresponding to satellite galaxies.

\begin{figure}
    \centering
    \includegraphics[width=1\columnwidth]{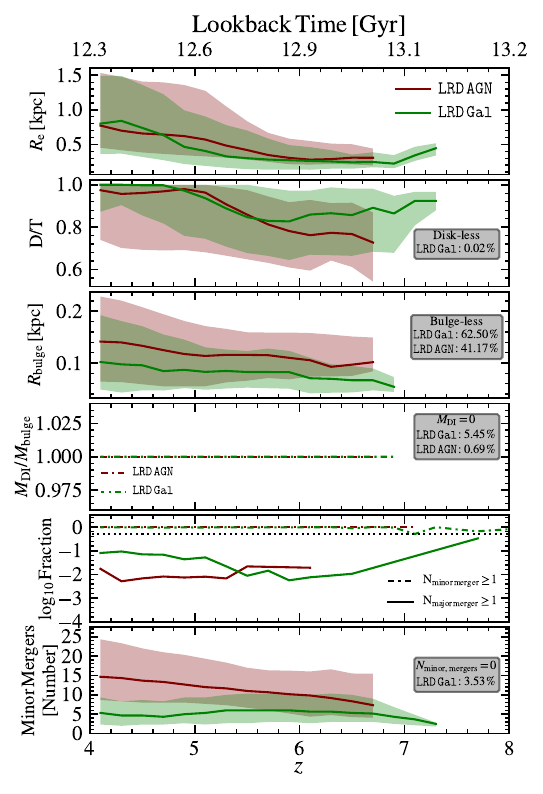}
    \caption{Morphological properties of \lrdAGN{} and \lrdGal{} (red and green, respectively). From top to bottom: galaxy size, Disk-to-Total ratio, bulge size, fraction of bulge mass assembled via disk-instabilities, fraction of objects which underwent major/minor mergers and total number of minor mergers. The lines and shaded areas represent the same as in Fig.~\ref{fig:lrd_agn_vs_lrd_gal_props}. The text inside gray boxes shows the fraction of galaxies that lack a disk, lack a bulge, do not have bulge-mass coming from disk-instabilities, and have not undergone any minor mergers.}
    \label{fig:lrd_agn_vs_lrd_gal_props_morph}
\end{figure}

Beyond the global properties of galaxies, it is also insightful to examine the morphological characteristics of the two \lrd{} samples. This is shown in Fig.~\ref{fig:lrd_agn_vs_lrd_gal_props_morph} where we show from top to bottom: (mass-weighted) galaxy radius ($R_e$), disk-to-total ratio ($D/T$), bulge radius ($R_{\rm bulge}$), bulge mass assembled via disk instabilities ($M_{\rm DI}/M_{\rm bulge}$), the fraction of major and minor mergers undergone by the objects and the median of number of minor mergers.  Regarding the galaxy radius, both samples exhibit similar extensions, with a general increase towards lower redshifts. Specifically, at $z\,{\sim}\,7$, the typical galaxy has a radius of $R_e \,{\sim}\, 500\, \rm pc$, while by $z\,{\sim}\,4$, this value increases to $R_e \,{\sim}\, 800\, \rm pc$. These results agree remarkably well with \cite{morishita2024}, which showed that the typical radius of $z\,{>}\,5$ galaxies ranges between 0.3 kpc and 1 kpc, depending on the JWST filter used. Regarding galaxy morphology, both samples are predominantly disk-dominated ($D/T \,{>}\, 0.6$), with the disk component becoming more prominent at lower redshifts. Despite these similarities, the \lrdAGN{} sample tends to exhibit slightly lower $D/T$ values, indicating a more prominent bulge component. This is further supported by the bulge size, with \lrdAGN{} galaxies featuring more extended bulges (${\sim}\, 150\, \rm pc$) compared to \lrdGal{} galaxies (${\sim}\, 100\, \rm pc$). Such a difference is not unexpected, as AGN activity in the \lgbh{} model is closely linked to the growth of galactic bulges \citep{david2020_bhgrowth}. The primary bulge formation scenario, and consequently the main triggering mechanism for MBHs in the \lrdAGN{} sample, becomes evident when examining the $M_{\rm DI}/M_{\rm bulge}$ ratio, which quantifies the fraction of the bulge formed via disk instabilities. As shown, 100\% of the bulge in both the \lrdGal{} and \lrdAGN{} samples is assembled this way, likely induced by minor mergers
% (indeed, we find a high incidence of disky galaxies),
regardless of redshift. Thus, our model indicates that the AGN activity observed at $z\,{>}\,4$ in \lrds{} is primarily fueled by internal processes, rather than by the more violent major merger mechanisms typically associated with AGN triggering at lower redshifts. The bottom panel of Fig.~\ref{fig:lrd_agn_vs_lrd_gal_props_morph} shows that ${<}\,10\%$ (${<}\,1\%$) of the \lrdGal{} (\lrdAGN) sample underwent a major merger, suggesting a relatively quiet merger history of \lrdGal{} and \lrdAGN{} populations. In terms of minor interactions, all objects in both samples have experienced at least one minor merger. However, \lrdAGN{} objects have a more active merger history, averaging 10-15 interactions compared to 2-5 for \lrdGal{} objects.

\begin{figure}
    \centering
    \includegraphics[width=1\columnwidth]{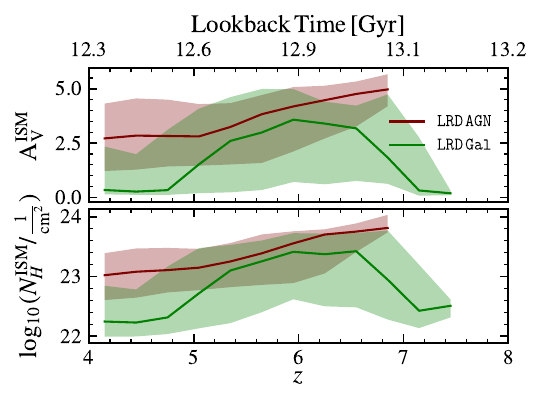}
    \caption{Hydrogen column density ($N_H^{\rm ISM}$) and dust extinction ($A_V^{\rm ISM}$) due to the ISM for the \lrdAGN{} and \lrdGal{} samples (red and green, respectively). The lines and shaded areas represent the same as in Fig.~\ref{fig:lrd_agn_vs_lrd_gal_props}.}
    \label{fig:Av_NH}
\end{figure}

Finally, in Fig.~\ref{fig:Av_NH}, we study the SED attenuation of the \lrdAGN{} and \lrdGal{} samples by presenting their hydrogen column density ($N_H^{\rm ISM}$) and dust extinction ($A_V^{\rm ISM}$) due to the ISM. In terms of $A_V^{\rm ISM}$, the \lrdAGN{} sample shows values in the range of ${\sim}\,2\,{-}\,5$, which are systematically higher (by a factor of ${\sim}\,1.3\,{-}\,2$) compared to those of the \lrdGal{} sample. A similar trend is observed for $N_H^{\rm ISM}$. While \lrdGal{} sources typically exhibit column densities of $N_H^{\rm ISM} \,{\sim}\, 10^{22-23}\,{\rm cm^{-2}}$, the \lrdAGN{} sample reaches higher values of $N_H^{\rm ISM} \,{\sim}\, 10^{23-24}\,{\rm cm^{-2}}$. These differences can be naturally attributed to the already described higher stellar masses of the AGN hosts, which imply larger cold gas and dust reservoirs in their ISM.
%\davcoment{David will correct/comment this paragraph: }Finally, in Fig.~\ref{fig:Av_NH} we provide the values of \lrdAGN{} and \lrdGal{} samples regarding the hydrogen column density, $N_H^{\rm ISM}$, and $A_V^{\rm ISM}$ values in the ISM. We want to stress that this predominantly affects nebular emission and continuum in the galactic disk, and the amount of MBHs affected by it is small (8.22\%). Nonetheless, we see how \lrdAGN{} have systematically higher values for $A_V^{\rm ISM}$. This difference is due to the fact that they are more massive than \lrdGal{}, since in our model $A_V^{\rm ISM}$ is defined by eq. \ref{eq:att_ism_lgalaxies}, which is proportional to $N_H^{\rm ISM} \sim M^{\rm gas}_{\rm cold}$ and metallicity. While we have not directly shown $M^{\rm gas}_{\rm cold}$, we do show in Fig. \ref{fig:lrd_agn_vs_lrd_gal_props} that they have similar values for sSFR, which coupled with more stellar mass means more mass in cold gas, since we are not dealing with quenched system but rather star forming ones. 

%Environment-wise, we also see in Fig. \ref{fig:lrd_agn_vs_lrd_gal_props_env} that indeed they are isolated central galaxies, with minimum contribution of satellites to any of the samples ({\tt AGN} or {\tt gal}. This is clear from the evolution of the mass of hot gas in their respective halos, that follow the same evolution as that of the halo mass. Therefore, no significant process of merger or galaxy interaction (i.e.: gas stripping) has occurred on average.
%\begin{figure}
%    \centering
%    \includegraphics[width=1\columnwidth]{figs/lrd_agn_vs_lrd_gal_props_env.pdf}
%   \caption{\footnotesize}
%    \label{fig:lrd_agn_vs_lrd_gal_props_env}
%\end{figure}

%As anticipated in the previous section, in Fig. \ref{fig:obscuration_props2} we see how \lrdAGN{} have higher dust grain number densities in the torus. We want to stress that a control sample in stellar mass was made for this comparison, hence demonstrating the clear difference across all redshifts. However, since the torus emission is missed in the NIRCam filters, this only implies that the obscuration from the torus will be higher for the \lrdAGN{} than for the \nolrdAGN{}.

\section{Caveats}
\label{sec:caveats}
Our model relies on several key assumptions, which can affect some of our results  and their interpretation. In the following paragraphs, we outline these assumptions, discussing their impact on the work and findings we presented.
\begin{itemize}
    \item The AGN emission framework we developed does not account for the reprocessing of rest-frame X-rays. Therefore, we cannot draw predictions nor gain insights about the X-ray weakness observed in LRDs and high-$z$ AGN \citep[][]{InayoshiAndIchikawa2024,Yue2024,Pacucci2024_Xrays}. We stress that the results we presented remain unaffected by this, since the NIRCam wavelength coverage does not extend into rest-frame X-rays for $z>4$ sources. We plan to address this point in future works, especially by modeling the spectra emerging from a dense gas obscurer surrounding the central source. Indeed, recent studies suggest that this approach could naturally account for the X-ray weakness observed in high-redshift AGN. \vspace{2mm}
    
    \item Our model does not include a treatment of the re-emitted light from the galaxy ISM. However, given that the ISM is expected to have temperatures on the order of a few Kelvin, its peak emission typically occurs at far-infrared or submillimeter wavelengths, well beyond the range covered by the NIRCam filters used in this study. Consequently, our results are not affected by the lack of modeling of the ISM emission. \vspace{2mm} %(the {\tt F444W} filter at $z=4$ can only capture the rise of the IR bump. 
    
    \item Our model for the galaxy compactness is a simplified one based on a single Sérsic profile. Furthermore, we do not model the JWST PSF nor its wavelength-dependent variations across different filters. These simplifications likely tend to over-estimate our sources compactness (as shown in the bottom panel of Fig.\ref{fig:Color_Color_Selection}).
    %(for instance: different JWST filters may trace distinct stellar populations with varying spatial distributions).
    As a consequence, we caution the reader to consider our compactness criterion as indicative. Nevertheless, we stress that the vast majority of our simulated sources satisfies the compactness cut we impose (see Table \ref{tab:lrd_numbers}), therefore its effect on our LRD samples is negligible.\vspace{2mm} %Finally, LRDs being unresolved sources, they fall in the limits of the JWST resolution. Including the modeling of the JWST PSF, which falls outside the scope of this paper, would likely cause the galaxies to appear more extended.
    
    %When computing the compactness, the underlying assumption, other than the light profile of high redshift galaxies is indeed a Sérsic profile, is that all filters have the same profile (same $r_e$), since the filter only comes into place when normalizing. Indeed, this is an oversimplification, but a correct modeling of the light profile of each filter is beyond the scope of this paper.
    
    \item In this work, we assume that the %ionizing radiation
    radiation reprocessed by dust
    spans the range $912 \, \AA \,{<}\, \lambda \,{<}\, 6564 \, \AA$. The lower limit of $912 \, \AA$ is a reasonable choice, as shorter wavelengths can potentially photodissociate dust. Defining the upper limit is more challenging, so we set it at the rest-frame wavelength of the $H\alpha$ line, since beyond this point, infrared emission begins to dominate at the sublimation temperature of 1500 K. While these assumptions are well-grounded, we emphasize that this choice could affect the location of the "V-shaped" feature for some objects. \vspace{2mm}
    %, as ionizing radiation influences the reprocessed emission from the torus. \vspace{2mm}
    %Another arbitrary choice is the definition of ionizing light. As we have explained, we have assumed that all light between $912 \ \AA < \lambda < 6564 \ \AA$ contributes to heating up the torus. This also implies that the obscuration will affect those wavelengths the same (since it is modeled as $e^{-\tau_{\rm torus}(r_2)}$), which may also affect the location of the v-shape for some objects. The lower limit, 912 $\AA$ is well argued taking into account that below it photodisociation effects may occur that destroy dust. For the upper limit, we were not able to find any strong argument to stop at any given wavelength given our model. Hence, we decided to stop at $H\alpha$ because beyond that the IR emission starts to dominate at the sublimation temperature of 1500 K. 
    
    \item  The gas density profile employed for the Narrow Line Region (NLR) is consistent with the expected hydrogen number densities. However, applying this same profile to the Broad Line Region (BLR) leads to unreasonably high ionization parameters. To obtain more physically plausible results, we have opted to scale the hydrogen density of the BLR by a factor of $10^{9.5}$ relative to that of the NLR. Alternative scaling factors could lead to different intensities for the AGN emission lines. However, we have verified that the emission from the BLR does not significantly impact our selection of \lrd{} population. \vspace{0.1mm}

    \item Our results indicate that at $z\,{>}\,7$, the emission from the stellar population dominates the spectral features used to classify objects as \lrd{}. This outcome is partly driven by the fact that, at these redshifts, our model predicts a lower number of massive MBHs compared to recent JWST observations. However, the {\it grafting} technique included in \citet{Bonoli2025} allows to sample a broader dynamical range in the BHMF and LF, alleviating the discrepancy at high redshift (although it persists 1-2 dex depending on the growth and seed model, favoring a light seed with Super Eddington growth scenario over one with a more abundant population of heavy BH seeds, see Figure 8 of \citealt{Bonoli2025}). Therefore the lack of bright and massive MBHs, potentially building up a more numerous AGN population at high redshift is partially a volume issue that arises because of using {\tt Millennium-II} as our background N-body simulation. Exploring alternative DM simulations, and extending the analysis with the {\it grafting} technique would yield a larger population of massive and actively accreting MBHs at $z\,{>}\,7$, enhancing the contribution of MBH emission in the identification of \lrds{}. We plan to explore these possibilities in upcoming works. \vspace{2mm}
    
    \item Recent models suggest that the "V-shaped" feature could be produced by a dense gaseous envelope around the central source \citep[sometimes referred to as black hole star model, BH$\star$, see][]{InayoshiAndMaiolino2025,Naidu2025_Bh_star,Ji2025_blackthunder}. In this scenario, the spectrum at shorter wavelengths of the Balmer break would be associated with an underlying stellar population \citep[such as a nuclear star cluster, see ][]{inayoshi2025b}, whereas the spectrum at longer wavelengths would correspond to AGN emission reprocessed by a dense, hydrogen-rich environment. This picture is opposite to what we present, unsurprisingly given the fundamental difference between our model and the BH$\star$ one. 
    %Since we adopt a more classical scenario in which the AGN is obscured by the torus, it is very unlikely that we can reach similar conclusions.  --> this was implicit, in my opinion
    We acknowledge that modeling a gaseous obscurer would significantly alter the results presented in this work. Nevertheless, this task should be treated carefully, not only taking into account the details of the BH$\star$ model, but also because of the fact that LRDs represent a complex and diverse class. Under this perspective, both schemes may be playing a role in shaping the observed LRD and (or) AGN populations at $z>4$. Therefore, while including the BH$\star$ model in our scheme is beyond the scope of this paper, we plan to undertake this task in detail in a future work.\vspace{2mm}

    % \item Our AGN SED model associates both broad and narrow lines to every MBH simulated by \lgbh{} (before accounting for inclination and obscuration). This is because all MBHs in our lightcone are within the $M_{\rm BH}$, $\dot{m}$, $U$, $Z$, and $n_H$ ranges sampled by the model of Su et al. (in prep). This implies that we cannot replicate the broad-line selection of AGN in LRDs presented in recent literature. Therefore the fraction of \lrdAGN{} fraction within our AGN sample
    % 10-20% LRDS in BLAGN
    % 70-80% LRDS in AGN
    % BLAGNS -> 50% AGNs    --->  threshold en FWHM  + inclination (obscuration) / flux threshold in BLR

\end{itemize}

%Finally, using the derived gas density profile to compute the number density of hydrogen of the BLR results in extremely high ionization parameter for the BLR. For that reason, we set the hydrogen density of the BLR to be $log(n_{\rm H,NLR}$)+9.5, which yields reasonable values. For $n{\rm H,NLR}$, we do use our gas density profile, as it agrees well with expected values.    
   
% \danitext{Finally, we do not attribute any specific process for which objects without an MBH are selected as \lrds{}. On the contrary, we acknowledge that object to object there may be differences in specific properties (sSFR, metallicity...), but {\it on average} both \lrds{} and {\tt non-LRDs} populate the same regions in each property-space. }

% \danitext{The findings presented in this section imply that the characteristics that give rise to the LRD photometric properties are just a matter of being observed ``at the right time in the right place''. In the sense that object to object there are differences that disappear when taking into account the whole population. This means that in our model, objects go into an ``LRD phase", that is common in galaxy evolution and it is the result of any combination of any number of properties that make an object meet the criteria for being selected as LRDs.\\}

\section{Conclusions}
\label{sec:conclusions}
In this work, we explored the properties of $z\,{\geq}\,4$ \lrds{}, photometrically selected within their cosmological context in a simulated lightcone. To this end, we used the \lgbh{} semi-analytic model \citep[][]{Bonoli2025} applied to the merger trees extracted from the high-resolution, N-body, cosmological simulation {\tt Millennium-II} \citep{millenium2_nbodysimul}. \lgbh{} was specifically designed to comprehensively capture several physical processes involved in the formation, growth, and dynamics of MBHs, a key feature which allows us to interpret \lrds{} populations within the context of current models for the formation and (co-)evolution of MBHs and galaxies.

To accurately model the emission from both galaxies and AGN, we developed a methodology within the \lgbh{} framework to obtain simulated SEDs. These include five distinct components: galaxy continuum, galaxy emission lines, AGN continuum, AGN torus emission and AGN emission lines (both broad and narrow). Additionally, for all these components, we accounted for attenuation due to absorption by dust and atomic gas along the line of sight.

Our simulated lightcone is designed to mimic JWST/NIRCam observations, enabling one of the most direct comparisons to date between simulated and observed LRDs. In an effort to mimic observational selections, we identify the population of \lrds{} within our lightcone applying the flux limits of the GOODS-S field along with the photometric color and compactness cuts presented in \cite{kokorev24_lrds}. Our main findings are as follows:
%In this paper we have introduced a novel, physically informed method to generate MBH spectra corresponding to a classical configuration of accretion disk + dusty structure + broad and narrow line regions. Moreover, we have also modeled the subsequent obscuration and attenuation from such light from the nuclear dust and ISM (respectively), in an orientation dependent scheme. 
%Then, we have applied this model to a series of accreeting black holes, that were generated with the \LGalaxies semi-analytic model. In particular, we did a mock tailored to mimick JWST NIRCam observations. Then, we computed the photometry of every object by taking into account both stellar and nuclear emission.
%In order to find our population of LRDs, we imposed realistic magnitude cuts to make our sample of detected NIRCam objects at $z>4$ (we replicate the depths of the GOOD-S field). Then, after modeling the compactness, we applied \cite{kokorev24_lrds} photometric cuts to select objects as LRDs. The findings are as follows:
\begin{itemize}
    \item The colors of our simulated \lrd{} population show remarkably good agreement with observations. However, our sample tends to be more compact than observed LRDs. We attribute this to the lack of PSF modeling and the little detail of our light-profile modeling. \vspace{2mm}% \danitext{(both being aspects which we did not focus on, in this work)}.\vspace{2mm}

    \item As expected, our simulated \lrds{} exhibit the characteristic ``V-shaped'' photometric profile observed with JWST NIRCam. In detail, our \lrd synthetic photometry shows a ``dip'' corresponding to filters associated to rest-frame optical-to-NIR wavelengths.
    %of the ${\tt F115W}$ and ${\tt F150W}$ filters (i.e. rest-frame UV-to-optical) shows lower flux densities compared to that of the ${\tt F277W}$, ${\tt F356W}$, and ${\tt F444W}$ filters (i.e. rest-frame optical-to-near-infrared).
    For low-mass \lrds{} ($M_\star\,{\sim}\,10^{7-9} \msun$), this spectral feature is entirely due to the Balmer break of stellar-continuum origin, with no contribution from AGN emission on average. \vspace{2mm}
    
    \item In the high-mass \lrds{} ($M_\star\,{>}\,10^9 \, \msun$), the "V-shaped" photometry arises from the combination of AGN emission (dominating the rest-frame UV-optical) and stellar emission (predominant at longer wavelengths). This preserves the characteristic ``V-shaped'' profile by combining a ``blue'' AGN with a ``red'' galaxy component. Interestingly, we find that this combination arises precisely because of the relatively small BH masses hosted by \lrds{}. Indeed, these produce low-luminosity AGN which are unable to outshine their galactic hosts at shorter wavelengths than the Balmer break. Brighter AGN (i.e. more massive BHs), on the other hand, can ``wash out'' the V-shape produced by the stellar Balmer break and be classified as \nolrds{}. We find that the torus obscuration plays a secondary role in the selection, recovering ${>}\,90\%$ of the selected \lrdAGN{} when removing the torus. \vspace{2mm}
    
    \item Overall, \lrds{} represent less than 40\% of the total detected sample, with a decreasing trend toward higher redshift. However, when considering only our AGN sample ($L_{\rm bol}^{\rm AGN}\,{>}\,10^{44}\, \rm erg/s$), \lrds{} make up 60–80\% of the entire population. Despite their relevance within the AGN class, we find that only 40\% (10\%) of \lrds{} at $z\,{=}\,4$ ($z\,{=}\,8$) can be classified as AGN, with the rest making up our galaxy population.\vspace{2mm}
    %LRDs are a subdominant class of objects in all the population, accounting for $<30\%$. However, when looking specifically in the AGN subset, they make up for $>70\%$ of them at all redshifts.

    \item \lrds{} dominate in the high end of the stellar ($M_{*}\,{\sim}\,10^{10}\,\msun$) and halo ($M_{\rm vir}\,{\sim}\,10^{11.5}\,\msun$) mass function, independently of redshift. On the other hand, \lrds{} tend to predominate at low-intermediate bolometric luminosity (i.e. $L_{\rm bol}^{\rm AGN}\,{<}\,10^{45.5}\, \rm erg/s$) and relatively small BH masses ($M_{\rm BH}\,{\lesssim}\,10^{7}\, \msun$). Taking the observed estimates for BH masses at face value, this may hint at the fact that currently discovered over-massive BHs hosted by LRDs may represent a biased sample of a larger and more complex \lrds{} population. At the same time, BH masses currently inferred from observations may be overestimated, hence alleviating the tension of our results with respect to current constraints.\vspace{2mm}
    %We find our LRD population consistently populates the high end of the stellar mass and halo mass functions, while it is the opposite for the AGN bolometric luminosity and black hole mass functions. However, we find good agreement, particularly in the latter, with observations.
    
    \item When splitting the \lrd{} class in AGN and galaxies, the latter (i.e. \lrdGal{}) systematically appear to be less massive systems ($M_\star\,{\sim}\,10^{8-9}\,\msun$, $M_{\rm vir}\,{\sim},10^{11}\, \msun$) than \lrds{} classified as AGN (i.e. \lrdAGN{}, $M_\star\,{\sim}\,10^{9.8}\,\msun$, $M_{\rm vir}\,{\sim}\,10^{11.8}\,\msun$). In addition, the \lrdGal{} sample tend to be less metal-enriched than \lrdAGN{} ($Z\,{\sim} \, 0.4 \, Z_{\odot}$ vs. $Z\,{\sim}\, Z_{\odot}$), and host smaller MBHs ($M_{\rm BH}\,{\sim}\,10^5 \msun$ vs. $M_{\rm BH}\,{\sim}\, 10^6\,\msun$).\vspace{2mm}
    
    \item The \lrdGal{} and \lrdAGN{} samples are typically disk-dominated galaxies ($D/T\,{>}\,0.6$) with characteristic sizes of 800 pc. Despite this similarity, the \lrdAGN{} population tends to host more extended bulges (${\sim}\,150$ pc vs. ${\sim}\,90$ pc) which assembled primarily via disk instabilities. We do not find significant differences between the two samples regarding their typical merger history.\vspace{2mm}
    %, as in both cases nearly all have experienced at least one major and one minor merger.
    
    %\item Our LRD sample is composed of galaxies and AGNs, that populate two completely different ranges of stellar and halo mass ($M_\star\sim 10^{10}\, {\rm M_odot};\, M_{\rm halo}\sim 10^{12}\, {\rm M_odot}$ for the AGNs and $M_\star\sim 10^{9}-10^{7}\, {\rm M_odot};\, M_{\rm halo}\sim 10^{11}-10^{10}\, {\rm M_odot}$ for the galaxies), but appear to have the same star formation activities. AGNs present more occurrence for bulges, and have bigger bulges, which can be explained for a more active minor merger history. LRDs in general are central galaxies with no major mergers.\\
    
    \item An increasing number of recent works suggested that heavy BH-seeds may be favoured to explain the origin of MBHs hosted in LRDs. Despite including a detailed model for BH-seeding which takes into account at the same time light, intermediate and heavy seeds, we do not find any indication of the prevalence of a heavy-seed origin for \lrds{} nor for the \lrdAGN{} classes. With only a few exceptions, nearly all of our simulated \lrds{} form as light seeds and reach $M_{\rm BH}\,{>}\,10^6 \, \msun$ by a combination of Eddington-limited and Super-Eddington accretion episodes.

    %\item The photometric properties of LRDs in our model are explained in two ways: i) For those without MBH contribution, no strong evidence has been found for any difference. Hence, we assume in our model that the LRD features are just an instantaneous outcome, but are not attributed to any specific property. ii) In the cases with MBH contribution, we see how the latter dominates at wavelengths lower than the Balmer break, and it ceases to do so beyond it. This implies that the blue slope of the v-shape is produced by the central engine, while the red slope is attributed to the underlying stellar population.  
\end{itemize}

This work represents an important step toward understanding the nature of photometrically selected \lrds{}, offering theoretical insight into their origin, evolutionary pathways, and physical and photometric properties. Our method allows us to pinpoint the role of accreting MBHs in shaping the photometry of high-redshift objects, building a picture in which \lrds{} appear as a rich and diverse class of objects. While this study provides foundational groundwork, it should be seen as part of a broader effort to interpret the nature of \lrds{}. Many open questions remain, particularly regarding the origin of their prominent absorption features near broad Balmer emission lines, the apparent deficit of hot dust emission and the lack of X-ray detections. These challenges do not appear to have a unique or universal explanation, underscoring the need for further development of theoretical models that jointly address galaxy and AGN emission, as well as the formation and growth of galaxies and MBHs.

%The work presented in this paper serves as a foundational ground for future studies of LRDs, giving strong background to test any possible variation. However, we acknowledge also the capabilities of \LGalaxies with AGN SEDs to be used to tackle also low-z AGNs.

\section*{acknowledgements}

D.H.C acknowledges Spanish Ministerio de Ciencia e Innovación through project PID2021-124243NB-C21 and the insightful comments from Dr. Alberto Torralba, Dr. Qi Guo and Dr. Jorryt Matthee. D.S. acknowledges support by the Fondazione ICSC, Spoke 3 Astrophysics and Cosmos Observations. National Recovery and Resilience Plan (Piano Nazionale di Ripresa e Resilienza, PNRR) Project ID CN\_00000013 "Italian Research Center on High-Performance Computing, Big Data and Quantum Computing" funded by MUR Missione 4 Componente 2 Investimento 1.4: Potenziamento strutture di ricerca e creazione di "campioni nazionali di R\&S (M4C2-19 )" - Next Generation EU (NGEU). D.I.V acknowledges the financial support provided under the European Union’s H2020 ERC Consolidator Grant ``Binary Massive Black Hole Astrophysics'' (B Massive, Grant Agreement: 818691) and the European Union Advanced Grant ``PINGU'' (Grant Agreement: 101142079). S.B. acknowledges support from the Spanish Ministerio de Ciencia e Innovaci\'on through project PID2021-124243NB-C21 and the Alexander von Humbold Foundation via a Research Fellowship for support during research stays at the Max Planck Institute for Astrophysics. P.R. acknowledges support from the Spanish Ministerio de Ciencia, Innovación y Universidades, through projects PID2022-138896NB; and the programme Unidad de Excelencia María de Maeztu, project CEX2020-001058-M.

\section*{Data Availability}
The data underlying this article will be shared on reasonable request to the corresponding author.
   
%\end{acknowledgements}

% Bibliography
\bibliographystyle{mnras}
\bibliography{./ref.bib}

\appendix

\section{AGN emission lines test} \label{appendix:EmissionLineTest}

%In this appendix, we explore the luminosity functions of the emission lines included in this work. To this end, we present in Fig.~\ref{fig:halpha_lf} the $H_\alpha$ luminosity function associated to the MBH emision of the broad component (BLR emision, top panel) and the total (BLR+NLR emision, bottom panel) at $4\,{<}\,z\,{<}\,5.5$. Compared to the data presented, the model underpredicts by a factor 2-3, either the number density of emitters. This can be caused by either the lack of emitters or the underprediction of the line luminosity (or potentially both). This is likely due to the assumptions made in the MBH line emission model. A more detailed analysis of line properties in LRDs and high-$z$ AGNs would require to revisit these asumptions, but given the aim of the paper and the role the BLR and NLR play in the model, we defer this to future work.

\begin{figure}
    \centering
    \includegraphics[width=1\columnwidth]{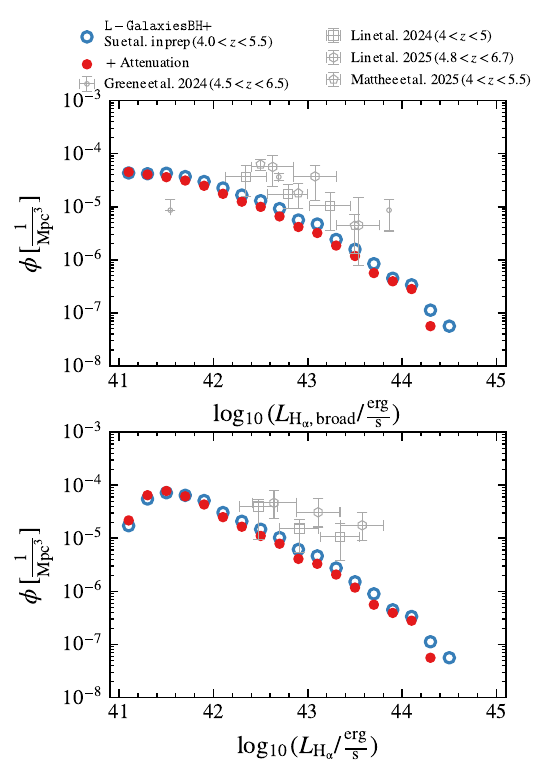}
    \caption{Luminosity functions of our simulated AGN $H_\alpha$ line. Blue empty circles show the luminosity obtained directly from the model of Su et al. (in prep), while red circles show the dust-corrected values. {\bf Upper panel}: luminosity function for the broad $H_{\alpha,\rm broad}$ component we model. {\bf Bottom panel}: luminosity function for the broad + narrow $H_\alpha$ components. We compare to data of \protect\citet{Lin2024}, \protect\citet{Lin2025}, \protect\citet{greene2024_lrds}, \protect\cite{Greene2025}, \protect\cite{matthee2024_lrds}, shown as gray simbols.}
    \label{fig:halpha_lf}
\end{figure}

In this appendix, we examine the luminosity functions of the emission lines considered in this study. Fig.~\ref{fig:halpha_lf} presents the $H_\alpha$ luminosity function associated with MBH emission for both the broad-line region (BLR, top panel) and the total emission (BLR+NLR, bottom panel) at $4\,{<}\,z\,{<}\,5.5$. When compared to the observational data, our model underpredicts the number density of emitters by a factor of approximately 2–3. This discrepancy may arise either from an underestimation of the number of emitters or from systematically lower predicted line luminosity (or a combination of both). Such differences likely stem from simplifying assumptions adopted in the MBH line-emission model. A comprehensive analysis of emission line properties in LRDs and high-$z$ AGN would require a reassessment of these assumptions. However, given the scope of this work and the specific roles of the BLR and NLR in our modeling framework, we leave this for future investigation.

\section{Redshift selection of {\tt red1} and {\tt red2}} \label{appendix:RedshiftSelection_red1_red2}

\begin{figure}
    \centering
    \includegraphics[width=1\columnwidth]{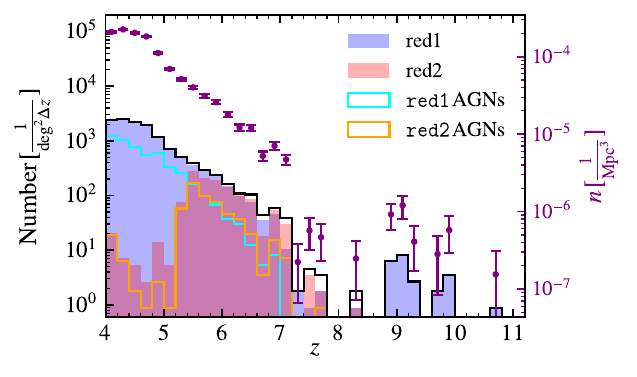}
    \caption{Redshift distribution of {\tt red1} (blue) and {\tt red2} (red). While the left y-axis corresponds to the number of objects per bin of redshift, the right one corresponds to the number density. For completeness, we have also added the distribution of AGN selected in each photometric selection.}
    \label{fig:redshift_distro_interloper}
\end{figure}

In this appendix, we present the redshift distribution of the {\tt red1} and {\tt red2} photometric selections. As shown in Fig.\ref{fig:redshift_distro_interloper}, the {\tt red1} selection primarily identifies galaxies at $z \,{<}\, 6$, although some objects at higher redshifts are also selected. These $z>6$ {\tt red1} objects make for $\sim 8.7\%$ of the total {\tt red1} population, and their number density (per unit area) is typically 1 to 1.5 dex lower than at $z \,{<}\, 6$. In contrast, the {\tt red2} selection predominantly selects objects at $z \,{>}\, 5.5$, with a pronounced peak at $z \sim 5.5$, in nice agreement with the intended behavior. While some low-redshift interlopers are also present, they correspond to $\sim 4.5\%$ of the whole {\tt red2} population and their number density is $\sim2$ orders of magnitude lower than that of the high-redshift population. 

We stress that, while these redshift distributions are those expected when applying the {\tt red1} and {\tt red2} color cuts \citep[see][]{kokorev24_lrds},  we did not impose any {\it a priori} redshift cut on our simulated sources. Therefore, generally recovering these distributions can be seen as an indirect test of our galaxy and AGN SEDs modeling.

\section{{\tt AGN} definition test} \label{appendix:alternative_agn_definition}

Recently, several works focusing on LRDs have reported high fractions of AGN within observationally selected samples \citep[e.g. $71\%$ in][]{Kocevski2025}. Therefore, it is interesting to understand how the definition of AGN we impose may affect the AGN fractions predicted by our model. As explained in the main body, we define as AGN those objects with $L_{\rm bol}^{\rm AGN}\geq10^{44}\, {\rm erg/s}$ to maximize the comparison with recent works, which generally detect AGN emission associated to LRDs above this luminosity \citep[e.g.][]{kokorev24_lrds,greene2024_lrds,Greene2025}. Nevertheless, it is interesting to explore how our results would change when lowering this threshold. Indeed, this would imply that fainter and lower massive MBHs could be classified as AGN. These objects would show AGN luminosity comparable to that of their host galaxy, effectively hindering the effects of the MBH emission onto the photometric selection.

\begin{figure}
    \centering
    \includegraphics[width=1\columnwidth]{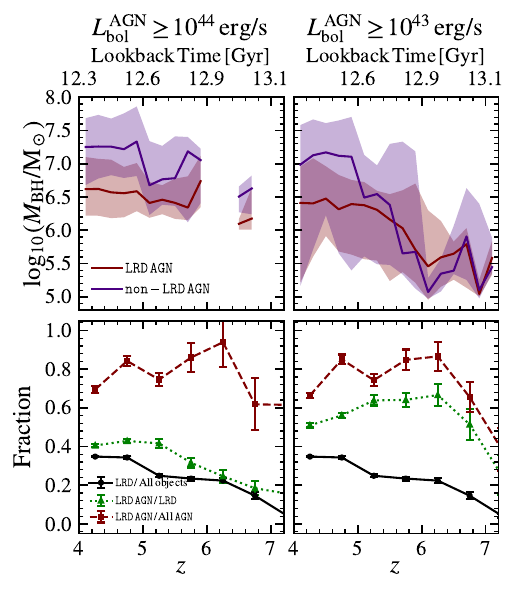}
    \caption{Black hole mass evolution (top panels) of the \lrdAGN{} and \nolrdAGN{} samples for. Bottom panels show the fractions of \lrds{} over the total number of detected systems (black points), of \lrds{} classified as AGN over the total number of detected objects classified as AGN (red points) and of \lrds{} classified as AGN over the number of \lrds{} (green points). Error bars correspond to the Poisson error. Left column correspond to the same AGN definition used in the main body of the paper ({\tt AGN}: $L_{\rm bol}^{\rm AGN}\geq10^{43}\, {\rm erg/s}$) and is shown for reference and ease the comparison. Right column correspond to AGN defined as those objects with $L_{\rm bol}^{\rm AGN}\geq10^{43}\, {\rm erg/s}$.} 
    \label{fig:alternative_agn_def}
\end{figure}

In Fig.~\ref{fig:alternative_agn_def} we show the main differences which arise in our results when lowering the threshold to $L_{\rm bol}^{\rm AGN}\geq10^{43}\, {\rm erg/s}$. In particular, we show the black hole mass evolution of \lrdAGN{} and \nolrdAGN{} (when the latter are matched in stellar mass to the \lrdAGN{}, top panels), and the fractions showing the contribution of \lrdAGN{} to the \lrd{} and AGN populations (bottom panels). These correspond to the top left panel of Fig.~\ref{fig:accreting_MBHs_props_in_lrdAGN_and_non-lrdAGN} and the bottom panel of Fig. ~\ref{fig:RedshiftDistribution}, respectively. To ease the comparison, we report these two panels in the left column of Fig.~\ref{fig:alternative_agn_def}, while the right column shows the updated results under our new AGN definition.

We note that the black hole mass evolution is now better sampled at all redshifts. In particular, owing to the larger number of sources selected by the lower $L_{\rm bol}^{\rm AGN}$ threshold, we observe continuous trends in the evolution of BH mass versus redshift. Regarding \lrdAGN{}, their typical mass appears to increase from $10^{5.5}\, {\rm M_\odot}$ to $10^{6.5}\, {\rm M_\odot}$ going from $z=7$ down to $z=4$. On the other hand, the typical mass of \nolrdAGN{} shows a more evident evolution, passing from $10^{5.5}\, {\rm M_\odot}$ at $z=7$ to $\gtrsim10^{7}\, {\rm M_\odot}$ at $z=4$. Interestingly, we observe a clear turning point of the MBH mass of \nolrdAGN{}, which increases by 1 order of magnitude between $z=5$ and $z=6$. Furthermore, we observe a brief redshift interval where the MBH mass of \nolrdAGN{} is on average smaller than that associated to \lrdAGN{}. This shift at $z\sim5.5$ is explained as follows: first, we note that at $z>6$, the contribution of the MBH to the photometry is generally $<50\%$ (see right columns of Fig.\ref{fig:lrd_photometry_LRDAGN_noLRDAGN}). Therefore the photometric selection of \lrdAGN{} and \nolrdAGN{} is driven by galaxy properties, hence the fact that MBHs in \lrdAGN{} appear more massive than those in \nolrdAGN{} can be seen as fortuitous. On the other hand, at $z<6$, MBHs start to become massive enough (i.e. their associated AGN become bright enough) to significantly affect the photometry. This can be seen by the build-up of the massive-end of our BH mass function (i.e. bottom row in Fig.\ref{fig:Properties_Mstellar_Mvir_Lbol_MBH_LRDs_no_LRDs}). This implies that MBHs start to play a more prominent role in the classification of an object as \lrd{} or \nolrd{}. In particular, we observe a clear effect of our color-based selection: in order to be classified as \lrd, an object must show a prominent V-shape. Therefore, as explained in the main body, faint AGN (i.e. less massive BHs) are more likely to be classified as \lrds{} than bright AGN (i.e.  more massive BHs). Consequently, as MBHs evolve in mass, our color selection tends to classify more massive BHs as \nolrdAGN{}, driving the increase of their average BH mass.

As for the fractions, the bottom panels of Fig.\ref{fig:alternative_agn_def} show the effect of our new AGN definition on the ratios between the number of objects within different classes.
%of: i) \lrds{} over the total number of detected objects (black points and solid line), ii) \lrdAGN{} over the total number of AGN (red points and dashed line) and iii) \lrdAGN{} over the total number of \lrds{} (green points and dotted line). 
As expected, the overall fraction of \lrds{} within the sample of detected objects is unaffected (black points and solid lines). This is because the \lrds{} definition does not depend in any way on our AGN definition. The most affected fraction is that of \lrdAGN{} within the \lrds{} sample (green points and dotted lines). At $z<5$ this gets boosted by $\sim10\%$, but at $z\sim6$ we see a $\sim30\%$ increase. This change is also expected, since a much larger number of \lrds{} is now classified as AGN under the new AGN definition. Interestingly, we observe a much smaller impact of the new AGN definition on the fraction of \lrdAGN{} within the entire AGN sample (red points and dashed line). In this case, we still measure fractions of about $\sim60-80\%$ at $4<z<7$, which is largely comparable to what we found under the more restrictive $L_{\rm bol}^{AGN}\geq10^{44}\, {\rm erg/s}$ cut. This shows that, independently of the AGN definitions we tested, the global AGN population in our model is dominated by \lrdAGN{}. That is: relatively faint AGN, whose emission cannot significantly balance the color gradient produced by the stellar Balmer-break.

Overall, this analysis suggests that the main results of our work are independent of the specific threshold we impose to define AGN. In particular: we find that \lrds{} tend to host fainter AGN (i.e. less massive BHs) than \nolrds{} also when applying the lower threshold of $L_{\rm bol}^{\rm AGN} \geq10^{43}\, {\rm erg/s}$ to define objects as AGN. This is because in our model only the contribution of bright AGN to wavelengths bluer than the stellar Balmer break can erase the V-shape feature targeted by the color-selection imposed to identify \lrds{}. Therefore, including fainter MBHs in the AGN class does not affect this result.

\section{The photometry of \lrdAGN{} and \nolrdAGN{} samples} \label{appendix:Photometry_LRDAGN_noLRDAGN}

\begin{figure*}
    \centering
    \includegraphics[width=2\columnwidth]{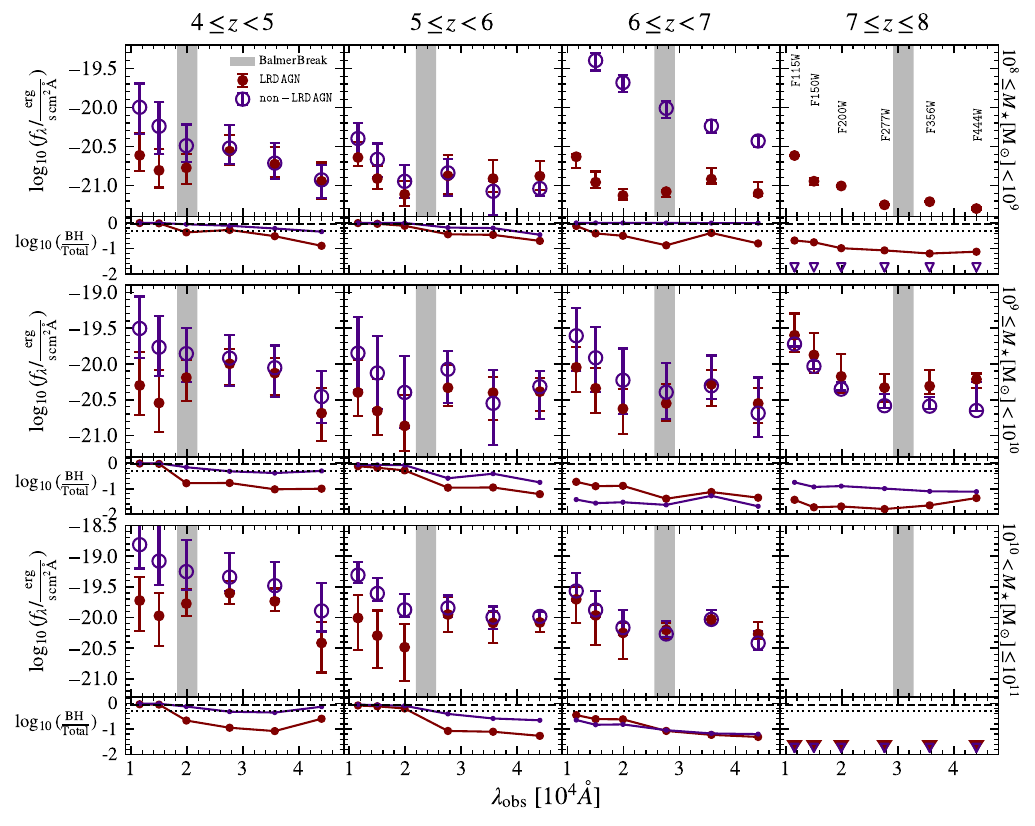}
    \caption{Median photometry of the \lrdAGN{} (dark red) and \nolrdAGN{} (purple) samples at different redshifts. The error bars correspond to the $\rm 16^{th}\,{-}\,84^{th}$ percentiles while the vertical gray shaded areas mark the Balmer break position. The small panels show the fractional contribution of the flux provided by AGN (i.e. accreting MBHs) within each JWST filter. The triangles represent the upper limits of this contribution. To guide the reader, the horizontal dashed lines highlight the 90\% and 50\% value (half of the flux is provided by the AGN). Different panels show different $M_\star$ bins, from top to bottom: $10^7 \,{<}\, M_{*} \,{<}\, 10^8 \, M_\odot$, $10^8 \,{<}\, M_{*} \,{<}\, 10^{9} \, M_\odot$ and $10^9 \,{<}\, M_{*} \,{<}\, 10^{11} \, M_\odot$.}
    \label{fig:lrd_photometry_LRDAGN_noLRDAGN}
\end{figure*}

In this appendix, we present the photometry of the \lrdAGN{} and \nolrdAGN{} samples, as shown in Fig.~\ref{fig:lrd_photometry_LRDAGN_noLRDAGN}. At $z<6$ the AGN emission dominates the flux in the {\tt F115W} and {\tt F150W} filters for all $M_\star$ bins, contributing with up to 80\%–90\% of the total flux. In contrast, the AGN contribution in the {\tt F227W}, {\tt F356W}, and {\tt F444W} filters shows a clear dependence on $M_\star$: at $M_\star \,{>}\, 10^9\, \msun$, AGN emission accounts for only ${\sim}\,10\%$ of the detected flux. However, in lower-mass systems, the AGN contribution increases significantly, varying from 30\% to 50\%. These trends reinforce the findings discussed in Section~\ref{sec:MBHs_in_LRDsAGNs_noLRDsAGNs}: AGN in low-mass galaxies (both \lrdAGN{} or \nolrdAGN{}) tend to be overmassive with respect to the overall $M_{\rm BH}-M_\star$ scaling relations of \lrds{} or \nolrds{}. Finally, at $M_\star \,{>}\, 10^9\, \msun$ we find very similar results as those presented in Fig.~\ref{fig:lrd_photometry}. This is expected, since the majority of the AGN population lies in this $M_\star$ bin.

\bsp	% typesetting comment
\label{lastpage}
\end{document}